\documentclass[11pt]{article}
\pdfoutput=1
\usepackage{jheppub}
\usepackage{bm}
\usepackage{amsmath, amsthm, color, graphicx}
\usepackage{tikz}
\usetikzlibrary{arrows, decorations.markings, shapes.arrows, patterns, calc}

\usepackage{mathrsfs}
\usepackage{amscd, amssymb, amsfonts, verbatim,subfigure, enumerate}
\usepackage[mathcal]{eucal}
\usepackage[super]{nth}

\graphicspath{{./figures/}}

\newcommand{\alg}{{\bm\Omega}}
\newtheorem*{conjecture} {Conjecture}
\newcommand{\Oo}{\mathcal{O}}
\def\sh{{\sf h}}

\newcommand{\cinfty}{C^{\infty}}
\newcommand{\CA}{{\mathcal A}}
\newcommand{\CN}{{\cal N}}
\newcommand{\CB}{{\cal B}}

\newcommand{\CT}{{\cal T}}
\newcommand{\CL}{{\cal L}}

\newcommand{\CJ}{{\cal J}}
\newcommand{\CO}{{\cal O}}
\newcommand{\CV}{{\cal V}}
\newcommand{\CX}{{\cal X}}
\newcommand{\CY}{{\cal Y}}
\newcommand{\CK}{{\cal K}}
\newcommand{\CC}{{\cal C}}
\newcommand{\CZ}{{\cal Z}}
\newcommand{\CH}{{\cal H}}

\newcommand{\bR}{{\mathbb R}}
\newcommand{\C}{{\mathbb C}}

\DeclareMathOperator{\Sym}{Sym}

\newcommand{\wt}{\widetilde}
\newcommand{\mc}{\mathcal}
 
\newcommand{\mf}{\mathfrak}

\newcommand{\zbar}{{\br{z}}}

\newcommand{\eps}{\epsilon}
\newcommand{\g}{\mathfrak{g}}

\newcommand{\what}{\widehat}

\newcommand{\til}{\widetilde}

\newcommand{\br}{\overline}

\newcommand{\Z}{\mathbb Z}

\newcommand{\op}{\operatorname}
\newcommand{\mbf}{\mathbf}
\newcommand{\mbb}{\mathbb}

\newcommand{\ip}[1]{\left\langle #1 \right\rangle}
\newcommand{\abs}[1]{\left| #1 \right|}

\newcommand{\R}{\mbb R}
\renewcommand{\d}{\mathrm{d}}

\newcommand{\dbar}{\br{\partial}}
\renewcommand{\c}{{\sf{c}}}

\newcommand{\be}{\begin{equation}}
\newcommand{\ee}{\end{equation}}
\newcommand{\pd}{\partial}
\newcommand{\ol}{\overline}

\title{Boundary Chiral Algebras and Holomorphic Twists}
\author[1]{Kevin Costello,}
\author[2]{Tudor Dimofte,}
\author[1]{Davide Gaiotto}

\affiliation[1]{Perimeter Institute for Theoretical Physics, Waterloo, ON N2L 2Y5, Canada}
\affiliation[2]{Department of Mathematics and Center for Quantum Mathematics and Physics (QMAP), University of California, Davis, CA 95616, USA}

\abstract{
We study the holomorphic twist of 3d ${\cal N}=2$ gauge theories in the presence of boundaries, and the algebraic structure of bulk and boundary local operators. In the holomorphic twist, both bulk and boundary local operators form chiral algebras (\emph{a.k.a.} vertex operator algebras). The bulk algebra is commutative, endowed with a shifted Poisson bracket and a ``higher'' stress tensor; while the boundary algebra is a module for the bulk, may not be commutative, and may or may not have a stress tensor. We explicitly construct bulk and boundary algebras for free theories and Landau-Ginzburg models. We construct boundary algebras for gauge theories with matter and/or Chern-Simons couplings, leaving a full description of bulk algebras to future work.
We briefly discuss the presence of higher A-infinity like structures.}

\begin{document}
\today
\maketitle

\section{Introduction}

Supersymmetric theories in flat space can be \emph{twisted} \cite{Witten-Donaldson, Witten-sigma} by selecting a nilpotent supercharge $Q$ and restricting one's attention to the subset of operators in $Q$-cohomology. In the BV-BRST formalism, this amounts to adding $Q$ to the BRST operator of the theory. Depending on the choice of supercharge $Q$, some or all translation generators may become exact,
rendering flat-space correlation functions independent of some combinations of coordinates of local operators. Furthermore, some or all components of the stress tensor may also become exact, so that the twisted theory can be defined on manifolds equipped with less structure than a full metric.

Most classic examples of twisted theories involve a fully \emph{topological} twist, \emph{i.e.} a choice of $Q$ such that all translations and all components of the stress tensor are exact. This includes the Donaldson-Witten \cite{Witten-Donaldson}, Vafa-Witten \cite{VafaWitten}, and Langlands \cite{KapustinWitten} twists of 4d supersymmetric gauge theories, the A and B twists \cite{Witten-sigma} of 2d $\CN=(2,2)$ gauge theories and sigma-models, and the Rozansky-Witten twist of 3d $\CN=4$ sigma-models \cite{RozanskyWitten} and its gauge-theory analogue \cite{BlauThompson2}.

Fully topological twists are somewhat special, and require a fairly large amount of supersymmetry to exist. In contrast, a generic nilpotent supercharge in Euclidean signature has cohomology that behaves \emph{holomorphically} with respect to most spacetime directions \cite{Costello2011, ElliottSafronov,EagerSaberiWalcher}. In even dimensions $d=2n$, this means correlation functions will depend holomorphically on coordinates of $\R^d \simeq \C^n$ (in a particular complex structure); while in odd dimensions $d=2n+1$, this means correlation functions will depend holomorphically on $n$ complex coordinates of 
$\R^d \simeq \C^n\times \R$ (in a particular splitting) and be independent of the final real coordinate.

The prototypical example of such a generic, \emph{holomorphic} twist is the so-called ``half-twist'' of 2d theories with at least $\CN=(0,2)$ supersymmetry \cite{Witten:1991zz, Kapustin-cdR, Witten-CDO, Nekrasov-betagamma, Gorbounov:2016oia}. 
The $Q$-cohomology of local operators in the 2d half-twist has the structure of a chiral algebra%
\footnote{We use the terms ``chiral algebra'' and ``vertex algebra'' interchangeably. \emph{A priori}, neither term implies the existence of a stress tensor or other additional structures.   We will carefully explain which structures are present in the twist of 3d $\CN=2$ theories below.},
which is related to chiral differential operators in the case of 2d (0,2) sigma models \cite{Witten-CDO} and the chiral de Rham complex \cite{cdR} in the case of 2d (2,2) sigma-models \cite{Kapustin-cdR}. Somewhat more recently, holomorphic and hybrid holomorphic-topological twists of 4d gauge theories were studied in (\emph{e.g.}) \cite{Kapustin-hol, Costello-Yangian, ElliottYoo-Langlands}.

In the current paper, our focus is on the holomorphic twist of 3d $\CN=2$ theories. The 3d $\CN=2$ algebra does not admit any topological twists, but it does have nilpotent supercharges. Every nilpotent supercharge looks essentially the same --- different choices are related by spacetime rotations and discrete symmetries --- which justifies referring to ``the'' holomorphic twist. 
In particular, for each nilpotent $Q$ there is a unique splitting of spacetime $\R^3\simeq \C_{z,\bar z}\times \R_t$ such that correlation functions of local operators in $Q$-cohomology depend holomorphically on $z$ and are independent of $t$.

The holomorphic twist of 3d $\CN=2$ gauge theories was discussed from a global perspective in \cite{ACMV}. It was explained there how the twisted theory may be defined on any 3-manifold with a transversely holomorphic foliation (THF) structure, in line with the supergravity analysis of \cite{CDFK-3d, CDFK-geometry}. It was also explained that some theories of this type arise from topological string setups involving both Lagrangian and coisotropic branes; and how partition functions may be computed via localization. Partition functions on a product spacetime $\Sigma\times S^1$ coincide with the twisted indices of \cite{BeniniZaffaroni-twisted, BeniniZaffaroni-Riemann, ClossetKim-twisted}.

Our present goal is complementary to \cite{ACMV}: we develop the algebraic structure of \emph{local operators} in the holomorphic twist of 3d $\CN=2$ theories, both abstractly and in specific examples of gauge theories with linear matter, Chern-Simons couplings, and superpotential interactions.

We will be especially interested in local operators on half-BPS boundary conditions that are compatible with the twist, and their interactions with bulk operators. The relevant boundary conditions for 3d $\CN=2$ gauge theories preserve 2d $\CN=(0,2)$ supersymmetry, and have been studied and classified in successive levels of generality by \cite{GGP-walls, OkazakiYamaguchi, GGP-fivebranes, YoshidaSugiyama, DGP-duality, BrunnerSchulzTabler}.

\subsection{General structure}
\label{sec:intro}

In Section \ref{sec:struc} of this paper we will review general arguments showing that bulk local operators in the holomorphic twist of any 3d $\CN=2$ theory with $U(1)$ R-symmetry have the structure of a chiral algebra $\CV$ that is
\begin{itemize}
\item $\Z\times \Z$ graded by rotations in the $\C$ plane (a ``spin'' or ``conformal'' grading) and by the R-symmetry (a cohomological grading)
\item commutative, meaning that OPE's are nonsingular
\item equipped with a Poisson bracket $\{\!\!\{\,,\,\}\!\!\}$ of cohomological degree $-1$ (more generally, an odd Lambda bracket)\,.
\end{itemize}
The Poisson bracket is a secondary operation that was defined in \cite{YagiOh} using topological descent. It is analogous to the secondary Poisson bracket of local operators that exists in general topological theories (\emph{cf.} \cite{Lurie}), realized by topological descent in \cite{Getzler, CostelloScheimbauer, descent}. Mathematically, $\CV$ generalizes the notion of a \emph{Poisson vertex algebra} \cite{Kac-book, FBZ-book}.

We further identify a new feature of the bulk algebra $\CV$: while $\CV$ cannot have a standard stress tensor that generates $z$-translations through the OPE (because $\CV$ is commutative), there exists a secondary stress tensor $G$ of cohomological degree $1$, which generates $z$-translations through the Poisson bracket:
\be \{\!\!\{G,\CO\}\!\!\} = \pd_z \CO \qquad \forall\, \CO\in \CV\,. \ee

The operators of the bulk chiral algebra $\CV$ are precisely those counted by the supersymmetric index, or $S^2\times_q S^1$ partition function, of 3d $\CN=2$ theories \cite{Kim-index, IY-index, KW-index}. Explicitly, the graded character of $\CV$ should coincide with the index:
\be \chi[\CV]:= \text{Tr}_{\CV} e^{i\pi R} q^J \;=\;  I(q)\,, \label{intro-bulkindex} \ee
where $J,R$ measure the spin and R-charge in $\CV$. In this sense, $\CV$ categorifies the index.

Boundary conditions $\CB$ that wrap the $\C$ direction and preserve 2d $\CN=(0,2)$ SUSY and $U(1)$ R-symmetry are compatible with the holomorphic twist and the gradings above. In the twisted theory, local operators on such a boundary condition also have the structure of a $\Z\times \Z$ graded chiral algebra $\CV_{\pd}[\CB]$. In contrast to the bulk algebra $\CV$, it is \emph{not} necessarily commutative, \emph{i.e.} there may be singular OPE's. We will explain that there is a bulk-boundary map of graded chiral algebras
\be \beta:\CV\to Z(\CV_\pd[\CB]) \hookrightarrow  \CV_\pd[\CB]  \ee
that maps bulk operators to the center of the boundary algebra, and equips $\CV_\pd[\CB]$ with the structure of a $\CV$-module. We prove that the kernel of $\beta$ is closed under the bulk Poisson bracket, which roughly amounts to the geometric statement that the support of $\CV_\pd$ in the bulk moduli space is coisotropic.
We also discuss the conditions under which the boundary algebra $\CV_\pd[\CB]$ may contain a standard stress tensor; a sufficient condition is  that the bulk algebra is fully topological, meaning that the (secondary) bulk stress tensor $G$ is $Q$-exact

The graded character of the boundary algebra $\CV_\pd[\CB]$ coincides with the 3d \emph{half-index}, or $D^2\times_q S^1$ partition function, which was defined and generalized in \cite{BDP-blocks, GGP-walls, GGP-fivebranes, YoshidaSugiyama, DGP-duality}. Explicitly,
\be \chi[\CV_\pd[\CB]] := \text{Tr}_{\CV_{\pd}[\CB]} e^{i\pi R} q^J \;=\; I\!\! I_\CB(q)\,. \label{intro-bdyindex} \ee
In this sense, boundary chiral algebras categorify the half-index.

\subsection{Examples}
\label{sec:intro-eg}

In the remainder of the paper, we aim to explicitly construct bulk and boundary algebras in a large class of Lagrangian 3d $\CN=2$ gauge (and matter) theories with Lagrangian 2d $\CN=(0,2)$ boundary conditions. Working in the BV-BRST formalism turns out to greatly simplify the analysis, and we devote Section \ref{sec:theories} to reviewing and expanding on the version of this formalism that was introduced in \cite{ACMV}.

In Section \ref{sec:freechiral} we consider the simplest bulk algebra, that of a free theory with matter valued in a complex vector space $V$. We show that the bulk algebra
\be \label{eq:freechiral} \CV \simeq \C[\CJ_\infty T^*[1]V] \ee
looks like functions on the infinite jet space of the shifted cotangent bundle of $V$. In more pedestrian terms: for a theory with free chiral multiplets $\Phi^i$, the bulk algebra is generated by the modes of complex bosons $\phi^i(z)$ in their bottom components, and modes of fermions $\psi_i(z)$ in their conjugates $\ol{\Phi}_i$. The Poisson bracket is $\{\!\!\{\phi^i,\psi_j\}\!\!\} = \delta^i{}_j$. Basic boundary conditions are labelled by complex subspaces $L\subseteq V$, and the corresponding boundary algebras are generated by the bosons $\phi^i(z)$ in $L$ and complementary fermions $\psi_i(z)$ in $L^\perp$. Geometrically, the boundary algebra consists of functions on the jet space of the conormal bundle to $L$,
\be \CV_\pd[L] \simeq \C[\CJ_\infty N^*[1]L]\,. \label{intro-freeL} \ee

The boundary algebra \eqref{intro-freeL} is commutative. In order to obtain non-commutative boundary algebras we must introduce further bulk and/or boundary interactions.

In Section \ref{sec:W}, we introduce a polynomial bulk superpotential $W:V\to \C$ (which is quasi-homogeneous of R-charge~2). We find that the bulk algebra $\CV$ is the cohomology of an algebra generated by same bosons and fermions $\phi^i(z),\psi_i(z)$ as before, with a new differential
\be Q \psi_i(z) = \frac{\pd W}{\pd \phi^i}(z)\,.  \label{intro-QW} \ee
Geometrically, this is the algebra of functions on the jet space of the derived critical locus,
\be \CV \simeq \C[\CJ_\infty \text{Crit}(W)]\,. \label{intro-CritW} \ee
The simplest boundary conditions are now labelled by subspaces $L\subseteq V$ on which $W$ vanishes; they support boundary algebras $\CV_\pd[L]$ generated by $\phi^i\in L,\psi_i\in L^\perp$, with differential \eqref{intro-QW} and a \emph{singular} OPE
\be \psi_i(z)\psi_j(0) \sim \frac{1}{z} \frac{\pd^2 W}{\pd \phi^i\pd\phi^j}(0)\,. \label{intro-WOPE} \ee
More interesting boundary conditions involve additional boundary matter and an analogue \cite{GGP-fivebranes} of the ``matrix factorizations'' that appear in B-type boundary conditions for 2d Landau-Ginzburg models \cite{KapustinLi}. We derive their boundary chiral algebras in Section \ref{sec:W-bdy-EJ}.


In Sections \ref{sec:gauge-N}--\ref{sec:gauge-D}, we add bulk gauge fields and Chern-Simons terms. The bulk chiral algebra $\CV$ of a gauge theory is nontrivial to describe due to the presence of monopole operators.%
\footnote{We expect that bulk chiral algebras in gauge theories could be constructed via a state-operator correspondence, analogous (on one hand) to the Braverman-Finkelberg-Nakajima construction \cite{Nak-I, BFN-II} in 3d $\CN=4$ theories, and (on the other hand) to the state-operator correspondence we eventually use to capture monopoloe operators on Dirichlet boundary conditions. However, we do not pursue bulk algebras further here.} %
Boundary algebras in gauge theory are, perhaps surprisingly, more tractable.

In particular, with Neumann boundary conditions on the gauge fields, there are no monopole operators at the boundary, so the boundary algebra may be computed perturbatively.%
\footnote{With some important caveats concerning the boundary degrees of freedom.} %
We find a simple result: If $\CV_\pd^{\rm matter}$ denotes the boundary algebra of the theory \emph{prior} to gauging a bulk $G$ symmetry, then $\CV_\pd^{\rm matter}$ has an action of the positive loop group $G_\C[\![z]\!]$. After gauging, the boundary algebra is obtained by taking derived $G_\C[\![z]\!]$ invariants
\be \CV_\pd = \big(\CV_\pd^{\rm matter}\big)^{G_\C[\![z]\!]}\,. \label{intro-N} \ee
Explicitly, this means adding to $\CV_\pd^{\rm matter}$ the modes $\pd_z\c,\pd_z^2\c,...$ of a $\c$-ghost, and adding an appropriate BRST differential to impose gauge invariance. We note that, in the presence of Neumann boundary conditions, bulk Chern-Simons terms are completely fixed by boundary anomaly cancellation, and do not otherwise affect the calculation.

With Dirichlet boundary conditions on the gauge fields, there \emph{are} interesting boundary monopole operators. In Section \ref{sec:gauge-D}, we begin by considering pure 3d $\CN=2$ $G$ gauge theory with (bare) Chern-Simons level $k$. Perturbatively, we find that the boundary algebra on a Dirichlet boundary condition is Kac-Moody at level $k-h$ (shifted by the dual Coxter number). Nonperturbatively, we use a state-operator correspondence to compute the boundary algebra. 

If $k \geq h$, we find that Kac-Moody is corrected to the WZW algebra

\be \CV_\pd \approx \text{KM}[G_{k-h}] \quad \leadsto \quad \text{WZW}[G_{k-h}] \quad (k\geq h)\,. \label{intro-D} \ee

For $|k|<h$ the boundary algebra is expected to be empty, because of the classic expectations that 3d $\CN=2$ $G_k$ Chern-Simons theory breaks supersymmetry for $|k|<h$ \cite{Witten-3dSUSY, BHKK,Ohta}. We give some indication of how this may come about.

In gauge theories with matter, the boundary chiral algebras on a Dirichlet boundary condition are more challenging to compute. We use a state-operator correspondence to give a precise (but not very explicit) mathematical proposal for boundary algebras with matter. We defer a more concrete analysis to future work.

In all the above examples, it is relatively straightforward to check that the characters of bulk and boundary chiral algebras agree with known 3d indices and half-indices, as in \eqref{intro-bulkindex}, \eqref{intro-bdyindex}. Indeed, some of the boundary algebras above already appeared in the literature, having been inferred from computations of boundary anomalies and half-indices. Examples include some Neumann algebras of the form \eqref{intro-N} for boundary conditions supporting 2d free fermions \cite{DHSV, GGP-fivebranes, ArmoniNiarchos}, and the WZW algebras on Dirichlet b.c. \cite{DGP-duality}.

\subsection{Bulk and boundary dualities}
\label{sec:intro-dualities}

Bulk and boundary chiral algebras are independent of energy scale.%
\footnote{A more precise statement is made in Section \ref{sec:RG}.} %
Indeed, they are some of the most sensitive observables of 3d $\CN=2$ theories and boundary conditions that have this property. They thus provide a powerful test of IR dualities --- stronger than any computations of numerical observables (partition functions, indices) performed so far.

For every pair of IR dual 3d $\CN=2$ theories $\CT,\CT'$, we expect an equivalence
\be \CV[\CT] \simeq \CV[\CT'] \ee
of bulk algebras. Similarly, given theories and boundary conditions $(\CT,\CB)$, $(\CT',\CB')$ that are IR dual, we expect an equivalence of boundary algebras
\be \CV_\pd[\CT,\CB] \simeq \CV_\pd[\CT',\CB']\,. \ee
The basic form of ``equivalence'' implied in these statements is an isomorphism of algebras, after taking $Q$-cohomology. A stronger expected form of equivalence is quasi-isomorphism of algebras \emph{prior} to taking $Q$-cohomology, which is discussed in Sections \ref{sec:dg} and \ref{sec:RG}.

In this paper, we discuss several simple examples of equivalences of boundary algebras, which are already quite nontrivial. In Section \ref{sec:flip1} and \ref{sec:W-bdy-EJ} we consider pairs of dual boundary conditions for a single bulk matter theory, related by ``flip'' operations that swap boundary conditions on bulk fields at the expense of adding boundary matter. We prove that the associated boundary algebras are quasi-isomorphic.

In Sections \ref{sec:SQEDXYZduality} and \ref{sec:SQEDXYZduality2} we propose an equivalence of boundary algebras resulting from the classic duality between 3d $\CN=2$ SQED and the XYZ model \cite{AHISS}, with two pairs of dual boundary conditions found in \cite{DGP-duality}. 
A brief summary of the first pair (Sec.  \ref{sec:SQEDXYZduality} ) is as follows. The boundary algebra in SQED is generated by bosons $\phi(z),\widetilde \phi(z)$ of $U(1)$ gauge charge $\pm1$, by boundary fermions $\Gamma(z),\widetilde \Gamma(z)$ of gauge charge $\pm 1$, and by modes of a ghost $\pd_z^n\c$, $n\geq 1$. The only nonvanishing OPE is the standard one for a 2d complex fermion,
\be \Gamma(z)\widetilde \Gamma(0) \sim \frac{1}{z}\,, \ee
and there is a BRST differential that imposes gauge invariance cohomologically. In the XYZ model, the boundary algebra is simply generated by a boson $X(z)$ and two fermions $\psi_Y(z),\psi_Z(z)$, with trivial differential, and an OPE
\be \psi_Y(z) \psi_Z(0) \sim \frac{1}{z} X(0) \ee
due to a bulk superpotential as in \eqref{intro-WOPE}. We propose, and prove, that the XYZ algebra is isomorphic to the BRST cohomology of the SQED algebra upon identifying
\be X(z)\,,\; \psi_Y(z)\,,\; \psi_Z(z)  \quad\leftrightarrow\quad    (\phi\widetilde \phi)(z)\,,\; (\Gamma\widetilde \phi)(z)\,,\; (\widetilde \Gamma \phi)(z)\,. \ee

We also propose some generalizations of this result to more complicated dual examples of theories and boundary conditions. 
There is a vast web of dualities of bulk 3d $\CN=2$ theories that has been developed in the literature (beginning decades ago in \cite{IS, dBHOO, dBHO1, dBHOY2, AHISS, Aharony-duality}), which extend far beyond the simple SQED/XYZ example. Dualities of boundary conditions were explored more recently in \cite{BDP-blocks, YoshidaSugiyama, GGP-walls, GGP-fivebranes, OkazakiYamaguchi, DGP-duality, Okazaki-abelian}. It should be extremely interesting (and highly nontrivial) to identify the chiral algebras and the equivalences among them that populate this vast web.

\subsection{Other connections and future directions}

We outline a few other motivations for studying bulk and boundary chiral algebras, analogues in other parts of the literature, and potentially exciting future directions.

\subsubsection{2d B-model}

If we compactify a holomorphically twisted 3d $\CN=2$ gauge theory along a circle in the holomorphic direction (\emph{e.g.} viewing 3d spacetime as $\C^* \times \R$), we obtain a 2d $\CN=(2,2)$ theory in the fully topological B-twist. Much of the structure of bulk and boundary chiral algebras discussed above may thus be interpreted as a ``chiral'' or ``loop space'' version of the 2d B-model.

For example, in the 2d B-model, bulk local operators form an ordinary graded-commutative algebra $\CA$, with Poisson bracket of degree $-1$ (the Gerstenhaber bracket). In a sigma-model with target $\CX$, $\CA$ is the algebra of polyvectorfields (with Schouten-Nijenhuis bracket), geometrically expressed as functions on the shifted cotangent bundle, $\CA\simeq \C[T^*[1]\CX]$. In a Landau-Ginzburg model with superpotential $W$, $\CA \simeq \C[\text{Crit}(W)]$ is the algebra of functions on the derived critical locus of $W$ \cite{Vafa-LG, Dyckerhoff2011}.
In 3d, we find in general that $\CV$ is a commutative \emph{chiral} algebra with Poisson bracket. The forms of $\CV$ for Landau-Ginzburg models \eqref{intro-CritW} are obvious loop-space generalizations of the B-model algebras.

The analogy extends to boundary conditions. Boundary algebras $\CA_\pd$ in the B-model are not necessarily commutative; they form modules for the bulk $\CA$ via a bulk-boundary map $\beta: \CA\to Z(\CA_\pd)$; and the kernel of the bulk-boundary map is closed under Poisson bracket.
In 3d we find chiral analogues of all these statements.  More so, many of our actual boundary chiral algebras, such as \eqref{intro-N} for Neumann b.c. in gauge theories, are straightforward chiral generalizations of familiar B-model results. 

(The relation between 2d and 3d is less direct in the case of bulk gauge theories, or Dirichlet boundary conditions for gauge theories such as \eqref{intro-D}, due to the presence of nonperturbative monopole operators in 3d.)

The analogy with the B-model moreover suggests that there are at least two additional pieces of higher structure present in the 3d holomorphic twist that go beyond the current paper:

\begin{enumerate} 
	\item In the 2d B-model, bulk operators (before taking cohomology) are endowed with an $E_2$ algebra structure which may include higher $L_\infty$ operations, and boundary algebras are endowed with an $E_1$ algebra structure which may include higher $A_\infty$ operations. We expect in general that bulk $\CV$ and boundary $\CV_\pd[\CB]$ chiral algebras of 3d theories also have such higher operations. On the boundary, the relevant structure would be an $A_\infty$ analog of a vertex algebra; this is a structure that has yet to be fully defined mathematically. We make some brief comments in Sections \ref{sec:dg} and \ref{sec:A-comment} as to how they may arise.

In the 2d B model with flat space target, a deep theorem of Kontsevich \cite{Kon97}, his formality theorem%
\footnote{Here the correct $L_\infty$ structure on bulk operators is the natural one on the Hochschild cochains of the algebra of functions on the target, or equivalently the transferred $L_\infty$ structure on the Hochschild cohomology.}, %
tells us that all higher $L_\infty$ operations vanish.    We do not know whether or not a similar ``chiral'' analog of Kontsevich's formality theorem can be expected to hold; it is certainly an interesting question. 

\item In the B-twist of 2d sigma-models or LG models, the bulk-boundary map $\beta: \CA \to Z(\CA_\pd[\CB])$ has a derived generalization that maps bulk operators onto the Hochschild cohomology (\emph{a.k.a.} derived center) of every boundary algebra \cite{Kon95a,KonSoi06, Costello-TCFT,KapustinRozansky, Lurie},
\be \beta_{\rm der} : \CA \longrightarrow HH^\bullet(\CA_\pd[\CB])\,. \label{intro-HH} \ee
More so, for sufficiently rich $\CB$,%
\footnote{namely: for a generator of the category of boundary conditions} %
the map $\beta_{\rm der}$ is an \emph{isomorphism}, allowing the bulk algebra $\CA$ to be fully reconstructed from the boundary algebra $\CA_\pd[\CB]$.  In axiomatic approaches to TFT \cite{KonSoi06, Costello-TCFT, Lurie} this is taken to be the definition of the algebra of bulk operators.  

		We expect an analogous statement to hold in the case of the 3d holomorphic twist, and comment on it briefly in Section \ref{sec:dg}.   K. Zeng \cite{Zeng-PSIthesis} has verified this proposal in a number of non-trivial cases. Such a statement would be particularly powerful in situations where the calculation of $\CV_\pd[\CB]$ is simpler than the calculation of $\CV$, \emph{e.g.} due to the absence of boundary monopole operators in gauge theories with Neumann b.c.. 
\end{enumerate}


\subsubsection{3d $\CN=4$}

The analysis and techniques of this paper extend immediately to 3d $\CN=4$ gauge theories, simply by viewing them as 3d $\CN=2$.

The holomorphic twist of a 3d $\CN=4$ theory moreover admits two deformations to either A-type or B-type topological twists of 3d $\CN=4$. These ``deformations'' should manifest via additional differentials $Q_A$, $Q_B$ in the bulk chiral algebra $\CV$ of a 3d $\CN=4$ theory, whose cohomologies $H^\bullet(\CV,Q_A)$ and $H^\bullet(\CV,Q_B)$ agree with the A-type and B-type topological algebras of 3d $\CN=4$. It would be interesting to explore these differentials in (say) 3d $\CN=4$ gauge theories, where $H^\bullet(\CV,Q_A)$ and $H^\bullet(\CV,Q_B)$ would be the Coulomb-branch and Higgs-branch  chiral rings.

3d $\CN=4$ theories may further admit 2d $\CN=(0,4)$ boundary conditions compatible with the holomorphic twist and either the A or B deformations. Boundary chiral algebras on these special $\CN=(0,4)$ boundary conditions were constructed in \cite{CostelloGaiotto-VOA}, and their Hochschild cohomology was used by \cite{VOAExt} to recover bulk chiral rings (implementing an analogue of \eqref{intro-HH}).


\subsubsection{Line operators}

The constructions of this paper should admit a further categorification in terms of line operators. 
3d $\CN=2$ theories admit a large collection of half-BPS line operators, which in gauge theories include Wilson lines, vortex lines, and combinations thereof \cite{DGG, KWY-Wilson, KWY-vortex, DOP-vortex}. Such line operators are compatible with the holomorphic twist if they extend in the real/topological direction%
\footnote{More precisely: on a 3-manifold with a transverse holomorphic foliation structure, the line operators should be supported on integral flows of the transverse vector field $\pd/\pd t$.} %
 \cite{ACMV}. More so, in the holomorphic twist, the line operators are expected to generate a \emph{chiral category} $\CC$ \cite{Gaitsgory-chiralcat, Raskin-chiralcat}. This chiral category will encode all bulk information: for example, the bulk algebra arises as the (derived) endomorphism algebra of the trivial line, $\CV \simeq \text{Hom}_\CC(1\!\!1,1\!\!1)$. 

For a free 3d $\CN=2$ theory whose fields are chirals living in a vector space $V$,  the category of lines appears to be the (derived) category of coherent sheaves on the loop space $V(\!(z)\!)$:
\be \CC \simeq \text{Coh}(V (\!(z)\!))\,, \ee
with the trivial line represented as the structure sheaf of the positive loops, $1\!\!1 = \CO_{V[\![z]\!]}  = \CO_{\CJ_\infty V}$. We invite the reader to recover \eqref{eq:freechiral} from this statement. In the case of a gauge theory with gauge group $G$ and matter in the representation $V$, we expect that the chiral category of bulk lines is the category of $G(\!(z)\!)$-equivariant coherent sheaves on $V(\!(z)\!)$.      It should be fascinating to concretely identify $\CC$ in more general examples.

An analogous discussion of categories of line operators in topologically twisted 3d $\CN=4$ theories was recently initiated in \cite{DGGH, HilburnYoo}. The category $\CC$ in the holomorphc twist of 3d $\CN=4$ theories (viewed as 3d $\CN=2$) also seems to have arisen in work of Aganagic-Okounkov \cite{AO-StringMath}.
  
%



\subsubsection{2d A-model and holomorphic blocks}

Let $\C\times_q S^1$ denote the $\C$ fibration over a topological $S^1$, with monodromy $z\to q z$. 
Upon sending the radius of $S^1$ to zero size in a careful scaling limit, 
the holomorphic twist of a 3d $\CN=2$ gauge theory in this geometry reduces to the Omega-deformed A-twist of a 2d $\CN=(2,2)$ gauge theory on~$\C$. (This much the same way that 5d ``K-theoretic'' instanton partition functions reduce to 4d instanton partition functions \cite{Nekrasov-Omega}).

Upon choosing a supersymmetric vacuum $\nu$ at infinity, one may define the partition function $\CZ_\nu$ of a holomorphically twisted 3d $\CN=2$ theory on $\C\times_q S^1$. These are known in the literature as ``K-theoretic vortex partition functions'' \cite{Shadchin,DGH} or ``holomorphic blocks'' \cite{BDP-blocks}.
For sigma-models with target $\CX$, the partition functions are related mathematically to equivariant quantum K-theory of $\CX$ \cite{GiventalLee} (see \emph{e.g.} \cite{JockersMayr} for recent developments on this relation). In the case of 3d $\CN=4$ gauge theories, partition functions $\CZ_\nu$ played a central role in Aganagic-Okounkov's construction of elliptic stable envelopes and applications to quantum K-theory \cite{AganagicOkounkov-elliptic, AganagicOkounkov-quasimaps}.

In the context of our current paper, we expect that, by state-operator correspondence, $\CZ_\nu$ coincides with the character of the boundary chiral algebra of a 2d $\CN=(0,2)$ boundary condition $\CB_\nu$ defined by the vacuum $\nu$,
\be \CZ_\nu = \chi[\CV_\pd[\CB_\nu]]\,. \label{intro-Bnu} \ee
In a 3d $\CN=2$ gauge theory, $\CB_\nu$ would be constructed by solving the 2d $\CN=(0,2)$ BPS equations on a half-space $\C\times [0,\infty)$ with vacuum $\nu$ at $\infty$.%
\footnote{Such half-BPS boundary conditions defined by vacua appeared in classic work of Hori-Iqbal-Vafa on the 2d A-model \cite{HoriIqbalVafa}. They are sometimes called ``thimble branes.'' They were described in the general  setting of massive 2d theories by \cite{GMW1, GMW2}. In 3d $\CN=4$ theories, boundary conditions defined by vacua were discussed in detail in \cite[Sec 4]{BDGH} and \cite{Dedushenko-gluing1, Dedushenko-gluing}. In 3d $\CN=2$ theories, many examples of $\CN=(0,2)$ boundary conditions defined by vacua have appeared in \emph{e.g.} \cite{GGP-walls, OkazakiYamaguchi, JockersMayr, DGP-duality}. }

It would be very interesting to use this circle of ideas to relate boundary chiral algebras to K-theoretic Gromov-Witten theory and to elliptic stable envelopes.


\subsubsection{2d half-twist}

We may also consider the straighforward compactification on a circle in the topological direction, \emph{i.e.} an untwisted product $\C\times S^1$; or compactification on an interval $\C\times [0,1]$ with two boundary conditions. A holomorphically twisted 3d $\CN=2$ theory reduces in these cases to a half-twisted 2d $\CN=(2,2)$ or $\CN=(0,2)$ theory, respectively

In principle, these compactifications can be used to relate bulk and boundary chiral algebras of 3d theories to the chiral algebras in the half-twists of 2d theories. In practice, this can be a subtle and difficult procedure, as the 2d algebras acquire additional contributions from line operators (and corrections from line-like instantons) that wrap $S^1$ or $[0,1]$. This should not be surprising, since (\emph{e.g.}) half-twisted $\CN=(0,2)$ models have famously subtle instanton corrections studied in many places including \cite{DSWW, SilversteinWitten-conformal, BasuSethi-instantons, BeasleyWitten-instantons, BertoliniPlesser} (see \cite{TanYagi} for an analysis of instanton effects specifically on the half-twisted chiral algebra). 

We will give an elementary example of interval compactification in a free theory in Section \ref{sec:interval}; even here, there are contributions from line operators, but no instantons. We explain there how singular OPE's in the compactified 2d algebras are induced from Poisson brackets in the 3d bulk.
Other occurrences of interval compactifications (with contributions from line operators) in the recent literature include \cite{GaiottoRapcak, ProchazkaRapcak} and \cite{GGP-fivebranes, FeiginGukov, DP-4simplex} in the context of 4d-2d correspondence (see below). It would be satisfying to conduct a more systematic analysis of such interval and circle compactifications.



\subsubsection{Hilbert spaces}

Hilbert spaces in the holomorphic twist of 3d $\CN=2$ gauge theories, on geometries of the form $\Sigma \times \R$ with $\Sigma$ a closed Riemann surface, were constructed in \cite{GPV, BullimoreFerrari}, categorifying the twisted indices of \cite{BeniniZaffaroni-twisted, BeniniZaffaroni-Riemann, ClossetKim-twisted}. These Hilbert spaces should have several interesting interactions with bulk and boundary chiral algebras.

For example, every Hilbert space $\CH(\Sigma)$ on a Riemann surface provides a representation of the bulk algebra $\CV$. Similarly, a half-space geometry $\Sigma\times \R_{t\geq 0}$, gives us a pairing between outgoing states $\langle \Psi|\in \CH(\Sigma)^*$ at $t\to \infty$ and boundary conditions at $t=0$. In particular, for each $\langle \Psi|\in \CH(\Sigma)^*$ and each boundary condition $\CB$, we can define correlation functions of operators $\CO_i\in \CV_\pd[\CB]$. This shows that there should be a map from $\CH(\Sigma)^*$ to the conformal blocks/chiral cohomology of $\CV_\pd[\CB]$. This should be further explored.


\subsubsection{3d-3d and 4d-2d correspondences}

A large part of the motivation for this paper comes from a large body of closely related work on supersymmetric theories associated to 3- and 4-manifolds. We hope there will be a rich interplay with constructions in this paper.

We recall that compactification of M5 branes on a closed manifold $M$ of dimension $d$ defines a $(6-d)$-dimensional theory $T[M]$ whose supersymmetry depends on the normal bundle geometry of $M$. If $M$ is noncompact, with asymptotic boundary, then $T[M]$ becomes a boundary condition for $T[\pd M]$.%
\footnote{One may also consider manifolds with corners, as in \cite{GGP-walls, GGP-dualitydefects, LSW-coupling, Gukov-trisecting, GaiottoRapcak}.} %
With appropriate choices of normal bundles, one finds:
\be \label{intro-M5} 
\begin{array}{lcl}
\text{Riemann surface $\Sigma$} && \text{4d $\CN=2$ thy $T[\Sigma]$ (complex) \quad \cite{Witten-solutions, GMN, Gaiotto-dualities}} \\
\text{closed 3-manifold $M^3$} && \text{3d $\CN=2$ theory $T[M^3]$ (diffeo=top)  \quad \cite{DGH,YamazakiTerashima,DGG,CCV}} \\
\text{3-manifold $M^3$, $\pd M^3=\Sigma$} &\leadsto\quad& \text{3d $\CN=2$ b.c. $\CB[M^3]$ for $T[\Sigma]$ (diffeo=top)} \\
\text{closed 4-manifold $M^4$} && \text{2d $\CN=(0,2)$ thy $T[M^4]$ (diffeo) \quad \cite{GGP-fivebranes}} \\
\text{4-manifold $M^4$, $\pd M^4 = M^3$} && \text{2d $\CN=(0,2)$ b.c. $\CB[M^4]$ for $T[M^3]$ (diffeo)} 
\end{array}
\ee
In the IR, the various theories $T[M]$ only depend on part of the geometry of $M$, as indicated. Moreover, all the theories above admit a holomorphic twist whose cohomology is a diffeomorphism invariant of $M$ (including the case $M=\Sigma$, which we touch on further below).

For $M=M^3$ a 3-manifold, one lands precisely on the class of 3d $\CN=2$ theories considered in this paper. The bulk chiral algebra $\CV[M^3]$ of $T[M^3]$ is a topological invariant whose character reproduces the ``3d index'' of 3-manifolds first discussed in \cite{DGG-index}.
It would be interesting to explicitly construct $\CV[M^3]$ in the original abelian Chern-Simons-matter theories  associated to ideal triangulations in \cite{DGG, CCV} and Dehn fillings \cite{GangTachikawaYonekura, GangYonekura}, or the newer classes of 3d abelian and nonabelian theories associated to graph manifolds in \cite{GPV, GPPV, EKSW} and to mapping tori in \cite{mapping-blocks}. In all these cases, bulk monopole operators will play an important role, and techniques beyond those of the current paper will be necessary to describe them.

For $M^4$ a 4-manifold with boundary $M^3$, one obtains a 2d $\CN=(0,2)$ boundary condition $\CB[M^4]$ for $T[M^3]$, and thus a boundary chiral algebra $\CV_\pd[\CB[M^4]]$. This boundary chiral algebra played a central role in the original 4d-2d constructions of \cite{GGP-fivebranes}; for $M^4$ an ALE space, it was also connected to classic work of Nakajima on instantons and Kac-Moody algebras \cite{Nakajima-ALE}.

For $M^4=  W\cup_{M^3} W'$ a closed 4-manifold obtained by gluing two 4-manifolds $W,W'$ along a common boundary $M^3$, the theory $T[M^4]$ should arise from interval compactification of $T[M^3]$ between boundary conditions $\CB[W],\CB[W']$ (of the same sort discussed abstractly above). 
Examples of such compactification appeared in \cite{GGP-fivebranes} and were generalized recently in \cite{FeiginGukov}. Interval compactifications related to triangulations of 4-manifolds also appeared recently in \cite{DP-4simplex}. A more direct construction of abelian $T[M^4]$ was given in \cite{DedushenkoGukovPutrov}.


\subsubsection{Homological blocks and modularity}

Another important relation between boundary chiral algebras and geometry involves the \emph{homological blocks} of Gukov-Putrov-Vafa  \cite{GPV}. Given a 3-manifold $M^3$, the homological blocks $\CZ_\nu[M^3]$ are $\C\times_q S^1$ partition functions of the 3d $\CN=2$ theory $T[M^3]$, labelled by a certain distinguished set of vacua $\nu$ at infinity. Thus, as in \eqref{intro-Bnu}, they are characters of boundary chiral algebras
\be \CZ_\nu[M^3] = \chi[\CV_\pd[\CB_\nu]]\,. \ee
The authors of \cite{GPV} proposed a simple, concrete way to combine homological blocks into the Witten-Reshetikhin-Turaev invariant of a 3-manifold \cite{Witten-Jones, RT}. The underlying spaces $\CV_\pd[\CB_\nu]$ then furnished (in principle) a categorification of the WRT invariant. 

Many examples of homological blocks have now been exhibited, for various classes of 3-manifolds, \emph{e.g.} \cite{GPPV, GukovManolescu, Park-Zhat, Chung-Seifert, mapping-blocks}. However, underlying chiral algebras $\CV_\pd[\CB_\nu]$ are only known in very few of these examples.
The structure of bulk and boundary chiral algebras developed in this paper could help inform further study of the $\CV_\pd[\CB_\nu]$.
It may also shed some light on the striking observations of \cite{3dmodularity, 3dmodularity2} that characters of many $\CV_\pd[\CB_\nu]$ are modular or modular-like; for example, one might hope that different types of modularity are linked with properties/existence of a boundary stress tensor.

\section{Chiral algebras in the holomorphic-topological twist}
\label{sec:struc}

We begin by reviewing the 3d $\CN=2$ and 2d $\CN=(0,2)$ SUSY algebras, and the structure of local operators that one expects to find in the holomorphic twist of 3d $\CN=2$ theories with $\CN=(0,2)$ boundary conditions. We do not yet specialize to a particular theory and boundary condition, and only make some general assumptions about the theories we will work with, such as the existence of an unbroken R-symmetry.

Many of the structures discussed here --- such as the existence of a shifted Poisson bracket on the algebra of bulk operators, the bulk-boundary map, and the conditions for existence of stress tensors --- are \emph{not} strictly necessary for understanding the constructions of boundary chiral algebras in the remainder of the paper. Some readers may want to move on after Section~\ref{sec:SUSY}. However, these structures put interesting constraints on the form of boundary chiral algebras, which we will revisit in examples, and which should be useful in generalizations.

\subsection{SUSY algebra and twisting}
\label{sec:SUSY}

We work on three-dimensional Euclidean spacetime $M$, which in this paper we usually take to be a flat space 
$M=\R^3\simeq \C_{z,\bar z}\times \R_t$, split as a product of a complex plane and a real direction $\R_t$. When we introduce boundary conditions, we will modify this to a half-space $\C_{z,\bar z}\times \R_{t\geq 0}$.

The 3d $\CN=2$ SUSY algebra in flat space $\C_{z,\bar z}\times \R_t$ has four odd generators $Q_\pm$, $\ol Q_\pm$ satisfying $\{Q_\alpha,\ol Q_\beta\} = i\sigma_{\alpha\beta}^\mu\pd_\mu$, or in components
\be \label{SUSY-3d} \begin{array}{c} \{Q_+,\ol Q_+\} = -2i \pd_{\bar z}\,,\qquad \{Q_-,\ol Q_-\} = 2i \pd_z\,,\\[.2cm] \{Q_+,\ol Q_-\}=\{Q_-,\ol Q_+\} = i\pd_t\,. \end{array} \ee
We are interested in the cohomology of the supercharge $\boxed{Q:=\ol Q_+}$.
Since the derivatives $\pd_{\bar z}$ and $\pd_t$ are $Q$-exact, the correlation functions of operators in $Q$-cohomology will be independent of $\bar z$, $t$; however, they may (and typically will) have nontrivial, holomorphic $z$ dependence. 
We thus refer to taking $Q$-cohomology as working in the holomorphic (or holomorphic-topological) twist --- holomorphic in $z$, topological in $t$.

The 3d $\CN=2$ algebra has a $U(1)_R$ R-symmetry, and we work in conventions such that the supercharges have half-integral R-charge. Under $U(1)_R$ and the $\text{Spin}(2)_E\subset \text{Spin}(3)_E$ subgroup of the Lorentz group that rotates the $\C_{z,\bar z}$ plane, the holomorphic coordinates and supercharges have charges
\be \label{SUSY-charges} \begin{array}{c|cccc|cc}
 & \ol Q_+ & Q_+ & \ol Q_- & Q_- & \d z & \d\bar z \\\hline
U(1)_R & 1 & -1 & 1 & -1 & 0 & 0 \\
\text{Spin}(2)_E &  \tfrac12 & \tfrac12 & -\tfrac12 & -\tfrac12 & 1 & -1 \\
U(1)_J & 0 & -1 & 1 & 0 & -1 & 1
\end{array}
\ee
Thus the supercharge $Q = \ol Q_+$ is a scalar under the anti-diagonal subgroup $U(1)_J \subset \text{Spin}(2)_E\times U(1)_R$ whose charge $J$ is related to charges $J_0,R$ for $\text{Spin}(2)_E\times U(1)_R$ as
\be J :=  \frac R2-J_0\,. \label{twist-hom} \ee

Henceforth, we will simply refer to the redefined $U(1)_J$ charge as ``spin.'' The R-symmetry also plays an independent and important role, giving rise to a cohomological grading. Specifically, defining
\be  \text{cohomological degree} := R\,,\ee
we see that the differential $Q$ has degree $+1$.

The holomorphic twist of a 3d $\CN=2$ theory may be defined more generally on a three-manifold $M$ with a transversely holomorphic foliation (THF) structure, \emph{i.e.} a manifold that looks locally like $\C_{z,\bar z}\times \R_t$. On such a manifold, the structure group of the tangent bundle is reduced to $\text{Spin}(2)_E$, and so one can use a homomorphism $\text{Spin}(2)_E\to U(1)_R$ as in \eqref{twist-hom} to redefine the Lorentz group so that $Q$ remains a scalar. Nevertheless, our focus in this paper will be on local structures of operator algebras, so restricting to flat Euclidean spacetime will suffice.

We would like to study boundary conditions for 3d $\CN=2$ theories, localized at $t=0$, which preserve $Q$ and $U(1)_R$ symmetry. We further assume that these are supersymmetric boundary conditions, so that they actually preserve a full 2d SUSY subalgebra of 3d $\CN=2$ that contains $Q$. The unique subalgebra with this property is the 2d $\CN=(0,2)$ algebra generated by $Q$ and $Q_+$,
\be \label{SUSY-2d} \{Q, Q_+\} = -2i\pd_{\bar z}\,.\ee
Thus, we are led to consider half-BPS 2d $\CN=(0,2)$ boundary conditions. With respect to the boundary SUSY algebra, the holomorphic twist reduces to the more familiar ``half-twist'' of 2d $\CN=(0,2)$ theories \cite{Witten:1991zz, Kapustin-cdR, Witten-CDO, Nekrasov-betagamma, Gorbounov:2016oia}.

\subsubsection{Nilpotence variety}
\label{sec:nilp}

It may be interesting to observe that the holomorphic twist discussed above is completely generic in 3d $\CN=2$ theories: every nilpotent supercharge in the 3d $\CN=2$ algebra is equivalent to $Q= \ol Q_+$, up to a spacetime rotation and/or a discrete symmetry. We briefly explain this.

A general analysis of nilpotent supercharges in various dimensions was carried out recently in \cite{ElliottSafronov, EagerSaberiWalcher}. In the 3d $\CN=2$ algebra, we find that nilpotent supercharges are all of the form $aQ_++bQ_-$ or $c\ol Q_++d\ol Q_-$.  Thus the ``nilpotence variety'' parametrized by $a,b,c,d$ has the structure of a cone over $\mathbb{CP}^1\sqcup \mathbb{CP}^1$. Overall scaling is unimportant for defining cohomology.

The cohomology of a supercharges $aQ_++bQ_-$ is topological along a line in $\R^3$ spacetime (determined by the ray $a/b\in \mathbb{CP}^1$), and antiholomorphic in the transverse plane. Similarly, the cohomology of a supercharge  $c\ol Q_++d\ol Q_-$ is topological along a line (determined by $c/d\in \mathbb{CP}^1$) and holomorphic in the transverse plane. Euclidean $SO(3)$ spacetime rotations act by rotating each copy of $\mathbb{CP}^1$. 

Once we fix a particular splitting $\R^3\simeq \C\times \R$, and require holomorphic (rather than anti-holomorphic) cohomology along $\C$, we are left with exactly two nilpotent supercharges up to scaling, \emph{e.g.} $\ol Q_+$ and $\ol Q_-$. They are related by `$PT$' symmetry, which acts as the antipodal map on $\mathbb{CP}^1$.

\subsection{Bulk chiral algebra}
\label{sec:chiral-bulk}

Now suppose we have a 3d $\CN=2$ theory with conserved $U(1)_R$ symmetry. We outline some of the properties of local operators expected to be present in its holomorphic twist.

Let $\text{Ops}$ denote the vector space of all local operators in the theory, supported at a given point. (The choice of point is not important, due to translation invariance.) The space $\text{Ops}$ is graded by $U(1)_R\times U(1)_J$, and endowed with the action of the differential $Q$. We will not assume that $U(1)_R$ and $U(1)_J$ charges take integral or half-integral values.%
\footnote{When $U(1)_J$ charges are non-integral, extra topological conditions are required in order to define the holomorphic twist globally, on 3-manifolds with THF structure. Any $3$-manifold with a THF structure has a canonical $SO(2)$ principal bundle, defined as the orthogonal complement of the real codimension $2$ foliation. This $SO(2)$ gets identified with $U(1)_J$. If the $U(1)_J$ charges of fields/operators can all be chosen to be rational, belonging to $\frac{1}{d}\Z$ for some $d$, then we must require the structure group of the canonical bundle to be equipped with a reduction to its $d$-fold cover. That is, we need a transverse $d$-spin structure. This will not be important in the current paper, since we are working in flat space. (In fact, even irrational charges are acceptable for us.)} %
We then denote the $Q$-cohomology of local operators as
\be \CV := H^\bullet(\text{Ops},Q)\,. \ee

We claim that the space $\CV$ has the structure of
\begin{itemize}
\item a chiral algebra (\emph{a.k.a.} vertex algebra),
\item with nonsingular OPE,
\item with a Poisson bracket (more generally a $\lambda$-bracket) of cohomological degree -1,
\item with no stress tensor in the usual sense (no operator $T\in \CV$ that generates $\pd_z$ derivatives via OPE), but instead endowed with an operator $G$ of degree $1$ that generates $\pd_z$ derivatives via the Poisson bracket.
\end{itemize}
The first three statements were recently derived in \cite{YagiOh}, and may be summarized by saying that $\CV$ is a ``commutative 1-shifted-Poisson vertex algebra.'' We will briefly explain why all these statements hold.

The space $\CV$ has the structure of a chiral algebra precisely because $\pd_t$ and $\pd_{\bar z}$ are exact in \eqref{SUSY-3d}. All correlation functions involving only $Q$-closed operators will thus be independent the $t,\bar z$ coordinates of insertion points, and depend holomorphically on $z$. For example,
\begin{align} \pd_{t} \langle \CO(z,\bar z,t)\CO'(z',\bar z',t')\cdots\rangle &=-i \langle  Q\big(Q_-\CO(z,\bar z,t)\big)\CO'(z',\bar z',t')\cdots\rangle \\ & = -i \langle Q\big(Q_-\CO(z,\bar z,t)\CO'(z',\bar z',t')\cdots\big)\rangle = 0\,, \notag \end{align}
noting that any correlation function of a $Q$-exact configuration of operators automatically vanishes, since $Q$ is a symmetry.
Similarly
$\pd_{\bar z}\langle \CO\CO'\cdots\rangle \sim \langle Q(Q_+\CO\CO'\cdots)\rangle  = 0$.

\begin{figure}[htb]
\centering
\includegraphics[width=4.5in]{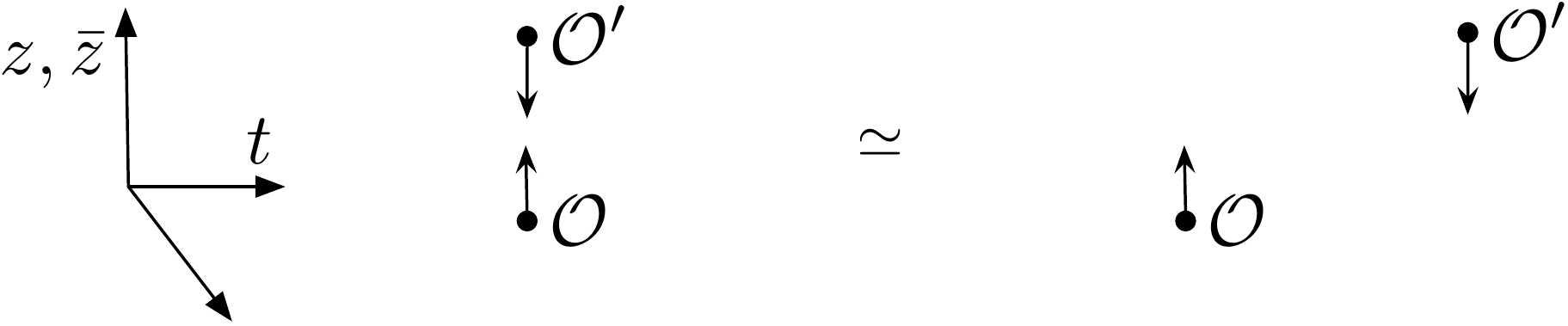}
\caption{Separating operators in the $t$-direction to show that their OPE must be nonsingular.}
\label{fig:bulkOPE}
\end{figure}

More so, $\CV$ is a \emph{commutative} chiral algebra, meaning that all the OPE's are nonsingular. To see this, take two $Q$-closed local operators $\CO$ and $\CO'$ and consider a correlation function of the form
\be f(z,\bar z,t,z',\bar z',t')= \langle \CO(z,\bar z,t)\CO'(z',\bar z',t')(...\text{other $Q$-closed ops}...) \rangle\,. \ee
In Euclidean spacetime, potential singularities involving $\CO$ and $\CO'$ can only occur when insertion points coincide,  $(z,\bar z,t) \to (z',\bar z',t')$. However, the correlator is independent of $\bar z,\bar z',t,t'$. 
Thus, prior to sending $z\to z'$ (and so also $\bar z\to \bar z'$, since we have not analytically continued), we may separate $t$ and $t'$ arbitrarily, making sure to avoid the insertion points of other operators. This is illustrated in Figure \ref{fig:bulkOPE}. We find
\be \lim_{(z,\bar z,t)\to (z',\bar z',t')}  f(z,\bar z,t,z',\bar z',t') = \lim_{(z,\bar z)\to (z',\bar z')} f(z,\bar z,t,z',\bar z',t')\big|_{t\neq t'\,\text{fixed}} = \text{nonsingular}\,.\ee
More succinctly: the presence of a topological $t$ direction ensures that correlators of $Q$-closed operators are nonsingular.

The fact that $\pd_t$ and $\pd_{\bar z}$ derivatives are only zero in cohomology suggests the existence of higher products in $\CV$, induced by topological descent \cite{Witten-Donaldson}. In the related context of (fully) topologically twisted theories in $d$ dimensions, it is has been known mathematically that local operators have the structure of an $E_d$ algebra \cite{Lurie}, which implies when $d\geq 2$ that the $Q$-cohomology of local operators is commutative algebra endowed with a Poisson bracket of degree $1-d$. The relation between this mathematical structure and topological descent was explained in \cite{descent} (see also \cite{CostelloScheimbauer}). In a holomorphic-topological twist in $d$ dimensions --- making one complex direction holomorphic and the remaining $d-2$ directions topological --- one similarly finds for $d\geq 3$ that the $Q$-cohomology of local operators is a commutative chiral algebra endowed with a $\lambda$-bracket \cite{Kac-book, FBZ-book} of degree $2-d$. This was explained physically, via descent, in the recent \cite{YagiOh}.

In the present case $d=3$, so we expect a bracket of degree $-1$. For any $Q$-closed local operator $\CO(z,\bar z,t)$, let
\be \CO^{(1)} = \tfrac{i}{2}Q_+\CO\,d\bar z - i Q_- \CO\, dt \label{def-descent} \ee
be its one-form descendant. By construction, it satisfies $Q\CO^{(1)} = d' \CO$, where $d' = \pd_{\bar z}d\bar z+\pd_t dt$ is the exterior derivative in $\bar z,t$ directions. Wedging with the holomorphic 1-form $dz$, we find
\be Q(dz\wedge \CO^{(1)}) = d(dz\wedge \CO)\,, \label{descent-d} \ee
where $d$ is the total exterior derivative on spacetime. Due to \eqref{descent-d} and Stokes' theorem, integrals of $dz\wedge \CO^{(1)}$ along a 2-cycle $\Gamma$ in spacetime are 1) $Q$-closed; 2) topological, in that they only depend on the homology class of $\Gamma$; and 3) dependent only on the $Q$-cohomology class of $\CO$, up to $Q$-exact terms.

Now, given two $Q$-closed local operators $\CO_1,\CO_2$, their bracket at $\lambda=0$ is defined by integrating $dz\wedge \CO_1^{(1)}$ around a small sphere $S^2$ surrounding the insertion point of $\CO_2$. Explicitly, if we insert $\CO_2$ at $(w,\bar w,s)$,
\be \label{bracket} \{\!\!\{ \CO_1,\CO_2\}\!\!\}(w,\bar w,s)  :=  \oint_{S^2} dz\wedge \CO_1^{(1)}(z,\bar z,t)\, \CO_2(w,\bar w,s)\,.  \ee
Note that, since the $S^2$ can be made arbitrarily small, the LHS is again a Q-closed local operator at the \emph{same} point as $\CO_2$. The generalization to arbitrary $\lambda$ is obtained by replacing $dz\to e^{\lambda z}dz$. 
This should be thought of as a generating function for an infinite collection of brackets $\{\!\!\{ \CO_1,\CO_2\}\!\!\}^{(n)}$ associated to one-forms $z^n dz$.
Various algebraic properties of the general bracket (\emph{e.g.} symmetry, and the fact that it's a derivation of the OPE) are derived in \cite{YagiOh}.


%
%
%
%

We can use the bracket \eqref{bracket} to understand the action of the stress tensor on $\CV$. The full physical 3d $\CN=2$ theory has a stress tensor $T_{\mu\nu}$ and supercurrents $G_{\pm \mu}, \ol G_{\pm \mu}$. Assuming that $T_{\mu\nu}$ is symmetric, all of its components involving a $\bar z$ or $t$ are $Q$-exact due to the SUSY algebra; explicitly,
\be T_{\mu\bar z} =  T_{\bar z \mu} = \tfrac i 2 Q (G_{+\mu})\,,\qquad T_{\mu t}  = T_{t \mu} = -i Q(G_{-\mu})\,. \ee
Therefore, given any $Q$-closed local operator $\CO$, its holomorphic derivative is obtained as
\be \label{d-T} \pd_w \CO(w) = \oint_{S^2} *(T_{z\mu}dx^\mu) \CO(w) = -i \oint_{S^2}  T_{zz}(z) dz\wedge dt\, \CO(w) + (\text{$Q$-exact})\,,  \ee
where $S^2$ is a small sphere surrounding the insertion point of $\CO$. In order for $\pd \CO$ to be picked up by the $S^2$ integral, we see that $\pd\CO$ must appear in the singular part of the OPE between $T_{zz}$ and $\CO$; but also that this OPE cannot be purely holomorphic. This show that $T_{zz}$ itself is \emph{not} a $Q$-closed local operator, and thus not an element of $\CV$.%
\footnote{\label{foot:Ttop}There is one (somewhat trivial) exception to this argument. It may be that all holomorphic derivatives $\pd\CO$ are zero in cohomology. In this case the bulk chiral algebra $\CV$ is fully topological, and reduces to an ordinary graded-commutative algebra. Then one can just define the stress tensor in $\CV$ to be zero. In the full physical theory, one would expect $T_{zz}$ to be $Q$-exact, though this is not strictly necessary.}

In fact, $T_{zz}$ appears as a descendant of a $Q$-closed local operator, and \eqref{d-T} can be neatly rewritten in terms of the bracket. Consider the component $G:=  -\tfrac i2\ol G_{-z}$ of the supercurrent. It is automatically $Q$-closed, and its descendant is
\begin{align} dz\wedge G^{(1)} &= dz\wedge\big(\tfrac{1}{4}Q_+(\ol G_{-z})d\bar z -\tfrac{1}{2}Q_-(\ol G_{-z})dt\big) = -iT_{zz} dz\wedge dt + (\text{$Q$-exact})\,.
\end{align}
Therefore, $G$ belongs to the chiral algebra $\CV$; and for any other $\CO\in \CV$ we have
\be \label{Gdz} \pd_z\CO = \{\!\!\{ G,\CO \}\!\!\}\,.   \ee

\subsubsection{The bracket and extended operators}

There is an alternative, perhaps more intuitive, description of the odd Poisson bracket. Deform the small sphere in the definition of the bracket to a very long, thin cylinder and do the integral along the topological direction first. 
The contour integral along the holomorphic direction is thus simply picking out the singular OPE coefficients between the local operator  $\CO_2$ and the integrated descendant of  $\CO_1$:
\be
\left[\int_{\bR} \CO_1^{(1)}(z,\bar z,t)\right]\, \CO_2(w,\bar w,s) \sim \sum_n \frac{ \{\!\!\{ \CO_1,\CO_2\}\!\!\}^{(n)}(w,\bar w,s)}{(z-w)^{n+1}} + \cdots
\ee
up to non-singular or $Q$-exact operators. 

Notice that an integrated descendant $\int_{\bR} \CO_1^{(1)}(z,\bar z,t)$ is not quite a ``line defect.'' It is simply an integrated local operator. 
On the other hand, a natural way to define a line defect is to deform the identity line defect by some defect action $\int_{\bR} \CO_1^{(1)}(z,\bar z,t)$,
possibly including a coupling to an auxiliary quantum-mechanical system. Then the odd bracket controls the leading term in the perturbative
expansion of a bulk-to-line defect OPE. Higher order terms in the perturbative expansion are associated to higher operations we discuss briefly in \ref{sec:dg}.

In a similar manner, we could deform the small sphere in the definition of the bracket to a very wide, short cylinder to relate the odd bracket to the 
bulk-to-interface OPE for two-dimensional interfaces deforming the identity interface.

\subsection{Boundary chiral algebra}
\label{sec:chiral-bdy}

Given a half-BPS boundary $\CN=(0,2)$ boundary condition that preserves $U(1)_R$, we may similarly consider the vector space of boundary local operators, denoted $\text{Ops}_\pd$. It is doubly graded by $U(1)_R\times U(1)_J$ and has an action of $Q$, so we may take its cohomology
\be \CV_\pd := H^\bullet(\text{Ops}_\pd,Q)\,.\ee

This is again a chiral algebra. In particular, due to \eqref{SUSY-2d}, correlation functions of $Q$-closed boundary local operators are holomorphic,
\be \pd_{\bar z} \langle \CO(z,\bar z) \CO'(z',\bar z')\cdots\rangle \sim \langle Q\big( Q_+\CO(z,\bar z) \CO'(z',\bar z')\cdots\big)\rangle =0\,. \ee
However the argument that showed bulk correlation functions to be nonsingular no longer holds, since boundary local operators are generally stuck at $t=0$, and cannot be separated in the topological $t$ direction. Thus, the chiral algebra $\CV_\pd$ may have singular OPE's.

We proceed to discuss two important (and related) features of $\CV_\pd$ that arise from the interactions of bulk and boundary local operators. 

\subsubsection{Bulk-boundary map}
\label{sec:bulkbdy}

\begin{figure}[htb]
\centering
\includegraphics[width=1.8in]{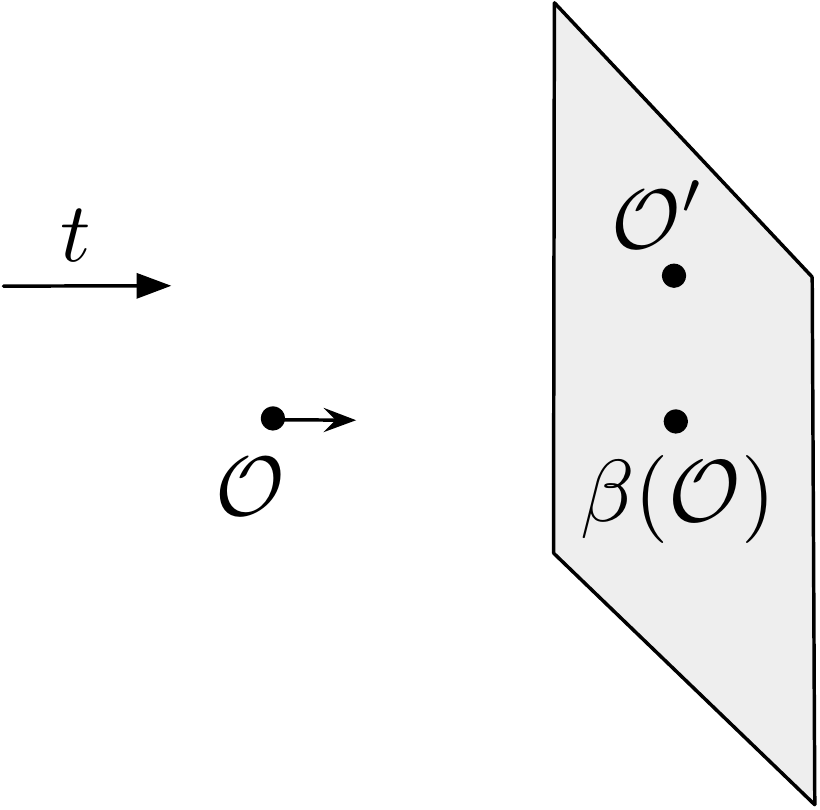}
\caption{Bringing a bulk operator $\CO$ to the boundary defines the bulk-boundary map $\beta(\CO)$. The OPE of $\beta(\CO)$ any other boundary operator $\CO'$ must be nonsingular, since the two operators can be separated in $t$.}
\label{fig:bulkbdy}
\end{figure}

Half-space correlation functions involving collections of $Q$-closed bulk \emph{and} boundary operators do not depend on the $t$ positions of the bulk operators. Bringing $Q$-closed bulk operators to the boundary as in Figure \ref{fig:bulkbdy} defines a ``bulk-boundary map'' of chiral algebras $\beta:\; \CV \to \CV_\pd$.
Moreover, since $\CV$ is commutative, the image of $\beta$ must lie in the center $Z(\CV_\pd)\subseteq \CV_\pd$,
\be \label{bulk-bdy} \beta:\; \CV \to  Z(\CV_\pd)\hookrightarrow \CV_\pd\,. \ee
By definition, $Z(\CV_\pd)=\{\CO\in \CV_\pd\;\text{s.t.}\, \text{the OPE $\CO(z)\CO'(w)$ is nonsingular $\forall$ $\CO'\in \CV$}\}$. 

In general, the map $\beta$ is neither injective not surjective. Surjectivity measures whether or not a boundary condition supports 2d degrees of freedom, localized exclusively on the boundary. 
Clearly, $\beta$ cannot be surjective if $\CV_\pd$ has operators with singular OPE's, since then $Z(\CV_\pd)\neq \CV_\pd$. One might wonder instead whether the map to the center $\beta:\CV\to Z(\CV_\pd)$ is surjective in some generality. This seems true for a large class of boundary conditions (including all the examples in this paper), but not universally. For example, one could trivially enrich any boundary condition by tensoring with a 2d TQFT, which will enlarge the center $Z(\CV_\pd)$ independently of the bulk local operators.

Injectivity is more interesting. The kernel $\ker\beta$ is a sub-chiral-algebra of $\CV$ (since \eqref{bulk-bdy} is compatible with the chiral algebra structures). We expect the identity operator in the bulk to agree with the identity operator on the boundary, $\beta(1)=1$, so $\ker\beta$ cannot contain all of $\CV$.%
\footnote{Technically, we are assuming that the identity on the boundary is not $Q$-exact. Otherwise, the entire chiral algebra $\CV_\pd$ would be zero --- indicating a spontaneous breaking of (0,2) SUSY.} %
Moreover, $\ker\beta$ must be closed under the bracket \eqref{bracket}, and its $\lambda$-generalization. These properties together can strongly constrain the size of $\ker\beta$. For example, we will often encounter theories in which the bracket is non-degenerate, forcing $\ker\beta$ to be (in rough terms) at most half the size of $\CV$.

\begin{figure}[htb]
\centering
\includegraphics[width=5.5in]{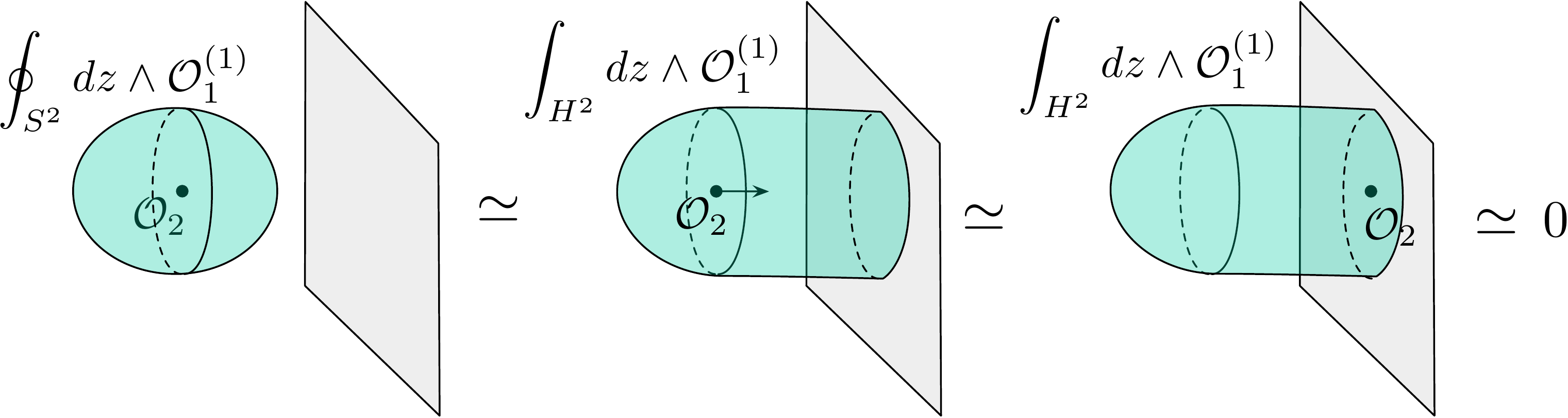}
\caption{Showing that $\beta(\CO_1)=\beta(\CO_2)=0$ implies $\beta(\{\!\!\{\CO_1,\CO_2\}\!\!\})=0$, up to $Q$-exact terms.}
\label{fig:bdybracket}
\end{figure}

To see that $\ker\beta$ is closed under the $\lambda$-bracket, suppose that $\CO_1$ and $\CO_2$ belong to $\ker\beta$, \emph{i.e.} they are both $Q$-closed bulk operators that become $Q$-exact when brought to the boundary,
\be \CO_1\big|_\pd = Q(\mathcal P_1)\,,\qquad \CO_2\big|_\pd = Q(\mathcal P_2)\,. \ee
Now consider the configuration in which $\CO_1^{(1)}$ is integrated around $\CO_2$,
\be \{\!\!\{ \CO_1 {}_\lambda \CO_2 \}\!\!\} := \int_{S^2}  e^{\lambda z}dz\wedge \CO_1^{(1)} \cdot \CO_2\,. \label{bracket-bdy} \ee
In the presence of the boundary, we may manipulate this configuration as shown in Figure \ref{fig:bdybracket}. We deform the $S^2 \simeq H^2\cup D^2$ into the union of a hemisphere and a flat disc lying along the boundary. On the disc $D^2$, we are integrating
\be \int_{D^2} e^{\lambda z} Q_+(\CO_1)\big|_\pd dz\wedge d\bar z\,. \label{disc-bdy} \ee
Using the fact that the boundary preserves (0,2) SUSY (including both $Q$ and $Q_+$), we write $Q_+(\CO_1)\big|_\pd = Q_+(\CO_1\big|_\pd) = Q_+Q(\mathcal P_1) = -Q(Q_+\mathcal P_1)-2i\pd_{\bar z}\mathcal P_1$, whence the integrand of \eqref{disc-bdy} is the sum of a $Q$-exact term and a total derivative. Therefore, the integral reduces to a boundary term,
\be   \int_{D^2} e^{\lambda z} Q_+(\CO_1)\big|_\pd dz\wedge d\bar z = -2i\oint_{S^1} e^{\lambda z}\mathcal P_1 dz + (\text{$Q$-exact})\,. \ee
Now that we have ``opened up'' the surface integral of $\CO^{(1)}$, we may freely bring  $\CO_2$ to the boundary and find that the entire configuration \eqref{bracket-bdy} is $Q$-exact, and thus vanishes in cohomology. In other words, $\beta\big(\{\!\!\{ \CO_1 {}_\lambda \CO_2 \}\!\!\} \big)=0$.

\subsubsection{Boundary stress tensor}
\label{sec:bdyT}


In the bulk, it was impossible to find an element $T\in \CV$ that acted like a standard chiral-algebra stress tensor --- in particular, generating $\pd_z$ derivatives via its OPE --- simply because all OPE's in $\CV$ were nonsingular. The exception to this is the case that the bulk algebra is topological, meaning that all $\pd_z$ derivatives vanish (in cohomology), allowing one to simply choose $T=0$.

We may similarly ask when a boundary algebra  $\CV_\pd$ can contain an operator $T\in \CV_\pd$ such that $T(z)\CO(w)\sim ... +\pd\CO(w)/(z-w)$ $\forall\,\CO\in \CV_\pd$. A necessary condition is clearly that the center $Z(\CV_\pd)$ is topological, meaning that $\pd_z \CO$ is $Q$-exact for all $\CO\in Z(\CV_\pd)$.

%
%

Some nontrivial examples satisfying this condition arise from theories in which \begin{enumerate}
\item the bulk algebra $\CV$ is topological (meaning $\pd_z\CO$ is $Q$-exact for all $\CO\in \CV$); and
\item the bulk-boundary map $\beta: \CV\to Z(\CV_\pd)$ is surjective (whence $Z(\CV_\pd)$ is topological).
\end{enumerate}
Later in Section \ref{sec:gauge-D} we will consider pure 3d $\CN=2$ Yang-Mills-Chern-Simons theories with pure Dirichlet boundary conditions, which are precisely of this type. The bulk theories flow to topological Chern-Simons theories, with trivial algebras $\CV$ of local operators (in cohomology); and the boundary conditions support chiral WZW models, with trivial centers and Sugawara stress tensors. Other simple examples include the boundary coset models discussed in \cite{DHSV, ArmoniNiarchos, GGP-fivebranes} and \cite[Sec. 7]{DGP-duality}.

More intricate examples of boundary stress tensors include the homological blocks of \cite{GPV} for plumbed 3-manifolds (graph manifolds), whose modularity properties were discussed in \cite{3dmodularity}. The boundary chiral algebras in these examples all appear to have stress tensors, and it would be interesting to investigate which general properties of the bulk/boundary allow their existence.

Another way to think about boundary stress tensors is the following.  Recall that, in the presence of a boundary, the $\pd_z$ derivative of a boundary local operator $\CO$ is given by the sum of a bulk hemisphere ($H^2$) integral and a boundary line integral,
\begin{align} \pd_w\CO(w) &= \int_{H^2} *(T_{z\mu}dx^\mu)\cdot \CO(w) + \oint_{S^1} *_\pd\big(T^\pd_{zz}dz+T^\pd_{z\bar z}d\bar z\big)\cdot \CO(w) \notag \\
 &= -i\int_{H^2} T_{zz} dz\wedge dt \cdot \CO(w) -i \oint_{S^1} T_{zz}^\pd\,dz \cdot \CO(w) + (\text{$Q$-exact})\,.    \label{boundary-dz} \end{align}
Here $T_{\mu\nu}$ is the physical bulk stress tensor as before (the Noether current for translations in the bulk action), while $T^\pd_{mn}$ ($m,n\in \{z,\bar z\}$) is the physical boundary stress tensor (the Noether current for $z,\bar z$ translations of an independent boundary action).
If the bulk is topological, the first term in \eqref{boundary-dz} becomes $Q$-exact. If in addition $T_{zz}^\pd$ is $Q$-closed, then $T_{zz}^\pd\in \CV_\pd$, and this operator can function as a stress tensor for the boundary chiral algebra.

Having $T_{zz}^\pd$ be $Q$-closed will likely require additional structure. By comparison with the analyses of \cite{SilversteinWitten-R, Witten-CDO, Nekrasov-betagamma} in the half-twist of 2d $\CN=(0,2)$ models, we suspect that having a non-anomalous boundary R-symmetry will play a role.

\subsection{Derived structure}
\label{sec:dg}

We saw above that the $Q$-cohomology of bulk local operators, denoted $\CV$, is not merely a chiral algebra, but is endowed with a Poisson bracket. The Poisson bracket is a piece of derived structure: it cannot be defined in terms of $Q$-cohomology classes alone, but requires information about descendants of local operators, which only exist in the underlying physical theory. There may exist other such operations on $\CV$, involving particular collections of local operators whose descendants are integrated around one another. 

As we mentioned in the Introduction, in the fully topological $d$-dimensional case the mathematical structure that encodes all the data of topological OPE and descent relations is called an $E_d$ algebra \cite{Lurie}. Its physical interpretation has been discussed in \cite[Ch. 5]{CostelloGwilliam-book} and \cite{descent}.
The $E_d$ algebra can be visualized as a large collection of operations associated to all possible cycles in the space of configurations of operator insertion points (or better, small neighborhoods thereof), 
such that operations associated to homologous cycles are quasi-isomorphic, with explicit homotopies given by operations associated to higher-dimensional cycles. 
With some work, in any given dimension one may attempt to distill this large amount of information to a more manageable ``minimal'' collection of operations that captures all physically relevant information.

In one topological dimension, the minimal collection can be given as an $A_\infty$ algebra. The most direct physical interpretation of the $A_\infty$ structure is that it controls the perturbative BRST invariance of 
deformations of the action: an action $\int_{\bR} \CO_1^{(1)}(z,\bar z,t)$ preserves the BRST symmetry iff $\CO_1$ satisfies a Maurer-Cartan equation 
\begin{equation}
Q \CO_1 + \CO_1 \cdot \CO_1 + (\CO_1,\CO_1,\CO_1)_3 + (\CO_1,\CO_1,\CO_1,\CO_1)_4 + \cdots = 0\,,
\end{equation}
where the degree $n$ term in the expansion defines the $n$-th operation in the $A_\infty$ algebra.

In two topological dimensions, the story is richer. The BRST invariance of deformations of the action defines a collection of operations that form a (degree shifted) $L_\infty$ algebra, 
but these do not exhaust all the data of the $E_2$ algebra. Indeed, they do not even include the actual OPE of local operators. A complete description can be obtained by 
working in a hierarchical manner, focussing first on the OPE in one topological ``vertical'' direction and associated $A_\infty$ structure and then on the extra structures involving the horizontal direction,
such as the fusion of vertical line defects. For example, we will have operations that give the defect action of the fusion of two defects as 
a perturbative sum of powers of the two respective defect actions
\be
O_{\ell_1 \circ \ell_2} = O_{\ell_1} + O_{\ell_2} +(O_{\ell_1} ;O_{\ell_2}) +(O_{\ell_1} ,O_{\ell_1} ;O_{\ell_2}) +(O_{\ell_1} ;O_{\ell_2} ,O_{\ell_2}) + \cdots 
\ee 
A rich physical example of these structures can be found in the ``web algebras'' of \cite{GMW2}.

The analog of $E_n$ algebras for the holomorphic case have not been yet fully developed. It should replace de Rham homology of configuration spaces of points with 
Dolbeault homology. It is also not clear what sort of ``minimal'' data would capture this information in the most economical way. 
It is likely that one can again proceed again in a hierarchical manner, adding one direction (topological or holomorphic) at a time. 

The most basic question would be what replaces an $A_\infty$ algebra in the holomorphic case, perhaps controlling the BRST invariance of deformations of the system. 
This structure should be related to a standard $A_\infty$ algebra upon reduction on a circle, in the same way as a vertex algebra is related to its mode algebra. 
We can dub it ``$A_\infty$-chiral algebra.'' The boundary chiral algebra $\CV_\pd$ should be endowed with such a structure. 

As for the bulk chiral algebra $\CV$, we could proceed in three alternative manners:
\begin{itemize}
\item We can deform the bulk theory and look at the deformation of the BRST symmetry. At the leading order, a deformation of the bulk theory by 
$\int\CO_1^{(2)} e^{\lambda z} dz$ should change the BRST differential by $\{\!\!\{\CO_1{}_\lambda \cdot\}\!\!\}$. Higher order deformations will 
be captured by chiral analogues of a (degree shifted) $L_\infty$ algebra. 
\item We can encode the OPE in the topological direction in an $A_\infty$ algebra, controlling topological line defects, and then study the OPE of topological lines 
along the holomorphic direction. The leading non-trivial terms in the OPE, bilinear in the defect actions, will be given by the usual $\lambda$ bracket. 
\item We can encode the OPE in the holomorphic direction in the appropriate derived analogue of a chiral algebra, and the study the OPE of chiral surface defects 
along the topological direction.
\end{itemize} 
The latter is presumably better suited to the study of the relation between $\CV_\pd$ and $\CV$.
Constructions in 2d TQFT suggest that the bulk-boundary map \eqref{bulk-bdy} will be much more interesting when we keep track of higher/derived structures. We would expect that the map extends to send bulk local operators to the derived center of the boundary algebra,
\be \beta_{\rm der}:\,{\CV} \to Z(\CV_\pd)_{\text{der}}\,, \ee
%
and that, for sufficiently rich boundary conditions, this map is actually an isomorphism.
As discussed in the Introduction, the analogous statement in 2d TQFT is that, given a generator $\CB$ for the category of boundary conditions, the derived center (= Hochschild cohomology) of the algebra $\text{End}(\CB)$ of local operators on $\CB$ is isomorphic to bulk local operators.
 Thus, we might expect to recover the all bulk local operators in $\CV$ from a boundary algebra $\CV_\pd$\,! We hope to explore this in future work.%
 \footnote{The expected relation between derived centers and bulk local operators was used by \cite{VOAExt} to compute Higgs and Coulomb branches of 3d $\CN=4$ theories from boundary chiral algebras. More directly relevant examples in 3d $\CN=2$ theories were recently studied by Zeng \cite{Zeng-PSIthesis}}

Even in the topological case, one should remember that derived structures such an $A_\infty$ algebras are only defined up to quasi-isomorphism. Physically, they depend on all sort of renormalization/regularization 
choices: different choices are related by operator re-definitions. The benefit of keeping track of the derived structures is that they simply contain more information than the OPE in the BRST cohomology. For example, 
it is easy to give examples of  2d topological theories with a very boring BRST cohomology of local operators, but a very rich category of line defects which can be fully recovered from the $E_2$ algebra structure. 

The derived structures are still expected to be RG flow invariants, albeit up to quasi-isomorphism (see Section \ref{sec:RG}). Expected dualities of 3d ${\cal N}=2$ theories and their boundary conditions 
will thus lead to non-trivial mathematical conjectures concerning the quasi-isomorphism of the derived structures associated to the respective $\CV$ and $\CV_\partial$ operator algebras. 

Starting in the next section, we will restrict our attention to 3d $\CN=2$ linear gauge theories, with Lagrangian boundary conditions. 
Within this class, we will construct conjectural \emph{dg models} for many boundary $A_\infty$-chiral algebras $\CV_\pd$, which make derived structures more explicit. By a dg model, we mean that we isolate a subspace of boundary local operators $\widehat \CV_\pd\subset \text{Ops}_\pd$ such that
\begin{itemize}
\item The correlation functions of operators in $\widehat \CV_\pd$ are all holomorphic (or meromorphic), so that $\widehat \CV_\pd$ itself is a chiral algebra;
\item $\widehat \CV_\pd$ is closed under the action of $Q$, and its cohomology is $H^\bullet(\widehat \CV_\pd,Q)\simeq \CV_\pd$;
\item $\widehat \CV_\pd$ has no higher operations, so that all putative higher operations in $\CV_\pd$ are induced from the chiral algebra structure and differential in $\widehat \CV_\pd$.
\end{itemize}
 In particular, the dg chiral algebras $\widehat \CV_\pd$ and $\widehat \CV_\pd'$ that we construct in dual pairs of gauge theories should be quasi-isomorphic. Such conjectural dg models should also be invaluable to 
 determine or constrain the higher operations in the bulk.

\section{Twist of $3d$ $\CN=2$ gauge theories}
\label{sec:theories}

In this section we introduce and review the class of 3d $\CN=2$ gauge theories whose boundary chiral algebras we wish to study. We quickly review a standard physical formulation of these theories, in terms of 3d $\CN=2$ superspace. 
%
We then rewrite the holomorphic-topological twist of 3d $\CN=2$ gauge theories in the twisted formalism of~\cite{ACMV}. 
This regroups the field content and simplifies the actions of the theories, while preserving the $Q$-cohomology of local operators (in fact, preserving all the derived structure from Section~\ref{sec:dg}). It makes many features of the chiral algebras $\CV$ and $\CV_\pd$ more transparent.

\subsection{Standard physics formulation}
\label{sec:conv-physics}

We wish to consider 3d $\CN=2$ gauge theories whose discrete data is given by
\begin{itemize}
\item a compact gauge group $G$
\item chiral matter in a unitary (linear) representation $V$ of $G$
\item a polynomial superpotential $W:V\to \C$
\item Chern-Simons terms for $G$, at either integral or half-integral level depending on $V$.
\end{itemize}
For the basic physical analysis of such theories, including restrictions on Chern-Simons levels, continuous parameters, and IR behavior, see the classic \cite{AHISS}.

In much of the following, it will be the complexification of the gauge group that plays a central role. Thus, we will adopt the convention that
\be \begin{array}{cl} G_c & = \text{compact gauge group (denoted $G$ above)}\,, \\[.1cm]  G &= \text{its complexification, a reductive algebraic group}\,. \end{array} \ee
%
The space $V$ becomes a complex-linear representation of $G$.

We require our theories to have an unbroken $U(1)_R$ symmetry, under which $V$ decomposes
\be V = \bigoplus_r V^{(r)}\,, \qquad \text{$V^{(r)}=$\;subspace of R-charge $r$}\,.\ee
We will further assume that the R-charges $r$ of the matter fields are all non-negative, though not necessarily integral.  Physically, this will be true for any theory that flows to a CFT in the infrared.
The superpotential $W$ must be quasi-homogeneous of R-charge two and spin zero; in terms of the twisted spin \eqref{twist-hom} this means
\be R(W) = 2\,,\qquad J(W)=1 \,. \label{W-charges}\ee

A chiral matter multiplet has a complex scalar field $\phi$ and four fermions $\psi_\pm,\bar\psi_\pm$. If $\phi$ has R-charge $r$, then the R-charges and twisted spins of the remaining fields are%
\footnote{In this table, we are actually giving the charges of the local operators called $\phi$, $\psi_+$, etc. This is standard and unspoken physics convention. Since the local operators are functionals on the space of fields, the fields technically have opposite charges.} %
\be 
 \begin{array}{c|cccccc}
 & \phi & \bar\phi &\psi_+ & \bar\psi_+ & \psi_- & \bar\psi_-  \\ \hline
U(1)_R & r & -r & r-1\, & \,1-r\, & \,r-1\, & \,1-r  \\
U(1)_J & \frac r2 & -\frac r2 & \frac r2-1& -\frac r2 & \frac r2 & 1-\frac r2
\end{array}
\ee
Given a collection of chiral multiplets in the representation $V$, we may jointly describe their scalar fields after the holomorphic twist as $\phi = \sum_r \phi_r dz^{r/2}$, where $\phi_r$ is a section of a $V^{(r)}$ bundle. Similarly, we can write $\bar\psi_- = \sum_r \psi_r dz^{1-r/2}$, where $\psi_r$ is a section of a dual $(V^{(r)})^\vee$ bundle.


The physical vector multiplet contains the $G_c$ connection  $A_\mu$, a real scalar $\sigma\in \mathfrak g_\R$, and $\mathfrak g$-valued gauginos  $\lambda_\pm,\bar\lambda_\pm$. They have canonical R-charges and spin, given by
\be 
 \begin{array}{c|ccccccc}
 & A_t,\sigma & A_z & A_{\bar z} & \lambda_+ &\bar\lambda_+ & \lambda_- & \bar\lambda_- \\ \hline
U(1)_R &  0&0&0 & 1 & -1 & 1 & -1 \\
U(1)_J &  0& 1 & -1 & 0&-1&1&0
\end{array}
\ee

The transformations of vector-multiplet fields under the holomorphic supercharge $Q = \ol Q_+$ are:
\be \label{SUSY-gauge} \begin{array}{l@{\;=\;}l@{\qquad} l@{\;=\;}l}
Q A_{\bar z} & 0 & Q A_t & \frac12 \lambda_+  \\[.1cm]
Q A_z & \frac12 \lambda_- & Q\sigma & -\frac{i}{2}\lambda_+   \\[.1cm]
Q \lambda_+ & 0 & Q \lambda_- & 0  \\[.1cm]
Q \bar\lambda_+ & 2i \mathcal F_{\bar zt} & Q \bar\lambda_- & -i(2F_{z\bar z}-i D_t\sigma-D) 
\end{array}
\ee
It is easy to see from this that the combination $\mathcal A_t = A_t-i\sigma$ is $Q$-closed. Altogether, the complexified connection
\be \mathcal A :=  A_{\bar z} d\bar z + \mathcal A_t dt\,, \label{calA} \ee
containing only components in the $\bar z$ and $t$ directions, is $Q$-closed.
 On the last line, the symbol $\mathcal F_{\bar z t} = i[\pd_{\bar z} -i A_{\bar z}, \pd_t-i \mathcal A_t] = F_{\bar z t}-i D_{\bar z}\sigma$ denotes the curvature of this complexified connection. Also on the last line, the symbol $D$ is the D-term, which is equal on-shell to a sum of the real moment map $\mu_\R(\phi,\bar\phi)$ for the matter scalars and a Chern-Simons contribution $\sim k\cdot \sigma$.
 
 It is also interesting to consider the complexified curvature component
\be \ol {\mathcal F_{\bar zt}}  
= F_{zt}+i D_z\sigma\,. \label{def-Fzt} \ee
A short calculation shows that $Q(\ol {\mathcal F_{\bar zt}})=0$ on shell, modulo the Dirac equations $D_t\lambda_-=2D_z\lambda_+$ and $D_t\lambda_+=-2D_z\lambda_-$. 
There is a good reason for this: in an abelian theory, $\ol {\mathcal F_{\bar zt}}$ coincides with the $\pd_z$-derivative of the complexified dual photon $\tilde\gamma:=\gamma+i\sigma$, where $d\gamma = *F$. The dual photon $\tilde \gamma$ itself is the bottom component of a chiral multiplet, and is $Q$-closed.

We may also compute descendants of various $Q$-closed local operators built from the above fields. Given a $Q$-closed  operator $\CO$, we set $\CO^{(1)} = \tfrac{i}{2}Q_+\CO\,d\bar z - i Q_- \CO\, dt$ as in \eqref{def-descent}.  If $\CO$ is not gauge invariant (so not strictly speaking a bulk local operator on its own), the descendant satisfies an $\CA$-covariant version of the descent equation
\be Q\CO^{(1)} = \d_\CA\CO\,, \label{A-descent} \ee
with covariant derivative $\d_\CA = \d'-i\CA = (\pd_{\bar z} -iA_{\bar z} )\d\bar z +(\pd_t-i\CA_t)\d t$. We find
\be  \begin{array}{l@{\;=\;}l}
A_{\bar z}^{(1)} & -\frac{i}{2}  \bar\lambda_+\,dt\,,\quad \mathcal A_t^{(1)} \,=\, \frac{i}2\bar\lambda_+ \,d\bar z \\[.1cm]
\mathcal A^{(1)} &  -i\bar\lambda_+dt d\bar z \\[.1cm]
\lambda_-^{(1)} & 2\mathcal F_{zt} dt + \big(F_{z\bar z} + \tfrac i2 D_t\sigma +\tfrac12 D\big)d\bar z \\[.1cm]
 \ol {\mathcal F_{\bar zt}} ^{(1)} & iD_z(\bar\lambda_-dt-\tfrac12\bar\lambda_+ d\bar z)
\end{array}
\ee
In particular, in an abelian theory, the descendant of the dual photon is $\tilde \gamma^{(1)} = i\bar\lambda_-dt-\tfrac i2\bar\lambda_+ d\bar z$. It follows from the definition of the secondary bracket \eqref{bracket} and the fact that $ \d(\tilde\gamma^{(1)}\wedge dz) = i(\pd_{\bar z}\bar\lambda_-+\tfrac12\pd_t\bar\lambda_+)\d t\d z\d \bar z \sim \delta S/\delta \lambda_-$ is the equation of motion for $\lambda_-$ (modulo supersymmetric Chern-Simons terms) that \cite{descent, YagiOh},
\be \{\!\!\{ \tilde\gamma,\lambda_- \}\!\!\} \sim 1\,. \label{gammabracket}\ee

Similarly, the on-shell transformations of chiral-multiplet fields are
\be \label{SUSY-chiral} \begin{array}{l@{\;=\;}l@{\qquad} l@{\;=\;}l}
Q \phi & 0 & Q \bar\phi & -i\bar\psi_+  \\[.1cm]
Q \bar \psi_+ & 0 & Q \psi_+ & -2D_{\bar z}\phi   \\[.1cm]
Q \bar \psi_- & -i\,\pd W/\pd \phi & Q \psi_- & \mathcal D_t \phi\,,
\end{array}
\ee
where $\mathcal D_t = \pd_t - i\mathcal A_t$ is the covariant derivative with respect to the complexified connection $\mathcal A_t$. Now the operators $\phi$ and $\bar\psi_-$ are Q-closed (as long as $W\neq 0$) but not exact. 
The $\mathcal A$-covariant descendants of these fields are
\be \phi^{(1)} = -\psi_- dt + \tfrac12 \psi_+ d\bar z\,,\qquad \bar\psi_-^{(1)} = -2i\ol{D_{\bar z}\phi}\,dt +\tfrac{i}{2}\ol{\mathcal D_t\phi}\,d\bar z\,, 
\ee
from which one computes
\be \{\!\!\{ \phi,\bar\psi_- \}\!\!\} = 1\,.\ee

\subsection{Twisted formalism}
\label{sec:twisted}

Given the SUSY transformations above, one could follow standard methods to analyze $Q$-preserving boundary conditions, as in \cite{GGP-fivebranes,DGP-duality,BrunnerSchulzTabler}. One could also begin to derive $Q$-cohomology of the algebras of bulk and boundary local operators, at least perturbatively.  A variant of this direct approach led to the construction of 3d $\CN=2$ indices \cite{Kim-index, IY-index, KW-index} and half-indices \cite{GGP-fivebranes,DGP-duality}.

We will follow a slightly different approach here, and first recast the SUSY gauge theory in the twisted formalism of \cite{ACMV}. The procedure delineated in \cite{ACMV} amounts to 1) rewriting the gauge theory in a first-order formalism (and more precisely, in the BV formalism); and 2) removing $Q$-exact terms to simplify the field content and action. The removal of $Q$-exact terms (and pairs of fields related by $Q$) means that we will get a new QFT whose full algebra of local operators will \emph{differ} from that in the original theory, but will nevertheless be quasi-isomorphic to it. In particular, the $Q$-cohomology of local operators and all higher operations (brackets, etc.) will be preserved.

Here we will review the twisted formalism for the bulk gauge theory. In subsequent sections we will introduce boundary conditions.


The Chern-Simons-matter gauge theory of Section \ref{sec:conv-physics} in the twisted formalism contains:
\begin{enumerate} 
	\item A complexified $2$-component gauge field $\CA = \CA_t\d t + \CA_{\zbar} \d \zbar$, just as in \eqref{calA}.
	
	\item A co-adjoint ($\mathfrak g^*$-valued) field $B = B_z \d z$. 
	
	In the physical theory, this field arises from writing the Yang-Mills action in a first-order formalism; on-shell, $B_z$ is identified with $\frac{1}{g^2}\ol {\mathcal F_{\bar zt}}= \frac{1}{g^2}F_{zt}+...$, up to Chern-Simons terms.

	\item A field $\displaystyle \phi = \sum_r \phi_r \d z^{r/2}$ where $\phi_r$ is a section of $ V^{(r)}$.
	
	This is the standard physical $\phi$, transforming as a section of a power of the canonical bundle according to its twisted spin.
	
	\item A field $\displaystyle \eta = \sum_{r} \big(\eta_{r,t} \d t + \eta_{r,\zbar}\d \zbar\big)\, \d z^{1-r/2} $
		where $\eta_{r,t}$ and $\eta_{r,\zbar}$ are sections of the dual representation $(V^{(r)})^\vee$.
		
	This arises from writing the scalar action in a first-order formalism. In the physical theory, in the absence of superpotential, $\eta_t = 2\ol{D_{\bar z}\phi}$ and $\eta_{\bar z}=-\tfrac 12 \ol{\mathcal D_t\phi}$, so that $\eta =  i\bar\psi_-^{(1)}$.
\end{enumerate}
The twisted action functional so far is
\begin{equation} \label{taction}
\int B F(\CA) + \int \eta\, \d_\CA \phi + \text{ superpotential  and CS terms}\,,
\end{equation}
where $F(\CA) = \d'\CA-i \CA\wedge \CA = \mathcal F_{\bar z t}\d\bar z \d t$ and $\d_\CA = d'-i\CA$, with $\d' = \pd_t\,\d t + \pd_{\bar z}\,\d\bar z$.
We will discuss the superpotential term shortly. The twisted spins of the fields are such that this action makes sense globally on any $3$-manifold $M$ with THF structure.

The action \eqref{taction} would be equivalent to the standard action for the bosonic fields in the physical theory, if we added quadratic terms $\sim B^2,\, \eta^2$ to set $B$ and $\eta$ to their on-shell values as indicated above. However, these quadratic terms are $Q$-exact and have been removed. Of course, we are still missing the original fermions. They arise as ghosts.

The action  \eqref{taction}  has two kinds of gauge symmetry.  First, there are \emph{complexified} $G$ gauge transformations. We introduce a $\c$ ghost (an odd, $\mathfrak g$-valued scalar) that generates infinitesimal $G$ gauge transformations, in the BRST formalism:
\begin{align} 
	\delta_{\c} \CA &= \d_\CA\c = \partial_{\bar z} {\c}\, \d\bar z + \partial_t {\c}\,\d t + i [{\c},\CA] \notag \\
	\delta_{\c} \phi &= -i \c \cdot \phi \label{gauge-c} \\
	\delta_{\c} \eta &= -i \c \cdot \eta\, \notag \\
	\delta_{\c} B &= i[\c,B] \notag 
\end{align}
where ``$\c\cdot \phi$'' and ``$\c\cdot\eta$'' schematically indicates the action of $\c\in \mathfrak g$ on $\phi,\eta$ in the appropriate representation. The transformation of $B$ acquires an additional correction in the presence of Chern-Simons term, discussed below \eqref{Qsuperfields}.

There is also a \emph{second} kind of local symmetry that leaves the twisted action invariant. We introduce a new ghost field 
\begin{equation} 
	\psi = \sum_r \psi_r \d z^{1-r/2}\,,
\end{equation}
where $\psi_r \in (V^{(r)})^\vee$.
Then the infinitesimal transformations of the fields are encoded in
\be \label{gauge-psi} \begin{array}{r@{\;=\;}l}
	\delta_{\psi} \CA & 0  \\[.1cm]
	\delta_{\psi} \eta & \d_\CA \psi\\[.1cm]
	\delta_{\psi} \phi & 0 \\[.1cm]
	\delta_\psi B &  \mu(\phi,\psi)\,,\end{array}
\ee
where $\mu:V\otimes V^\vee \to \mathfrak g^*$ is the moment map for the $G$ action on $V\otimes V^\vee\simeq T^*V$ (which is automatically Hamiltonian).%
\footnote{Explicitly, given any element $\tau\in \mathfrak g$, we have $\langle \mu(\phi,\psi),\tau\rangle = \psi \tau\phi$, where the RHS involves the action of $\tau$ on $\phi\in V$, and a contraction with $\psi\in V^\vee$. More schematically, we have $\mu(\phi,\psi) = \frac{\pd}{\pd\tau} (\psi\tau\phi) \sim \phi\psi$.}

The $z$-derivative of the $\c$ ghost roughly corresponds to the physical gaugino $\lambda_-$. To understand this, we note that when writing the full physical theory in the BV-BRST formalism, and introducing a $\c$ ghost to implement gauge invariance cohomologically, the BRST differential must be added to the holomorphic-topological supercharge, giving a total differential
\be Q = \ol Q_+ +Q_{\rm BRST}\,. \label{Qsum} \ee
Before simplifying the action and field content, we still have the component $A_z$ of the gauge connection, on which the total differential acts as
\be Q A_z= \ol Q_+ A_z +Q_{\rm BRST}A_z = \tfrac12\lambda_- + D_z c\,. \label{clambda} \ee
Therefore, $D_z c= \pd_z c-i[A_z,c]$ is cohomologous with $\frac12\lambda_-$.

In a simpler way, the ghost $\psi$ for the exotic symmetry \eqref{gauge-psi} of the twisted action corresponds to the physical fermion $i\bar\psi_-$. As evidence, we may compare the BRST transformation $\delta_\psi\eta=\d_\CA\psi$ from \eqref{gauge-psi} with the original SUSY transformation of $\eta \sim 2\ol{D_{\bar z}\phi}\d t-\tfrac12\ol{\mathcal D_t\phi}\d\bar z$, and see that they coincide on shell. When thinking of $\bar\psi_-$ as a ghost, we reinterpret the SUSY transformation as a BRST transformation.

We also note that the theory in the twisted formalism has a cohomological ghost-number symmetry/grading, which coincides with the R-symmetry of the original physical theory. At intermediate steps (see \cite{ACMV}), when the physical theory is written in the BV-BRST formalism but the action/field content have not yet been simplified, the cohomological grading is the sum
\be \text{coh. degree} = \text{R-charge} + \text{ghost number}\,.\ee
With this definition, the total differential \eqref{Qsum} has degree $+1$.

The twisted theory also has a non-cohomological grading $J$ by twisted spin, which is the same as twisted spin of the original physical theory.

\subsection{Twisted superfields}
\label{sec:superfields}

If we further recast the twisted theory in the BV formalism --- introducing new anti-fields and anti-ghosts $\CA^*,V^*,\phi^*,\eta^*,\c^*,\psi^*$ for each of the fields and ghosts above --- the field content and action have a concise representation in terms of ``superfields.'' These are analogous to (but not the same as) the more familiar superfields that show up in supersymmetry.

%
%
%
%
%
%
%
%
%

To proceed, we need to introduce an auxiliary graded algebra $\alg^\bullet$, which is the quotient of the de Rham complex of our spacetime $M$ by the subspace of forms that 
are divisible by $\d z$.  In local coordinates $t,z,\zbar$,
\be \label{def-alg} \alg^\bullet = \cinfty(\R^3)[\d t, \d \zbar]\,, \ee
where $\d t$, $\d \zbar$ are treated as odd (Grassmann) variables.
The grading on the algebra, which will match cohomological degree in our theories, is such that functions $C^\infty(\R^3)$ are in degree zero, while $\d t$ and $\d\bar z$ are both in degree $+1$. Explicitly, the graded components are
\be \alg^0 = \cinfty(\R^3)\,,\qquad  \alg^1=\cinfty(\R^3)\d t \oplus \cinfty(\R^3)\d \bar z\,,\qquad \alg^2 = \cinfty(\R^3)\d t\d\bar z\,. \ee

The de Rham operator on forms on $M$ descends to a differential on $\alg^\bullet$ that, locally, is just $\d'=\pd_t\,\d t+\pd_{\bar z}\,\d\bar z$ on $\alg^\bullet$.  Locally, its cohomology consists of those functions that are independent of $t$ and holomorphic in $z$.


There are natural variants $\alg^{\bullet, (j)}$ of $\alg^{\bullet}$ that in local coordinates take the form $\cinfty(\R^3)[\d t, \d \zbar] \d z^j$.  That is, we modify the definition of $\alg^{\bullet}$ so that elements transform as sections of the $j$-th power of the canonical line bundle in the $z$-direction. This introduces the twisted-spin ($J$) grading.
The de Rham operator and its covariantization on $\alg^{\bullet}$ continue to be well-defined on $\alg^{\bullet,(j)}$.  The definition of $\alg^{\bullet,(k)}$ does not depend on the chosen coordinates, but is intrinsic to a $3$-manifold with THF structure.


There are natural product maps 
\begin{equation} 
\alg^{r, (j)} \otimes \alg^{r',(j')} \to \alg^{r+r',(j+j')}\,,
\end{equation} 
and an integration map
\begin{equation} 
\int :\quad  \alg^{2,(1)} (M) = \Omega^3(M,\C) \; \to \; \C\,. 
\end{equation}
The cochain complexes $\alg^{\bullet,(j)}$ are locally well-defined even if $j$ is not an integer.
In this paper, we will typically work on $M=R_t\times \C$, so a local definition suffices.

Let us now return to holomorphically twisted $3d$ $\CN=2$ gauge theory, with complexified gauge group $G$ and matter representation $V$. Once we introduce both ghosts and anti-fields (and anti-ghosts), the full field content can be grouped into the four superfields
\be \label{superfields}
\begin{array}{rcl}
\mbf{A} & \in & \alg^{\bullet} \otimes \g[1]\,,  \\ 
	\mbf{B}& \in & \alg^{\bullet,(1)} \otimes \g\,, \\
\mbf{\Phi}& \in & \oplus_r \alg^{\bullet,(r/2)} \otimes V^{(r)}\,, \\
\mbf{\Psi}& \in& \oplus_r \alg^{\bullet,(1-r/2)} \otimes \left( V^{(r)} \right)^\vee [1].  
\end{array}
\ee
The symbol $[1]$ indicates a shift of cohomological degree by one.


We will adopt the convention throughout that fields written in bold font are superfields.  Fields in non-bold fond equipped with an index $i$ from $0$ to $2$ will indicate the component of the superfield that lives in $\alg^{i,(j)}$.  
 Altogether, the components of the superfields are related to the original (twisted) fields, ghosts, and anti-fields as
\be \label{components}
 \begin{array}{r@{\qquad}l@{\qquad}r@{\qquad}l@{\quad}|@{\quad}r@{\qquad}l@{\qquad}l}
  r+1& & -r+1 &   \mbf{\Psi}^0 = \psi & 1 & \mbf{A}^0 = \c \\[.2cm]
r&\mbf{\Phi}^0 = \phi & -r &  \mbf{\Psi}^1 = \eta  & 0 &   \mbf{A}^1 = \CA &    \mbf{B}^0 = B \\[.2cm]
r-1&\mbf{\Phi}^1 = \eta^\ast & -r-1 &  \mbf{\Psi}^2 = \phi^\ast  & -1 &  \mbf{A}^2 = B^\ast  & \mbf{B}^1 = \CA^\ast \\[.2cm]
r-2&\mbf{\Phi}^2 = \psi^* & -r -2 & & -2 & & \mbf{B}^2 = \c^\ast \end{array}
\ee
The labels to the left of each row indicate cohomological degree of the components.

For example, in $\mbf A$ we find an expansion
\be \mbf A = \c + (\CA_t\, \d t+ A_{\bar z}\,\d \bar z) + B^*_{\bar zt} \d\bar z\d t \ee
where the entire superfield $\mbf A$ has cohomological degree $1$, implying that $\c$ has degree 1 (coming from ghost number), $\CA_t,\CA_{\bar z}$ have degree 0, and $B^*_{\bar zt}$ has degree $-1$. This superfield also has spin $J=0$, consistent with
\be J(\c) = 0\,,\quad J(\CA_t)=0\,,\quad J(A_{\bar z}) = -1\,,\quad J(B^*_{\bar zt}) = -1\,;\qquad J(\d t)=0\,,\quad J(\d\bar z) = 1\,.\ee
Similarly, given a component of $V^{(r)}$ of fixed R-charge $r$, the superfield $\mbf \Psi$ has cohomological degree $1-r$ and spin $0$, and an expansion
\be \mbf \Psi = \psi \,\d z^{1-\frac r2} + (\eta_t\,\d t+\eta_{\bar  z}\,\d\bar z)\d z^{1-\frac r2} + \phi^*_{\bar z t}\,\d \bar z \d t \d z^{1-\frac r2}\,,\ee
consistent with the component charges
\be \begin{array}{c|cccc|ccc}
& \psi & \eta_t & \eta_{\bar z} & \phi^*_{\bar z t} & \d t & \d \bar z & \d z \\\hline
\text{coh} & 1-r & -r & -r & -1-r & 1 & 1 & 0 \\
J  & 1-\frac r2 & 1-\frac r2 & -\frac r2 & -\frac r2 & 0 & 1 & -1
\end{array} \ee

We emphasize again that, throughout this paper, we are using ``physics conventions'' in describing the degrees of fields. All the charges above really refer to charges of \emph{local operators} defined by evaluating the fields at a point. We are ultimately interested in local operators anyway. Technically speaking, local operators and fields belong to dual spaces, and the degrees of the actual  fields are the negatives of what's given above.

The action functional, now including the superpotential and a Chern-Simons term, is
\begin{equation}  \label{actionWk}
	S= \int \mbf{B} F(\mbf{A}) +  \int \mbf{\Psi} \d_{\mbf{A}} \mbf{\Phi} -  \int W(\mbf{\Phi}) + \frac{k}{4\pi} \int \mbf{A} \partial \mbf{A}.  
\end{equation}
Here the covariant derivative is $\d_{\mbf A} = \d' - i{\mbf A}$, and its curvature is
\be \label{FA} F(\mbf A) = i(\d'-i \mbf A)^2 = \d' \mbf A -i \mbf A^2 = -i\c^2 + \d_{\CA}\c + F(\CA)-i\{\c,B^*\}\,. \ee
The Chern-Simons term is $k \int \mbf{A} \pd \mbf{A}$, where $\pd = \pd_z\,dz$ is the holomorphic exterior derivative, and we have left the Cartan-Killing form implicit.%
\footnote{It is useful to think of the Chern-Simons level `$k$' as a normalization of the Cartan-Killing form. For classical groups, the Cartan-Killing form at $k=1$ correspond to the trace in the fundamental representation.}

The SUSY/BRST transformation acts on superfields in a simple way:
\be \label{Qsuperfields} \begin{array}{l@{\qquad}l} Q\,\mbf A = F(\mbf A)\,, & Q\,\mbf B = \d_{\mbf A}\mbf B -i \mu(\mbf \Phi,\mbf \Psi) - \tfrac{k}{2\pi}\pd \mbf A\,,\\[.1cm] Q\,\mbf\Phi = \d_{\mbf A}\mbf \Phi\,, & Q\,\mbf \Psi = \d_{\mbf A}\mbf \Psi  +  \pd W(\mbf  \Phi)/\pd \mbf \Phi\,. \end{array}\ee
We encourage the reader to check that $Q^2=0$ \emph{off shell}, and that the action is in fact invariant. (Invariance of the action requires using a Bianchi identity $d_{\mbf A} F(\mbf A) = 0$, and removing some total derivatives.)
From $Q\,\mbf A = F(\mbf A)$ and \eqref{FA} we quickly recover the standard BRST transformations of the $\c$-ghost and connection, $Q(\c) = -i\c^2$, $Q(\CA) = \d_{\CA}\c$. From $Q\,\mbf\Phi = \d_{\mbf A}\mbf \Phi$ we find that $\eta^*$ must be the covariant descendant $\phi^{(1)}$, corresponding in the physical theory to $-\psi_-\d t+\frac12 \psi_+ \d\bar z$. Similarly, from transformations of $\mbf B$ and $\mbf \Psi$, we see that $\CA^*$ and $\eta$ must be the descendants of $B$ and $\psi$, respectively, up to a slight mixing with the moment map and superpotential. (We already knew that $\eta = \psi^{(1)}$ in the absence of superpotential.)

Recall that the superpotential $W$ is a polynomial in the chiral fields $\phi_i$ ($i=1,...,\text{dim}\,V$), now promoted to superfields $\mbf \Phi_i$. The fact that $W$ is homogeneous of cohomological degree (R-charge) 2 and twisted-spin 1 ensures that the integral $\int W(\mbf\Phi)$ is non-vanishing. Explicitly, given chirals
\be \mbf \Phi_i = (\phi_i + \eta^*_{t,i}\,\d t+\eta^*_{\bar z,i}\,\d \bar z + \psi^*_{t\bar z,i}\,\d t \d\bar z)dz^{r_i/2}\,, \ee
the superpotential term is
\be \label{W-expanded} \int W(\Phi) = \int \big(\pd_i W(\phi) \psi^*_{t\bar z,i} - \pd_i\pd_j W(\phi) \eta^*_{t,i}\eta^*_{\bar z,j}   \big)\d t\d\bar z \d z \ee
This is clearly reminiscent of the holomorphic superspace integral of the superpotential in the original physical theory. The anti-field $\psi^*_{t\bar z,i}$ corresponds to the physical F-term $F_i$; though the twisted Lagrangian is missing the $|F|^2$ that would set $F_i \sim \ol{\pd_i W}$ on shell. The anti-fields $\eta^*_i$, which we know correspond to the physical $-\psi_{-,i}dt+\tfrac12\psi_{+,i}d\bar z$, appear in a familiar (holomorphic) Yukawa coupling.

We finally note that whenever the Chern-Simons level $k$ is nonzero, the gauge kinetic terms in the action above are actually equivalent to those of a pure, physical Chern-Simons theory. 
The point is that we can form a new gauge field
\begin{equation} 
	\til{\CA} = -\frac{2\pi}{k} B_z\, \d z + \CA
\end{equation}
that now contains components in all three spacetime directions.
The Lagrangian for $\til{\CA}$ is the ordinary physical Chern-Simons Lagrangian $k\, \text{CS}(\CA)$.  Due to of the presence of the anti-field for $B$ and of the ghost $\c$ in the superfield $\mbf{A}$, the term $k \int \mbf{A} \partial \mbf{A}$ modifies the gauge transformations of $B$ (as in \eqref{Qsuperfields}) so that $\til{\CA}$ transforms as an ordinary gauge field. 

Therefore, at non-zero level, the twisted theories are standard Chern-Simons theories coupled to an unusual type of matter. 

\subsection{Brackets and superfields}
\label{sec:brackets}

In Section \ref{sec:chiral-bulk}, we gave a very general argument (following \cite{descent,YagiOh}) that local operators in the holomorphic twist of a 3d $\CN=2$ theory are endowed with a Poisson bracket $\{\!\!\{-,-\}\!\!\}$ of degree -1. The bracket was defined by a descent procedure \eqref{bracket}, and we refer to it as the secondary bracket.

In the BV formalism, any QFT is endowed with \emph{another} Poisson bracket, of degree +1. We will refer to it as the BV bracket, and denote it $\{-,-\}_{\rm BV}$. It acts on the space of all operators, both local and nonlocal. Two key properties are that the bracket of the action with any operator defines the BRST differential
\be Q\,\CO =  \{\CO,S\}_{\rm BV}\,, \ee
and that field/anti-field pairs have canonical brackets
\be \{\varphi(x),\varphi^*(y) \} = \delta(x-y) \d\text{Vol}\,. \ee
These together imply that the differential acts on fields and anti-fields to produce equations of motion
\be Q\, \varphi^* = -\frac{\delta S}{\delta \varphi}\,,\qquad Q\,\varphi = \frac{\delta S}{\delta \varphi^*}\,.\ee

Given a 3d $\CN=2$ theory written in the BV formalism, it turns out that the secondary bracket and the BV bracket are closely related. 
To motivate this, suppose we have a $Q$-closed field $\varphi$ and an operator $\CO$ whose second descendant coincides with the anti-field $\varphi^*$,
\be \varphi^* =  \CO^{(2)}\,dz  \;+\,(\text{$Q$-exact})\,. \ee
(The second descendant satisfies $Q\CO^{(2)}=\d'\CO^{(1)}$ if $\CO$ is gauge-invariant, and may be computed as $\CO^{(2)} =  \big( \tfrac{i}{2}Q_+\,\d\bar z - i Q_- \, \d t  \big)^2\CO =  Q_+Q_-\CO\,\d t \d\bar z$.) Then the secondary bracket of $\CO$ and $\varphi$ is
\begin{align} \{\!\!\{\CO,\varphi\}\!\!\}(x) &= \oint_{S^2} \CO^{(1)}dz \cdot \varphi(x)  \notag \\
 &= \int_B d'\CO^{(1)} dz \cdot \varphi(x)\qquad \text{(by Stokes)} \notag \\
 &= \int_B Q(\varphi^*) \cdot \varphi(x) \notag \\
 &= -\int_B \frac{\delta S}{\delta\varphi} \cdot \varphi(x) \;\;= \; -1\,.
\end{align}
Here $B$ is a 3-ball that contains the point $x$, and in the last line we use the canonical correlation function $\frac{\delta S}{\delta\varphi}(y) \varphi(x) \sim \delta(y-x) \text{dVol}$ that follows from integration by parts in the path integral. We have thus found that $\{\!\!\{\CO,\varphi\}\!\!\}(x) = \big\{ \int_B \CO^{(2)}dz,\CO(x)\big\}_{\rm BV}$. More generally, we expect that for any pair of $Q$-closed local operators $\CO_1$ and $\CO_2$,
\be \{\!\!\{\CO_1,\CO_2\}\!\!\}(x) =  \Big\{\int_B\CO_1^{(2)}dz,\CO_2(x)\Big\}_{\rm BV}\;+\,(\text{$Q$-exact})\,, \ee
for any choice of 3-ball $B$ containing the point $x$. This should be related to the idea that a deformation of the bulk theory by $\int\CO_1^{(2)}dz$ changes the BRST differential by $\{\!\!\{\CO_1, \cdot\}\!\!\}$.

This relation between the secondary bracket and BV bracket has a particularly nice reformulation in terms of superfields. The superfields introduced in Section \ref{sec:superfields} for 3d $\CN=2$ gauge theories have canonical BV brackets,
\be \{ \mbf \Phi(x),\mbf \Psi(y)\}_{\rm BV} =  \{\mbf A(x),\mbf B(y)\}_{\rm BV} = \delta(x-y)\text{dVol}\,. \label{super-brackets} \ee
This is easy to see by writing out the fields in components, and using the BV brackets between fields and anti-fields. All the SUSY/BRST transformations \eqref{Qsuperfields} then follow by taking appropriate derivatives of the action,
\be Q\mbf \Phi = \frac{\delta}{\delta \mbf \Psi}S\,,\quad Q\mbf \Psi = - \frac{\delta}{\delta \mbf \Phi}S\,,\qquad
 Q\mbf A = \frac{\delta}{\delta \mbf B}S\,,\quad Q\mbf B = - \frac{\delta}{\delta \mbf A}S\,. \ee
$Q$-invariance of the action is equivalent to the classical BV master equation $\{S,S\}_{\rm BV}=0$.

More interestingly, since taking a second descendant relates the bottom component of each superfield to its top component, we find that the canonical BV brackets \eqref{super-brackets} get related to the secondary brackets of bottom components
\be \{\!\!\{ \phi,\psi\}\!\!\} = 1\,,\qquad \{\!\!\{\c,B\}\!\!\} =1\,. \ee
(Strictly speaking, $\psi$, $\c$, and $B$ are not $Q$-closed in the presence of a superpotential, nonabelian gauge symmetry, and matter/CS-terms, respectively; but these elementary brackets may nevertheless be used to generate the brackets of actual $Q$-closed and local operators.) The bracket $\{\!\!\{\c,B\}\!\!\} =1$ may look unfamiliar. In \eqref{gammabracket} we found instead that an abelian gauge theory has a bracket $\{\!\!\{\tilde\gamma,\lambda_-\}\!\!\} \sim 1$ between the dual photon and $\lambda_-$ gaugino. The two expressions are related by removing a $\pd_z$ derivative from $\lambda_-$ (recalling that $\pd_z \c \sim \lambda_-$) and placing it instead on the dual photon (recalling that $\pd_z\tilde\gamma \sim B$).

\subsection{Shifted geometry}
\label{sec:geometry}

The superfield formalism also makes manifest a shifted symplectic structure on the space of fields, dual to the shifted Poisson structure on operators given by the BV bracket (or the secondary bracket, depending on how shifts/descendants are counted).

Given an algebraic variety $\CX$, we denote its shifted cotangent bundle as
\be T^*[1](1)\CX\,, \ee
where $[1]$ and $(1)$ are shifts in the cohomological and twisted-spin gradings, respectively. Explicitly, linear functions on the cotangent fibers of $T^*[1](1)\CX$ have cohomological degree $+1$ and spin $J=1$. The notion of shifted cotangent bundle can be naturally extended to super-varieties, or more generally to graded dg schemes $\CX$, meaning spaces $\CX$ whose rings of functions have both cohomological and spin gradings, and are equipped with a differential $Q$.

Here we are interested in the case that $\CX$ is a graded dg vector space. In a gauge theory, the pair of superfields $\mbf A$ and $\mbf B$ from \eqref{superfields} can be naturally combined into a single superfield taking values in a shifted cotangent bundle
\be \label{cotan-gauge} (\mbf A,\mbf B) \in T^*[1](1) \big( \alg^\bullet \otimes \g[1]\big)\,. \ee
Similarly, the two superfields associated to chiral matter can be combined into
\be \label{cotan-chiral} (\mbf \Phi,\mbf \Psi) \in T^*[1](1) \big(  \oplus_r \alg^{\bullet,(r/2)} \otimes V^{(r)} \big)\,. \ee
In each case, the BV/secondary bracket on operators is directly induced from the cotangent-bundle geometry.

%
%


\subsection{Boundary conditions in the holomorphic twist}
\label{sec:bdy-sum}

Several basic classes of supersymmetric boundary conditions for 3d $\CN=2$ gauge theories were studied systematically in \cite{DGP-duality}, extending previous work of \cite{GGP-walls, YoshidaSugiyama, GGP-fivebranes, OkazakiYamaguchi}. These were half-BPS boundary conditions preserving 2d $\CN=(0,2)$ SUSY and a $U(1)_R$ symmetry, which were automatically compatible with the holomorphic twist.
We can now reinterpret the basic boundary conditions of \cite{DGP-duality} in terms of superfields, in the twisted formalism.

Each 3d $\CN=2$ chiral multiplet gives rise to a pair of twisted superfields $\mbf \Phi,\mbf \Psi$.
 In the absence of superpotential, $Q$ acts independently on each superfield, $Q\,\mbf \Phi=d'\mbf \Phi$ and $Q\,\mbf\Psi = d'\mbf \Psi$. The two basic $Q$-invariant boundary conditions at $t=0$ are\footnote{We use the slash notation ``$\big|$'' as shorthand for evaluation at $t=0$}
\begin{enumerate} 
	\item Dirichlet b.c., abbreviated $D$, which sets $\mbf{\Phi}\big| = 0$\,. \\
	In components, this fixes $\phi\big|=0$ (whence the name Dirichlet) and leaves the fermion $\psi$ (\emph{i.e.} $\bar\psi_-$) unconstrained. This b.c. also leaves unbroken the exotic gauge symmetry generated by $\psi$.
	
	The Dirichlet b.c. has a natural deformation to $\mbf \Phi\big| = \mbf C$, where $\mbf C$ is a constant boundary superfield. At generic $\mbf C$, this will break $U(1)_R$ symmetry unless $\mbf \Phi$ has R-charge zero.
		
	\item Neumann b.c., abbreviated $N$, which sets $\mbf{\Psi}\big| = 0$\,. \\
	In components, this fixes $\psi\big|=0$ as well as $\eta\big|=0$, which physically corresponds to $\pd_t\phi\big|=0$, whence the name Neumann. The boundary value of $\phi$ is unconstrained. This b.c. explicitly breaks the exotic gauge symmetry generated by $\psi$. \end{enumerate}
Note that both N and D b.c. satisfy the primary purpose of boundary conditions: they ensure vanishing of the boundary variation of the half-space action $S = \int_{\R_-\times \C} \mbf \Psi \d' \mbf \Phi$.

%

If there are multiple chiral multiplets, with scalars $\phi$ valued in $V$, we can mix D b.c. for some chirals and N b.c. for others. A geometric way to say this is that we choose a $U(1)_R$-invariant subspace $L \subset V$, denote its complement as $L^\perp \subset V^\vee$, and restrict
\be \mbf \Phi\big| \in  \oplus_r \alg^{\bullet,(r/2)} \otimes L^{(r)}\,,\qquad \mbf \Psi\big| \in  \oplus_r \alg^{\bullet,(1-r/2)} \otimes L^{\perp (-r)} [1]\,. \label{L-bc} \ee
Taken together, the pair of superfields is restricted to the shifted conormal bundle
\be \label{N*bc} (\mbf \Phi,\mbf \Psi)\big| \,\in\, N^*[1](1) \big( \oplus_r \alg^{\bullet,(r/2)} \otimes L^{(r)}\big)\,,\ee
within the shifted cotangent bundle \eqref{cotan-chiral}.
Note that \eqref{N*bc} defines a Lagrangian subspace of the bulk fields.

When there is a superpotential, these basic boundary conditions on the chiral multiplets may be anomalous. This ``Warner problem'' is familiar from the 2d B-model \cite{Warner, KapustinLi}, and its resolution requires the addition of boundary matter, leading there to matrix factorizations. Similarly, boundary conditions for 3d $\CN=2$ theories with superpotentials may require extra boundary matter \cite{GGP-fivebranes, DGP-duality, BrunnerSchulzTabler}. We'll discuss the implications in Section \ref{sec:W}.

Similarly, the basic boundary conditions for the gauge multiplet are
\begin{enumerate}
	\item Dirichlet b.c., abbreviated $\mc{D}$, which sets $\mbf{A}\big| = 0$\,. \\
	This breaks gauge symmetry explicitly at the boundary (to a flavor symmetry), setting to zero both the ghost $\c\big|$ and the connection $A_{\bar z}\big| =0$.
	\item Neumann b.c., abbreviated $\mc{N}$, which sets $\mbf B\big|=0$. \\
	This sets $B_z\big| \sim F_{zt}\big| = 0$, which is (part of) the physical Neumann boundary condition. Gauge symmetry is preserved, at least classically.
\end{enumerate}
In the presence of matter and/or Chern-Simons terms, the gauge symmetry on a Neumann b.c. is anomalous, and additional boundary matter must be added to cancel the anomaly \cite{GGP-fivebranes, DGP-duality}. We'll discuss this further in Section \ref{sec:gauge-N}.

More generally, one can introduce a mixed boundary condition for gauge theory that breaks the group $G$ to a subgroup $H$. Letting $\mathfrak h\subset \g$ denote the unbroken subalgebra, this boundary condition is succinctly described as 
\be (\mbf A,\mbf B)\big| \,\in\, N^*[1](1)\big( \alg^\bullet\otimes \mathfrak h[1]\big)\,. \label{GHgeom}\ee
This is manifestly a Lagrangian subspace of the cotangent bundle \eqref{cotan-gauge}. 
The case $H=G$ is full Neumann and the case $H=1$ is full Dirichlet.
If additionally there is chiral matter present, valued in a $G$-representation $V$, then boundary conditions for the chiral multiplets must be chosen as well. The basic b.c. for chirals are again of the form \eqref{N*bc}, labelled by subspaces $L\subset V$; however, to be compatible with \eqref{GHgeom}, one must require that $L$ is $H$-invariant.

\subsection{RG flow independence}
\label{sec:RG}

Let's now revisit the claim from Section \ref{sec:dg} that the $Q$-cohomology of bulk and boundary operator algebras is constant under RG flow.

For the class of 3d gauge theories we consider, all parameters in the twisted action \eqref{actionWk} are either complex or (in the case of the Chern-Simons couplings) discrete. Notably, some of the continuous real parameters of physical 3d $\CN=2$ gauge theory (such as gauge couplings) have entirely disappeared in the twisted action because their variations are $Q$-exact. Others have been complexified; for example, real masses have combined with the $t$-components of background flavor connections. The boundary conditions we consider also depend exclusively on complex parameters.

Discrete parameters cannot vary continuously with energy scale, so we will disregard them. Continuous complex parameters can of course vary.  The (full) algebras of bulk and boundary local operators in the twisted formalism, as well as the action of $Q$, will depend on them. However, complex conjugates never appear in the twisted action \eqref{actionWk}, so dependence on the complex parameters is meromorphic. Any potential jumps in the $Q$-cohomology of local operators will occur at loci in parameter space of \emph{complex} codimension $\geq 1$.%
\footnote{These sorts of complex-codimension jumping loci are distinct from wall crossing, and featured in recent generalizations of the elliptic genus \cite{KachruTripathy-Hodge,KachruTripathy}.} %
Generic RG flows will avoid complex codimension $\geq 1$ loci, so the chiral algebras $\CV$ and $\CV_\pd$ will be constant along them.
Non-generic RG flows can be deformed to generic ones by slight variations of the ``bare'' parameters in the action; we always assume this has been done.

This analysis does not guarantee the absence of jumping right at the limit points of RG flow, \emph{e.g.} at IR fixed points. If IR fixed points are disjoint from jumping loci (which is our working optimistic assumption), then chiral algebras of IR-dual theories will be isomorphic \emph{and} the strong expectations of quasi-isomorphism discussed in Section \ref{sec:dg} should hold. Even if there is jumping at an IR fixed point, the algebras $\CV,\CV_\pd$ of IR-dual theories should (at least) still be isomorphic, by the following argument.

Consider a pair of 3d theories and boundary conditions $(\CT,\CB)$, $(\CT',\CB')$ that flow to the same fixed point $(\CT_{\rm IR},\CB_{\rm IR})$. We may flow into an arbitrarily small neighborhood of the fixed point, along trajectories $\gamma, \gamma'$, while avoiding jumping loci:
\be \raisebox{-.5in}{\includegraphics[width=2in]{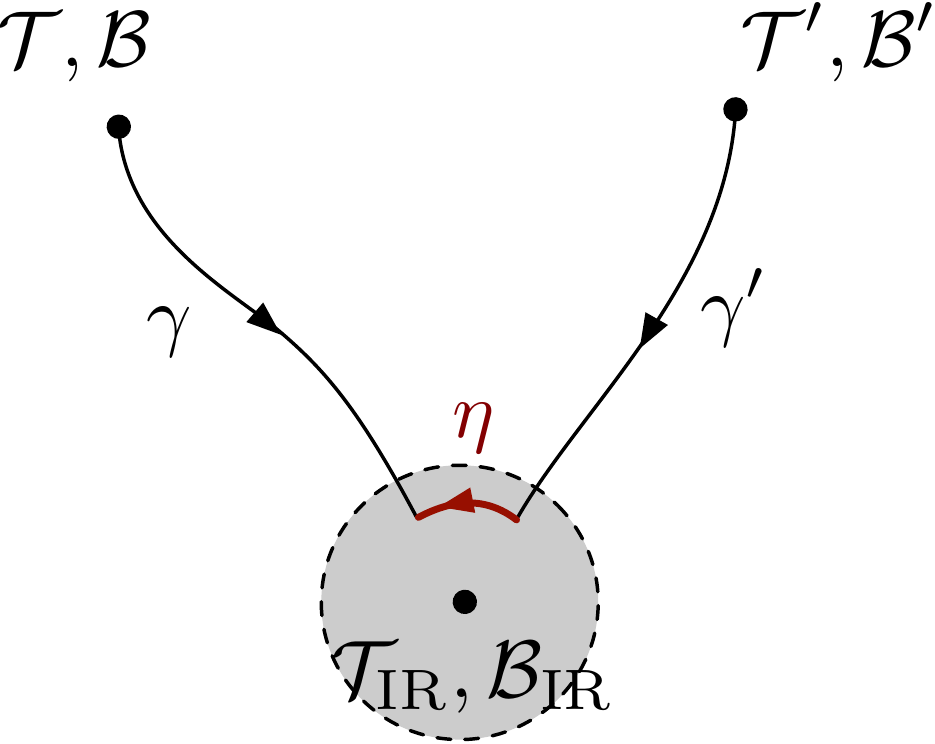}} \ee
Within this neighborhood, we may then connect the theories `by hand' by varying parameters along a path $\eta$, again avoiding jumping loci. The cohomologies $\CV,\CV_\pd$ are constant along the continuous path $\gamma^{-1}\circ\eta\circ\gamma$, establishing an isomorphism $(\CV,\CV_\pd)\simeq (\CV',\CV_\pd')$. If the variation $\eta$ is sufficiently small, we might still hope for a quasi-isomorphism of underlying dg  chiral algebras as well, though its existence is not obvious.


\section{Chiral algebras for free chirals}
\label{sec:freechiral}

We now begin to construct examples of bulk and (especially) boundary chiral algebras. The simplest examples arise in theories of free chiral multiplets.

\subsection{Single chiral}
\label{sec:singlechiral}

Consider a single free chiral multiplet. In notation of previous sections, this means
\be V = V^{(r)} \simeq \C \ee
has a single graded component of dimension one. We are free to choose its $U(1)_R$ charge `$r$' arbitrarily. Then there are two superfields $\mbf \Phi \in \alg^{\bullet,(r/2)}$, $\mbf \Psi \in \alg^{\bullet,(1-r/2)}$.

The bulk algebra was described in \cite{YagiOh}. The $Q$-closed local operators in the bulk are polynomials (or ``words'') in the bottom components $\phi,\psi$ of the two superfields, and their $\pd_z$ derivatives, for example
\be \CO(x) =  \phi(x)^2\pd_z \phi(x) \pd_z\psi(x)+\cdots \ee
(where $x= (t,z,\bar z)$ jointly denotes the bulk spacetime coordinates). The basic operators $\{\pd_z^n\phi\},\{\pd_z^m\psi\}$ all have non-singular OPE's with each other; there are no nontrivial Feynman diagrams contributing to their correlation functions. Thus, we find that the bulk algebra is strongly generated (in the sense of chiral algebras, or, equivalently, vertex algebras) by the two ``fields''%
\footnote{Here we mean ``fields'' in the standard sense of chiral algebras as well. Physically, they are really operators.} $\phi(z)$, $\psi(z)$, which we write as
\be \CV[\text{free chiral}] = \langle\!\langle \phi(z),\psi(z)\rangle\!\rangle\,, \ee
with necessarily trivial OPE. The Poisson bracket of the generators is
\be \{\!\!\{\phi,\psi\}\!\!\} = \{\!\!\{\psi,\phi\}\!\!\} = 1 \qquad\text{(identity operator)}\,, \ee
and the supercurrent that generates $\pd_z$ derivatives via the bracket \eqref{Gdz} is
\be G = \phi\pd_z\psi\,. \ee

There is a simple geometric description of this algebra as well, in terms of the jet scheme of an odd-shifted-symplectic variety. It generalizes the well-known Poisson vertex algebra structure on jet schemes of even-shifted-Poisson varieties \cite{BeilinsonDrinfeld2004, Arakawa-PVA}.

Let us endow the space $V$ with an additional twisted-spin grading, so that $V=V^{(r,r/2)}$ has $R=r$ and $J=r/2$; and let $T^*[1](1)V$ denote its shifted cotangent bundle as in \eqref{GHgeom}. Note that the fields $(\phi,\psi)$ are naturally valued in $T^*[1](1)V$. To capture their derivatives, we further consider the infinitesimal neighborhood $D$ of the origin in the $\C_{z,\bar z}$ plane, such that the algebraic functions on $D$ are formal series, $\C[D] \simeq \C[\![z]\!]$. Then the derivatives of $\phi,\psi$ provide coordinates on the space of algebraic functions $\text{Maps}\big(D,T^*[1](1)V\big)\simeq \CJ_\infty T^*[1](1)V$, otherwise known as the ``infinite jet space'' of $T^*[1](1)V$. Therefore, the bulk chiral algebra $\CV$, which consists of polynomials in $\phi,\psi$ and their derivatives, is simply the algebra of functions on the infinite jet space,
\be \CV[\text{free chiral}] \simeq \C\big[\CJ_\infty T^*[1](1)V\big]\,. \label{jet} \ee
The Poisson bracket is canonically induced from the shifted symplectic structure on $T^*[1](1)V$.
\footnote{We should stress that even the free chiral theory may admit higher operations. For example, our calculations in Section \ref{sec:W} may be seen as evidence for a non-trivial higher operation 
involving three operators colliding along the topological direction: two fermions and a polynomial in the bosons of at least cubic degree.}.

Now let's describe the boundary algebras. On a Dirichlet b.c. $\mbf \Phi\big| = 0$, the bulk operator $\psi$ (and its derivatives) survives whereas $\phi$ (and its derivatives) is set to zero. Thus
\begin{align} \CV_\pd[\text{free chiral},D] &= \langle\!\langle \psi(z)\rangle\!\rangle \\ 
 &\simeq  \C\big[\CJ_\infty N^*[1](1)\,0\big]\,. \notag \end{align}
On the second line, `0' denotes the origin of $V$,  \emph{i.e.} the subspace to which D b.c. restricts $\phi$. The bulk-boundary map is obviously
\be \rho: \begin{array}{ccc} \CV &\to& \CV_\pd[D] \\
 (\phi,\psi) &\mapsto & (0,\psi) \end{array}\,. \ee
Moreover, the kernel of $\rho$, generated by $\phi$, is clearly preserved by the Poisson bracket. (This was a general requirement from Section \ref{sec:chiral-bdy}.)

Since the boundary operator $\psi$ is in the image of the bulk-boundary map, it must have a trivial OPE with itself. (The entire boundary algebra $\CV_\pd$ has trivial OPE's.) This property could also be derived from symmetry principles: $\psi$ has spin $J=1-r/2$ and R-charge $R=1-r$, and there is no way to build a singular $\psi(0)\psi(z)$ OPE that respects these charges.

Similarly, on a Neumann b.c. $\mbf \Phi\big|=0$, the bulk operator $\psi$ is killed while $\phi$ survives. The boundary chiral algebra is 
\begin{align} \CV_\pd[\text{free chiral},N] &= \langle\!\langle \phi(z)\rangle\!\rangle \\ 
 &\simeq  \C\big[\CJ_\infty N^*[1](1)\,V\big]\,. \notag \end{align}
The $\phi(0)\phi(z)$ OPE is necessarily trivial, for the same reasons as in the case of D b.c.; for example, $\phi$ is in the image of the bulk-boundary map
\be \rho: \begin{array}{ccc} \CV &\to& \CV_\pd[N] \\
 (\phi,\psi) &\mapsto & (\phi,0) \end{array}\,, \ee
and also has charges $J=r/2,R=r$ that are incompatible with a singular OPE.

\subsubsection{Characters}

For any chiral algebra $\mathcal Y$ with homological grading $R$, twisted-spin grading $J$, and (potentially) other non-homological gradings $e$, we define the character 
\be \chi[\mathcal Y] = \text{Tr}_{\mathcal Y}\, e^{i\pi R} q^J x^e\,. \label{defchi} \ee
For a bulk chiral algebra $\CV$ associated to a 3d $\CN=2$ theory in the holomorphic twist, this coincides with the standard supersymmetric index%
\footnote{Being more careful: the standard supersymmetric index is usually defined following the original \cite{Witten-index} as $\text{Tr}(-1)^Fq^J...$ rather than $\text{Tr}e^{i\pi R}q^J...$. The two expressions are equivalent, up to a redefinition of $q$. Namely, physical fermion number is related to \emph{untwisted} spin as $F=2J_0$ (mod 2); thus, keeping in mind that $J=R/2-J_0$, we see that shifting $q\to e^{-2\pi i}q$ relates the character \eqref{defchi} to the standard index.} %
 of \cite{Kim-index,IY-index}. (It was further generalized in \cite{KW-index} to include flavor monopole sectors, which we do not consider here.)  For boundary chiral algebras, the character coincides with the ``half-index'' that was constructed in increasingly general contexts by 
\cite{GGP-walls, GGP-fivebranes, YoshidaSugiyama, DGP-duality}.

In the case of the bulk and boundary algebras associated to a free chiral above, there is an additional non-homological grading, corresponding to a $U(1)_e$ flavor symmetry that acts on $(\phi,\psi)$ with charges $e=(+1,-1)$. Let us choose the R-charge of $\phi$ to be $R(\phi)=r=0$. Then the bulk character is
\be \chi[\CV]_{r=0} = \frac{(qx^{-1};q)_\infty}{(x;q)_\infty}\,, \label{chi-free-bulk} \ee
where
\be (z;q)_\infty := \prod_{n=0}^\infty (1-q^n z)\,. \label{def-qPoch} \ee
The numerator of \eqref{chi-free-bulk} counts operators in the fermionic Fock space generated by the operators $\{\pd_z^n\psi\}_{n=0}^\infty$ with $R=1,J=n+1$; whereas the denominator counts operators in the bosonic Fock space generated by $\{\pd_z^n\phi\}_{n=0}^\infty$ with $R=0,J=n$. Also note that the character for general R-charge $R(\phi)=r$ is obtained from the one above by redefining $x\to e^{i\pi r}q^{\frac r2}x$ (since different choices of R-charge come from mixing $U(1)_R$ with $U(1)_e$ flavor symmetry).

The characters of the boundary algebras for a free chiral are
\be \chi[\CV_\pd[D]]_{r=0} = (qx^{-1};q)_\infty\,,\qquad \chi[\CV_\pd[N]]_{r=0}=\frac{1}{(x;q)_\infty}\,,\ee 
counting either $\pd_z^n\psi$ or $\pd_z^n\phi$ operators, respectively. This agrees with \cite{GGP-walls, GGP-fivebranes}.

\subsection{Interval compactification vs. boundary conditions}
\label{sec:interval}

Another way to think about $\CV_\pd[\text{free chiral},N]$ is as \emph{half} of a $\beta\gamma$ system. The standard 2d $\beta\gamma$ system (\emph{a.k.a.} symplectic bosons) has an OPE
\be \beta(z)\gamma(0) \sim \frac{1}{z}\,. \label{bgOPE} \ee
We can actually expect to obtain the full $\beta\gamma$ system by compactifying a 3d $\CN=2$ free chiral multiplet on an interval $[0,1]\times \C$ with Neumann b.c. on both sides. As the interval is squeezed to zero size (or as we flow to the IR), this becomes a purely two-dimensional $\CN=(0,2)$ theory with a single chiral multiplet, whose chiral algebra in the holomorphic twist is well known to be a $\beta\gamma$ system \cite{Witten-CDO, Nekrasov-betagamma}. In contrast, with a single Neumann b.c., we only get (say) the field $\phi = \beta$, but are missing `$\gamma$.'

It is instructive to analyze interval compactification from the perspective of the holomorphic twist. Consider the setup suggested above: a 3d chiral sandwiched between two Neumann b.c. There is a chiral algebra generated by $\phi(z)$ on each boundary. However, these algebras are not independent:  the operator $\phi(z)$ can be moved off one boundary, into the bulk, and onto the other boundary. Thus, interval compactification ultimately provides a \emph{single} copy of $\phi(z)$. This is puzzling, since we expected a second boson `$\gamma$' to appear.

The puzzle is resolved by adding line operators stretched between the two boundaries.%
\footnote{We thank N. Paquette for discussions related to this idea \cite{WIP-Paquette}.} %
Consider the bulk operator $\psi$, which is set to zero at the boundaries. The integral of its first descendant
\be \gamma(z) := \int_0^1 \psi^{(1)} \ee
is $Q$-closed by Stokes' theorem, since $Q\gamma = \int_0^1 Q\psi^{(1)} = \int_0^1 d'\psi = \psi\big|^{t=1}_{t=0} = 0$. 
Therefore, the full chiral algebra on the interval is generated by both $\beta(z):=\phi(z)$ and $\gamma(z)$. More so, the pole in the 2d OPE between $\beta$ and $\gamma$ is induced from the bulk Poisson bracket:
\be \label{sandwichOPE} \oint_{S^1} \beta(0)\gamma(z)\,dz = \int_{S^1\times[0,1]} \psi^{(1)}dz\cdot \phi(0)
 \overset{Fig. \ref{fig:bdybracket}}{\simeq}  \oint_{S^2}  \psi^{(1)}dz\cdot \phi = \{\!\!\{ \psi,\phi \}\!\!\} = 1\,.
 \ee
In the first equality, we rewrite a circle integral (in the 2d compactified theory) as a cylinder integral in the 3d interval setup.
For the second equality here, we pull the cylinder off the boundaries (just as in Figure \ref{fig:bdybracket}), to obtain a purely bulk configuration in which $\psi^{(1)}dz$ is integrated on a two-sphere surrounding $\phi$. This defines the bulk Poisson bracket. More generally, the complete 2d OPE can be extracted by inserting additional powers of $z$ in \eqref{sandwichOPE}, and using the bulk $\lambda$-bracket, \emph{cf.} \cite{YagiOh} and Section \ref{sec:chiral-bulk}. The result, as expected, is \eqref{bgOPE}. 


Similarly, if we consider a 3d free chiral sandwiched between two Dirichlet b.c., we find find a 2d chiral algebra generated by $b(z):= \psi(z)$ (identified with the boundary operators, on either of the boundary conditions) together with the integrated descendant
\be c(z) := \int_0^1 \phi^{(1)}\,. \ee
The OPE, induced from the bulk Poisson bracket, is $b(z)c(0)\sim 1/z$, so this is a standard $bc$ system. This agrees with the expectation that the holomorphic twist of a purely 2d $\CN=(0,2)$ fermi multiplet is a $bc$ system. On a \emph{single} Dirichlet b.c., we only have \emph{half} of the $bc$ system, generated by $\psi$.

These examples, already nontrivial, only illustrate some of the features (and complications) of general compactifications. The general procedure should look roughly as follows. Suppose we have a 3d $\CN=2$ theory with bulk chiral algebra $\CV$, and two boundary conditions with algebras $\CV_\pd^L$ and $\CV_\pd^R$. There are bulk-boundary maps
\be \rho^L:\CV\to \CV_\pd^L\,,\qquad \rho^R:\CV\to \CV_\pd^R\,. \ee
inducing actions of the bulk algebra on each boundary algebra. 
Naively, an interval compactification between the boundary conditions would produce
\be \CV_{I}^{(0)} = \CV_\pd^L \otimes_\CV \CV_\pd^R\,, \label{interval-naive} \ee
where the tensor product over $\CV$ means we identify operators in $\CV_\pd^L$ and $\CV_\pd^R$ related by the action of the bulk algebra. (Explicitly: given bulk $\CO$ and left and right boundary operators $v$ and $w$,  $\rho^L(\CO) v\otimes w \simeq v\otimes \rho^R(\CO)w$.) The naive answer \eqref{interval-naive} must then be extended by line operators that stretch between the boundaries; this includes
\begin{itemize}
\item integrals of all descendants from one boundary to another (mathematically, these are included by turning \eqref{interval-naive} into a derived tensor product)
\item other line operators, such as Wilson and vortex lines, that belong to the category of (bi)modules for $\CV_\pd^L \otimes_\CV \CV_\pd^R$ and are mutually local with respect to one another.
\end{itemize}
Recent examples of extension by other line operators include \cite{GaiottoRapcak, ProchazkaRapcak, FeiginGukov}. Finally, the action of $Q$ and the OPE in the resulting algebra may be corrected by finite-action instantons (dynamical vortices) that stretch between the two boundaries.

\subsection{Boundary couplings and flips}
\label{sec:flip1}

It was argued in \cite{DGP-duality} that Neumann b.c. for a 3d chiral can be ``flipped'' to Dirichlet b.c. by adding a 2d $\CN=(0,2)$ boundary fermi multiplet, and an extra boundary superpotential interaction.%
\footnote{The term ``flip'' originates in \cite{DGG}, where a similar mechanism was used to relate half-BPS boundary conditions for 4d hypermultiplet.} %
Similarly, Dirichlet b.c. can be flipped to Neumann by adding a 2d $\CN=(0,2)$ chiral multiplet, and a boundary superpotential. We would like to describe the effect of flips on boundary chiral algebras.

\subsubsection{2d degrees of freedom}
\label{sec:2d}

We first need to review some simple facts about the holomorphic twist of 2d $\CN=(0,2)$ chiral and fermi multiplets, and their superpotential deformations. (For the standard superspace formulation of 2d $\CN=(0,2)$ GLSM's, see the classic \cite{DineSeiberg-02, Witten-phases} or the many recent reviews, \emph{e.g.} \cite{McOrist-02}.) In the BV formalism, a twisted 2d chiral multiplet of R-charge $r$ can be re-organized into a pair of superfields
\be \mbf C \,\in\, \alg_{2d}^{\bullet,(r/2)}\,,\qquad \mbf {\wt C} \,\in\, \alg_{2d}^{\bullet,(1-r/2)}\,, \ee
where $\alg_{2d}^\bullet = \Omega^\bullet(\C)/(\d z) \simeq C^\infty(\C)[\d\bar z]$ is the quotient of the de Rham complex of 2d spacetime $\C$ by $(\d z)$ (this is just the 2d analogue of the 3d algebra $\alg^\bullet$ in \eqref{def-alg}); and $\alg_{2d}^{\bullet,(j)} = \alg_{2d}^\bullet dz^j$ is a further twist by the $j$-th power of the canonical bundle on $\C$. In components, we have
\be \mbf C = C + \wt C^* \d\bar z\,,\qquad \mbf {\wt C} = \wt C + C^* \d\bar z \ee
where $C$ is the complex boson in the physical chiral multiplet; $\wt C$ corresponds to $\pd_z\ol C$; and the fermionic anti-fields $\wt C^*, C^*$ correspond to right-handed physical fermions sometimes written as $\psi_+,\ol \psi_+$, respectively. The twisted action is
\be S_{\text{2d chiral}} = \int_\C  \mbf {\wt C}\, \bar\pd \mbf C\,, \ee
with anti-holomorphic exterior derivative $\bar\pd = \pd_{\bar z} \d \bar z$. The action of the SUSY/BRST operator  $Q(\mbf C) = \bar \pd \mbf C$, $Q(\mbf{\wt C}) = \bar\pd \mbf{\wt C}$ is induced by the BV anti-bracket
\be \{\mbf C(x),\mbf{\wt C}(y) \}_{\text{BV}} = \delta(x-y)\text{dVol}\,.\ee
The Q-cohomology of the 2d operator algebra is a chiral algebra strongly generated by the bosonic components $C,\wt C$, with an OPE
\be C(z)\wt C(0) \sim \frac{1}{z} \ee
that can be calculated from the propagator between $C$ and $\wt C$. This is simply the $\beta\gamma$ system from \eqref{bgOPE}, with $C$ and $\wt C$ playing the roles of $\beta$ and $\gamma$.

The BV description of a 2d fermi multiplet in the holomorphic twist is almost identical. A fermi multiplet of R-charge $r$ gets rewritten in terms of a pair of superfields
\be \mbf \Gamma \,\in\, \alg_{2d}^{\bullet,(1/2+r/2)}\,,\qquad \mbf{\wt \Gamma} \,\in\, \alg_{2d}^{\bullet,(1/2-r/2)}\,, \ee 
with \emph{fermionic} bottom components $\Gamma,\wt \Gamma$, corresponding physically to a left-handed complex fermion and its conjugate. There is an action
\be S_{\text{2d fermi}} = \int_\C \mbf {\wt\Gamma}\,\bar\pd \mbf \Gamma\,, \ee
anti-bracket $\{\mbf \Gamma,\mbf{\wt \Gamma}\}_{\text{BV}} = \delta(x-y)\text{dVol}$, and SUSY/BRST operator $Q(\mbf\Gamma) = \bar\pd\Gamma$, $Q(\mbf{\wt\Gamma}) = \bar\pd\mbf{\wt\Gamma}$. Now the $Q$-cohomology of local operators is a chiral algebra strongly generated by $\Gamma,\wt \Gamma$ with
\be \Gamma(z)\wt\Gamma(0) \sim\frac{1}{z}\,. \ee
In other words, this is the $bc$ ghost system of the previous section.

Given a collection of chiral multiplets $\mbf C^i,\mbf{\wt C}_i$ and fermi multiplets $\mbf \Gamma^\alpha,\mbf{\wt \Gamma}_\alpha$, superpotential interactions among them can be introduced. In standard $\CN=(0,2)$ superspace, the interactions are usually encoded in two sets of holomorphic functions $E^\alpha(C)$, $J_\alpha(C)$, the so-called E and J terms. Only one set of functions (say, the J terms) can be made manifest in the $\CN=(0,2)$ superspace Lagrangian. In the twisted formalism, the role of E and J terms is beautifully symmetric; the complete twisted action takes the form
\be \label{SEJ} S = \int_\C \sum_i  \mbf {\wt C}_i\, \bar\pd \mbf C^i + \int_\C \sum_\alpha \Big[ \mbf {\wt\Gamma}_\alpha\,\bar\pd \mbf \Gamma^\alpha + \mbf \Gamma^\alpha J_\alpha(\mbf C) + \mbf{\wt \Gamma}_\alpha E^\alpha(\mbf C)\Big]\,.\ee
The BV master equation is
\be \label{EJ2d} \{S,S\}_{\rm BV} = 2\int_\C \sum_\alpha J_\alpha(\mbf C) E^\alpha(\mbf C) = 0\,, \ee
which reproduces the familiar physical requirement for unbroken SUSY: the sum of products $\sum_\alpha J_\alpha E^\alpha$ must be a constant \cite{Witten-phases}.%
\footnote{A constant is allowed because the integral of a constant superfield vanishes. This is equally true in the original physical 2d $\CN=(0,2)$ theory. However, preserving $U(1)_R$ symmetry forces $\sum_\alpha J_\alpha E^\alpha=0$.}

In the presence of E and J terms, the Q-cohomology of local operators can be analyzed perturbatively as follows. The
OPE among $C^i,\wt C_i,\Gamma^\alpha,\wt\Gamma_\alpha$ is still
\be \label{OPE2d-gen} C^i(z)\wt C_j(0) \sim \frac{\delta^i{}_j}{z}\,,\qquad \Gamma^\alpha(z)\wt\Gamma_\beta(0) \sim \frac{\delta^\alpha{}_\beta}{z}\,, \ee
because the E and J terms in \eqref{SEJ} do not contain $\mbf{\wt C}$, and so cannot introduce any Feynman diagrams deforming the OPE from that of a free theory. However, now we have
\be \label{Q2d-superfields} \begin{array}{l@{\qquad}l}
 Q(\mbf C^i) = \bar\pd \mbf C^i\,, & Q(\mbf \Gamma^\alpha) = \bar\pd \mbf \Gamma^\alpha+ E^\alpha(\mbf C)\,, \\[.2cm]
 Q(\mbf {\wt C}_i) = \bar \pd \mbf {\wt C}_i + \sum_\alpha\big[\mbf \Gamma^\alpha \pd_iJ_\alpha(\mbf C) + \mbf{\wt \Gamma}_\alpha \pd_iE^\alpha(\mbf C) \big]\,, & Q(\mbf {\wt \Gamma}_\alpha) = \bar\pd \mbf {\wt \Gamma}_\alpha+ J_\alpha(\mbf C)\,. \end{array}
\ee
In particular, the SUSY/BRST operator acts on leading components as
\be \label{Q2d-gen} Q(C^i)=0\,,\quad Q(\wt C_i) = \sum_\alpha\big(\Gamma^\alpha\pd_i J_\alpha+\wt\Gamma_\alpha\pd_i E^\alpha\big)\,;\qquad
 Q(\Gamma^\alpha) = E^\alpha\,,\quad Q(\wt\Gamma_\alpha) = J_\alpha\,.\ee
Note that the structure of these transformations is highly constrained in order to ensure both $Q^2=0$ and compatibility with the OPE \eqref{OPE2d-gen}. For example, acting on $\wt C_i(z) \wt \Gamma_\beta(0) \sim 0$,
\begin{align} Q\big(\wt C_i(z) \wt \Gamma_\alpha(0) \big) &= Q(\wt C_i(z))  \wt \Gamma_\alpha(0) + \wt C_i(z) Q( \wt \Gamma_\alpha(0))  \notag \\
 &\sim \Gamma^\beta(z)\pd_i J_\beta(C(z)) \wt \Gamma_\alpha(0)  + \wt C^i(z) J_\alpha(C(0)) \\
 &\sim \frac{\pd_i J_\alpha(z)}{z} - \frac{\pd_i J_\alpha(0)}{z}  \sim 0  \notag
\end{align}
by virtue of a cancellation.

Altogether, we expect that with E and J terms, the Q-cohomology of local operators is a chiral algebra $\CV_{\rm 2d}$ that may be modeled as a tensor product of free chiral and fermi algebras (\emph{i.e.} $\beta\gamma$ and $bc$ systems), deformed by the differential \eqref{Q2d-gen}:
\be \CV_{\rm 2d} \simeq H^\bullet  \big\langle\!\!\big\langle C^i(z),\wt C_i(z),\Gamma^\alpha(z),\wt\Gamma_\alpha(z)\,\big|\, \text{OPE \eqref{OPE2d-gen}}, \text{$Q$ as in \eqref{Q2d-gen}}  \big\rangle\!\!\big\rangle \ee

If the E and J terms are not linear, this description may miss some higher $A_\infty$-like operations. We will comment on this later on, in the context of boundary chiral algebras.
Finally, we note that there may be instanton corrections to the structure above, when $E$ and $J$ terms are nonlinear. The powerful arguments of \cite{SilversteinWitten-conformal, BasuSethi-instantons, BeasleyWitten-instantons} do not apply here, since the classical moduli space is noncompact. 
We will not need to consider instantons further in this paper, since the boundary conditions we introduce only involve $E$ and $J$ terms that depend on 3d bulk (rather than 2d boundary) chirals, which preclude instanton corrections. 

\subsubsection{Implementing flips}

Now, suppose we have a 3d chiral multiplet with Neumann b.c.  $\mbf \Psi\big| = 0$, and boundary chiral algebra $\CV_\pd[N]= \langle\!\langle \phi(z) \rangle\!\rangle$.
 Following \cite{DGP-duality}, we expect that this can be ``converted'' to Dirichlet b.c. by introducing a boundary fermi multiplet $\mbf \Gamma,\mbf{\wt\Gamma}$, coupled to the bulk via a J-term,
\be \label{flipND} S' = \int_{\C\times \R_-} \mbf{\Psi}\d'\mbf{\Phi}  + \int_\C \big[\mbf {\wt\Gamma}\bar\pd \mbf\Gamma + \mbf \Phi\big|\,\mbf \Gamma\big]\,.\ee
Here $J(\mbf \Phi) =  \mbf \Phi\big|_{t=0} = (\phi + \eta^*_{\bar z}\d\bar z)\big|_{t=0}$ is a holomorphic function of the bulk superfield $\mbf \Phi$, pulled back to the boundary.

To see the equivalence of \eqref{flipND} with a pure Dirichlet b.c. (in cohomology), first note that the original boundary condition $\mbf \Psi\big|=0$ must be modified to $\mbf \Psi\big| = \mbf \Gamma$ in the presence of the bulk-boundary coupling above, in order to cancel the boundary variation of the action:
\begin{align} \delta S' 
 &= \int_{\C\times \R_-}(\delta\mbf\Psi\d'\mbf\Phi + \d'\mbf\Psi \delta\mbf\Phi) + \int_\C\delta\mbf\Phi|(-\mbf\Psi+\mbf\Gamma)+ (\text{$\delta\Gamma$ terms})
\end{align}
We take $\mbf \Psi\big| = \mbf \Gamma$ as the definition of the boundary condition for the theory \eqref{flipND}.

The new boundary chiral algebra $\CV_\pd'$ is generated by the bottom components $\psi(z) = \Gamma(z)$, $\phi(z)$, and $\wt\Gamma(z)$, with OPE $\psi(z)\wt\Gamma(0) \sim 1/z$ and differential
\be Q(\psi) = Q(\phi) = 0\,,\qquad Q(\wt\Gamma) = \phi\,. \ee
There are no additional corrections to other OPE's, since the interaction vertex $\mbf\Phi\mbf\Gamma$ cannot be used to construct any new Feynman diagrams that would contribute to them. The $Q$-cohomology is trivial to compute: $\tilde\Gamma$ and $\phi$ are eliminated, and we are left with
\be \CV_\pd' \simeq \langle\!\langle \psi(z) \rangle\!\rangle \simeq \CV_\pd[D]\,,\ee
isomorphic to the boundary chiral algebra for Dirichlet b.c.

The inverse flip, which converts Dirichlet b.c. back to Neumann, is almost identical. Starting with Dirichlet b.c. $\mbf \Phi\big| = 0$, we can add a boundary chiral multiplet $\mbf C,\mbf{\wt C}$ with a boundary superpotential interaction $\int_\C \mbf\Psi|\,\mbf C$. In the presence of the interaction, the boundary condition must be modified to $\mbf \Phi\big| = -\mbf C$. The new boundary chiral algebra is generated by $\phi(z)=-C(z)$, $\psi(z)$, and $\wt C(z)$, with OPE
\be \phi(z)\wt C(0) \sim \frac{-1}{z} \ee
and differential
\be Q(\phi) = Q(\psi) = 0\,,\qquad Q(\wt C) = \psi\,.\ee
In cohomology, $\wt C$ and $\psi$ are eliminated, and we are left with a chiral algebra generated by $\phi(z)$ alone --- which is equivalent to $\CV_\pd[N]$.

\subsection{Multiple chirals} 
\label{sec:multiplechirals}

The generalization of bulk and boundary algebras to multiple free chiral multiplets is straightforward, but has a nice geometric formulation in terms of shifted symplectic geometry.

Suppose we have a collection of 3d chiral multiplets, whose physical complex scalars are valued in a vector space $V$. The space $V$ is naturally graded by R-charge $V = \oplus_r V^{(r)}$, and in the twisted formalism the chiral multiplet is encoded in  $(\mbf \Phi,\mbf \Psi) \in T^*[1](1) \big(  \oplus_r \alg^{\bullet,(r/2)} \otimes V^{(r)} \big)$ as described in \eqref{cotan-chiral}. For discussing chiral algebras it is convenient, just as in Section \ref{sec:singlechiral}, to ``move'' the twisted spin grading from $\alg^\bullet$ to $V$; we do this by giving each component $V^{(r)}$ of fixed R-charge $r$ the twisted spin $J=r/2$,
\be V^{(r)} \leadsto V^{(r,r/2)}\,. \label{RJregrade} \ee
Now $V = \oplus_r V^{(r,r/2)}$ has become doubly graded. After this modification, the bulk algebra is
\be \CV = \langle\!\langle \phi(z),\psi(z) \rangle\!\rangle \simeq \C[\CJ_\infty T^*[1](1)V]\,, \ee
with Poisson bracket induced from the shifted-symplectic structure on $T^*[1](1)V$.

Basic boundary conditions are labelled by $U(1)_R$-invariant (and thus also $U(1)_J$-invariant) subspaces $L\in V$. Given such an $L$, the bottom components of  $(\mbf \Phi,\mbf \Psi)$ are restricted on the boundary to
\be \phi\big| \in L \subset V\,,\qquad \psi\big| \in L^\perp \subset V^*[1](1)\,. \ee
Altogether, the boundary algebra may be described as functions on the jet space of the conormal bundle,
\be \CV_\pd[L]  = \langle\!\langle \phi(z),\psi(z) \,\text{s.t.}\, \phi\in L,\psi\in L^\perp \rangle\!\rangle
 \simeq \C[\CJ_\infty N^*[1](1)L]\,. \ee
The OPE is of course trivial. Rewriting $\CJ_\infty N^*[1](1)L \simeq N^*[1](\CJ_\infty L)$, we also see that the embedding $N^*[1](\CJ_\infty L)\subset T^*[1](\CJ_\infty V)$ is shifted-Lagrangian. Dually, the kernel of the bulk-boundary map $\rho:\CV\to \CV_\pd$ is preserved by the bulk Poisson bracket.

\section{Chiral fields with a superpotential}
\label{sec:W}

Boundary conditions and boundary chiral algebras for chiral multiplets get much more interesting in the presence of a bulk superpotential. Very generally, bulk superpotentials will lead to the presence of boundary operators that do not appear in the image of the bulk-boundary map, and can have singular OPE's. In some cases, this will happen because certain operators are $Q$-closed on the boundary but not in the bulk. In other cases, preserving SUSY/BRST symmetry will require the addition of new boundary degrees of freedom.

\subsection{Bulk algebra}
\label{sec:W-bulk}

We first derive the modifications to the bulk algebra in the presence of a superpotential. Suppose we have bulk chiral multiplets $(\mbf \Phi,\mbf \Psi)$ with scalars $\phi$ valued in a vector space $V$ as usual. The superpotential
\be W: V\to \C \ee
is a quasi-homogeneous polynomial of R-charge 2, and twisted spin $J=1$. It deforms the action of the supercharge $Q$ as in \eqref{Qsuperfields}. To be more explicit, let us choose a complex basis for $V$ (and a dual basis for $V^*$), and correspondingly label the individual chirals as $\{\mbf \Phi^i,\mbf \Psi_i\}_{i=1}^n$, $n=\text{dim}_\C V$. Then we have
\be Q\,\mbf\Phi^i = \d'\,\mbf\Phi^i\,,\qquad Q\,\mbf\Psi_i = \d'\,\mbf\Psi_i + \pd W(\mbf \Phi)/\pd\mbf \Phi_i\,. \ee

As in the case of free chirals, only the bottom components $\phi^i,\psi_i$ and their $\pd_z$ derivatives are potentially $Q$-closed. Higher components cancel in cohomology against the $\pd_{\bar z}$ and $\pd_t$ derivatives of lower components. Thus we expect to generate the bulk algebra with $\phi^i(z)$ and $\psi_i(z)$. The $\phi^i(z)$ are always $Q$-closed, while
\be Q\,\psi_i = \pd_iW := \pd W(\phi)/\pd \phi^i\,.  \ee

In principle, the $\psi_i$'s that aren't $Q$-closed could now have a singular bulk OPE. We will compute the OPE momentarily and find this is indeed the case! Nevertheless, after we restrict to $Q$-closed combinations of bulk local operators, the general results of Section \ref{sec:chiral-bulk} \emph{guarantee} that the singular terms in the bulk OPE are $Q$-exact. Therefore, the bulk algebra $\CV$ is simply isomorphic to the cohomology of the commutative chiral algebra
\be \label{Vbulk-W} \CV \simeq H^\bullet\big\langle\!\big\langle \phi^i(z),\psi_i(z)\,\big|\, Q\psi_i = \pd_iW\big\rangle\!\big\rangle\,.\ee
The Poisson bracket on the cohomology induced from the bracket $\{\phi^i,\psi_j(z)\} = \delta^i{}_j$.

As one might hope, this has a natural geometric description as functions on an infinite jet space. Let us promote $V = \oplus_r V^{(r)}$ to a doubly-graded space by giving each component $V^{(r)}$ a twisted-spin degree $r/2$, exactly as in \eqref{RJregrade}. We think of the values of the fields $(\phi^i,\psi_i)$ as coordinates on the shifted cotangent bundle $T^*[1](1)V$. Given a polynomial superpotential $W: V\to \C^{(2,1)}$, we may construct a finite-dimensional dg algebra of polynomial functions on $T^*[1](1)V$, \emph{i.e.} polynomials $\C[\phi,\psi]$, deformed by a differential $\d=dW\cdot \frac{d}{d\psi} = \sum_i \pd_i W(\phi) \frac{\pd}{\pd \psi_i}$. This dg algebra is (by definition) the algebra of functions on an affine dg scheme known as the derived critical locus of $W$, denoted $\text{dCrit}(W)$. The chiral algebra \eqref{Vbulk-W} arises as the algebra of functions on the infinite jet scheme of the derived critical locus,
\be \CV \simeq \C[\CJ_\infty\, \text{dCrit}(W)]\,. \ee

\subsubsection{Bulk OPE}
\label{sec:W-bulk-OPE}

It is a useful exercise to compute the bulk OPE of the fermionic field $\psi_i$ (the only ones for which the OPE is potentially nonsingular). It will prepare us for the computation of boundary OPE's further below. The result is also interesting in its own right, demonstrating the subtleties one encounters in defining holomorphic-topological operations beyond the Q-cohomology. 

We begin by noting that bulk correlators can be computed entirely in perturbation theory. Nonperturbative corrections would arise from instantons. However, there are no bulk instantons localized in both the complex $(z,\bar z)$-plane and the real $t$ direction, since $Q$-fixed field configurations are necessarily constant in $t$. Any nontrivial instanton configuration would be infinitely extended in the $t$ direction, and carry infinite action.

Perturbatively, we work with twisted superfields. (The computation is very similar to \cite[Sec 5.2]{CYW}.) The propagator associated with the $\int \mbf \Phi \d'\mbf \Psi$ kinetic term connects $\mbf \Psi_i$ and $\mbf \Phi^i$ insertions. To write it down explicitly, we must gauge-fix the `exotic' gauge symmetry acting on $\eta$, the one-form component of $\mbf \Psi$. We choose a gauge that sets
\be \d * \eta = \pd\eta_t + 2\pd_z\eta_{\bar z} = 0\,.  \ee
The propagator in this gauge is
\be \begin{array}{c} \mbf \Psi_i(x)\raisebox{.2cm}{\underline{\phantom{xxxxxxxxx}}}\,\mbf \Phi^j(x') \\[.1cm]
 \displaystyle P_i{}^j(x;x') = \frac{3}{16\pi i} \delta_i{}^j \frac{1}{|x-x'|^{3}}\big[ (\bar z-\bar z')\d(t-t')-(t-t')\d(\bar z-\bar z')\big]\d z \end{array} \label{prop-bulk} \ee
This formula is interpreted as follows. First, $x=(z,\bar z,t)$ are the spacetime coordinates of a $\mbf \Psi_i$ insertion, and $x'= (z',\bar z',t')$ are the coordinates of a $\mbf \Phi^j$ insertion. We expand out the wedge product of forms. 
The $\d t,\d \bar z$ forms indicate the superfield components of $\mbf \Psi_i$, while $\d t',\d \bar z'$ indicate the superfield components of $\mbf \Phi^j$. For example, the term $-\frac{3}{16\pi i}\delta_i{}^j|x-x'|^{-3} (\bar z-\bar z')\d t'\d z$ indicates that the two-point function between $\psi_i(x)$ (the 0-form component of $\mbf \Psi_i$) and $\eta^{i,*}_t(x')$ (the $\d t$ component of $\mbf \Phi^i$) is given by $-\frac{3}{16\pi i}\delta_i{}^j|x-x'|^{-3} (\bar z-\bar z')$. In the underlying physical theory, this is the two-point function between fermions $\ol\psi_{-,i}(x)$ and $\psi_{-}{}^j(x')$.

Vertices for Feynman diagrams in the twisted formalism are given by the superpotential $W(\mbf \Phi)\d z^{-1}$. (The factor of $\d z^{-1}$ comes from the twisted spin of $W$.) The fact that this interaction involves only $\mbf \Phi$, whereas the propagator connects $\mbf \Phi$ with $\mbf \Psi$, severely restricts the available diagrams. In particular, there cannot be any diagrams containing loops, as there is no way to connect multiple $W(\mbf \Phi)$ vertices!

We may compute the OPE $\psi_i(0)\psi_j(z,\bar z,t)$ by evaluating the two-point function of these operators in the presence of an arbitrary smooth background for the $\mbf \Phi$ and $\mbf \Psi$ fields. A single Feynman can contribute, in which the $\mbf \Psi_i(0)$ and $\mbf \Psi_j(0)$ are each connected by propagators to a $W(\mbf \Phi)$ vertex:\bigskip
\be \bigskip \raisebox{-.5in}{\includegraphics[width=2.7in]{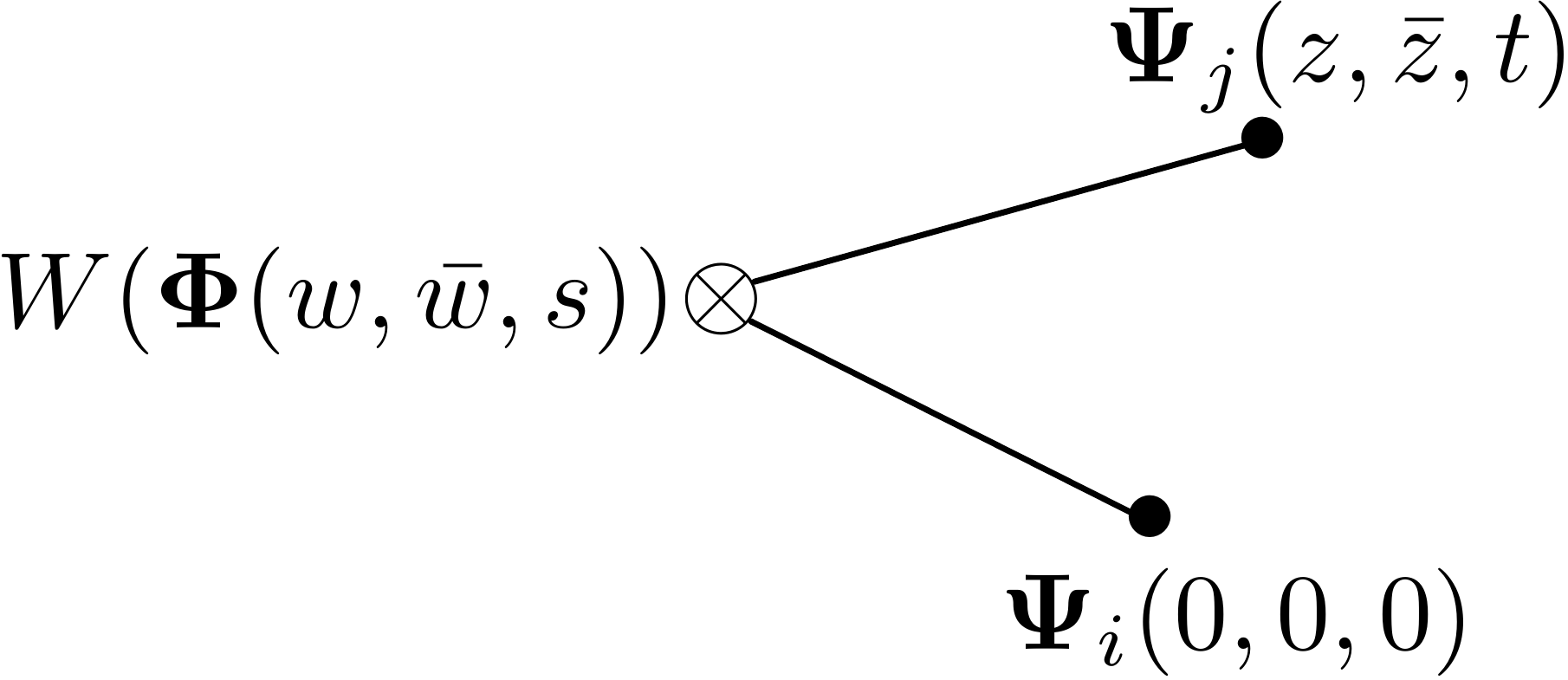}} \ee 
Placing the vertex at $(w,\bar w,s)$ and integrating over all values of this insertion point, we find that the diagram evaluates to
\begin{align} & \int_{\C_{w,\bar w}\times \R_s} P_i{}^k(0;w,\bar w,s)P_j{}^\ell(z,\bar z,t;w,\bar w,s) \pd_k\pd_\ell W(\mbf \Phi(w,\bar w,s)) \d w^{-1} \notag \\
 & \quad = -\frac{9}{2^8\pi^2}  \int_{\C_{w,\bar w}\times \R_s} \frac{\pd_i\pd_j W(\phi(w,\bar w,s))}{(|w|^2+s^2)^{\frac32}(|z-w|^2+(t-s)^2)^{\frac32}}
 \big[\bar w \d s-s\d\bar w\big]\big[(\bar w-\bar z)\d s+(t-s)\d\bar w\big]\d w \notag \\
 & \quad = -\frac{9}{2^8\pi^2}  \int_{\C_{w,\bar w}\times \R_s}   \frac{\pd_i \pd_j W(\phi(w,\bar w,s))}{(|w|^2+s^2)^{\frac32}(|z-w|^2+(t-s)^2)^{\frac32}}
   (t\bar w-s \bar z) \d s \d \bar w \d w\,. \label{bulk-OPE-int}
\end{align}
The $\phi(w,\bar w,s)$ appearing in the numerator here is the background in which the two-point function is being evaluated. Note that in order to obtain a non-zero integral, with the correct integration measure, only the 0-form component of $\pd_i\pd_jW$ can appear.

Observe that the integral \eqref{bulk-OPE-int} converges and is generically nonzero, as long as $(z,\bar z,t)\neq (0,0,0)$, and the background field $\phi$ does not grow too fast near infinity in spacetime.
We are interested in singularities of the integral as $(z,\bar z,t)\to (0,0,0)$. We can analyze this by expanding
$W(\phi(w,\bar w,s))$ as a series near $(w,\bar w,s)=(0,0,0)$. Derivatives in $\bar w$ and $s$ will contribute $Q$-exact terms to the OPE, so we only need consider the expansion in $w$,
\be \pd_i\pd_j W(\phi(w,\bar w,s)) = 
\pd_i\pd_j W(\phi(0,0,0)) + \pd_i\pd_j\pd_k W(\phi(0,0,0)) \pd_w \phi^k w + \ldots \ee


The leading singularity in the 2-point function comes from the constant term:
\be -\frac{9}{2^8\pi^2} \pd_i \pd_j W(\phi(0,0,0))  \int_{\C_{w,\bar w}\times \R_s}   \frac{1}{(|w|^2+s^2)^{\frac32}(|z-w|^2+(t-s)^2)^{\frac32}}  (t\bar w-s \bar z) \d s \d \bar w \d w\,. \label{bulk-OPE-int2} \ee
%
However, the integrand $\frac{t\bar w-s\bar z}{(|w|^2+s^2)^{\frac32}(|z-w|^2+(t-s)^2)^{\frac32}}$ is odd under the reflection $(w,\bar w,s)\mapsto (z-w,\bar z-\bar w,t-s)$, so this integral vanishes identically.

The next term in the expansion is
\be  -\frac{9}{2^8\pi^2} \pd_w\pd_i \pd_j W(\phi(0,0,0))  \int_{\C_{w,\bar w}\times \R_s}   \frac{t|w|^2-s w \bar z}{(|w|^2+s^2)^{\frac32}(|z-w|^2+(t-s)^2)^{\frac32}}   \d s \d \bar w \d w\,.  \label{bulk-OPE-int3} \ee
This also converges.
The integrand (and integration measure) here is invariant under the rescaling $(z,\bar z,t;w,\bar w,s)\mapsto (\lambda z,\bar \lambda \bar z,|\lambda|t;\lambda w,\bar\lambda\bar w,|\lambda|s)$ for complex $\lambda$. We deduce from this that the integral evaluates to a function $f(t/|z|)$ that depends only on the ratio $t/|z|$.

Subleading terms in the expansion of $\pd_i\pd_j W$ cannot contribute to a singular OPE. Consider the integral
\be  \int_{\C_{w,\bar w}\times \R_s}   \frac{w^n(t\bar w-s \bar z)}{(|w|^2+s^2)^{\frac32}(|z-w|^2+(t-s)^2)^{\frac32}}   \d s \d \bar w \d w\,, \qquad n \geq 2\,. \label{bulk-OPE-int4} \ee
The integrand (and measure) scales with a factor of $\lambda^{n-1}$ under $(z,\bar z,t;w,\bar w,s)\mapsto (\lambda z,\bar \lambda \bar z,|\lambda|t;\lambda w,\bar\lambda\bar w,|\lambda|s)$, so one might expect this integral to evaluate to $g(t,|z|) z^{n-1}$ for some scale-invariant (and therefore bounded) two-variable function $g$. Then the limit $\lim_{(z,t)\to 0} g(t,|z|) z^{n-1} = 0$ is well defined, and the result is nonsingular at $(z,\bar z,t)=(0,0,0)$. Of course, the integral \eqref{bulk-OPE-int4} does not actually converge, so we must be more careful. We can impose an IR regulator, integrating (say) over a region $|w|^2+s^2\leq R^2$ for large $R$. The actual size of the region is irrelevant for analyzing the behavior as $(z,\bar z,t)\to 0$. For every $\epsilon > 0$ there exists a constant $C(R,\epsilon)$ such that
\be \bigg |  \int_{|w|^2+s^2\leq R^2}   \frac{w^n(t\bar w-s \bar z)}{(|w|^2+s^2)^{\frac32}(|z-w|^2+(t-s)^2)^{\frac32}}   \d s \d \bar w \d w \bigg| \leq C(R,\epsilon) (|z|^2+t^2)^{n-1-\epsilon}\,.  \ee
Choosing $\epsilon$ so small that $n-1-\epsilon > 0$, this ensures that the integral has a well-defined, zero limit as $(z,\bar z,t)\to (0,0,0)$.

Altogether, we see that the only contribution to the OPE is from \eqref{bulk-OPE-int3}. A standard Feynman trick shows that the integral evaluates to $4\pi^2 t / \sqrt{|z|^2+t^2}$, whence the OPE is
\begin{align} \label{bulk-W-OPE} \psi_i(z,\bar z,t)\psi_j(0,0,0) &\sim -\frac{9}{64\pi} \pd_z \pd_i\pd_jW(\phi(0,0,0)) \frac{t}{\sqrt{|z|^2+t^2}} \\
 &= -\frac{9}{64\pi} \pd_i\pd_j\pd_k W\,\pd_z\phi^k \frac{t}{\sqrt{|z|^2+t^2}}  \,. \notag\end{align}

We can double-check the answer by computing the OPE of two somewhat general $Q$-closed operators, and making sure that the answer is $Q$-exact (and thus zero in cohomology, as discussed in \eqref{Vbulk-W}). Let's take operators $\CO_f = f^i(\phi)\psi_i$ and $\CO_g = g^i(\phi)\psi_i$, for two collections of functions $f^i(\phi)$ and $g^i(\phi)$.
In order for $\CO_f$ and $\CO_g$ to be closed, the functions must satisfy $f^i\pd_iW = g^i\pd_iW= 0$\,. Using \eqref{bulk-W-OPE}, we find
\begin{align} \left[f^i\psi_i\right](z,\bar z,t) \left[g^j\psi_j\right](0,0,0) &\sim -\frac{9}{64\pi} \big[f^i g^j \pd_z \pd_i\pd_jW\big](\phi(0,0,0)) \frac{t}{\sqrt{|z|^2+t^2}}\,. \end{align}
%
It is easy to construct an operator $\hat\CO$ satisfying $Q\hat\CO = f^ig^i\pd_z\pd_j\pd_jW$, thus making the RHS exact. For example, we may choose $\hat\CO =
+\pd_jf^i(\pd_zg^j)\psi_i  -f^i\pd_z(\pd_ig^j\psi_j)  $. Then
\begin{align} Q\hat\CO &= 
  \pd_jf^i(\pd_zg^j)\pd_iW - f^i\pd_z(\pd_ig^j\pd_jW)  \notag \\
  &= - f^i (\pd_z g^j)\pd_i\pd_j W + f^i\pd_z(g^j\pd_i\pd_jW)\qquad (\text{using $f^i\pd_iW = g^i\pd_iW=0$}) \\
  &= f^ig^j\pd_z\pd_i\pd_jW \qquad \text{as desired}\,.  \notag
\end{align}


The OPE \eqref{bulk-W-OPE} illustrates well the subtleties one may encounter in defining algebraic operations on spaces of local operators 
that go beyond the strict Q-cohomology. Here, we see that the topological OPE of local operators along the $t$ direction 
is not (graded) commutative, and differs from the OPE of local operators along the $z$ direction! 

The topological OPE will be the first element of a tower of $A_\infty$ operations defined on operators at $z=\bar z=0$.
The existence of nontrivial higher operations is manifest in the observation that the OPE in the $t$ direction  
\be \psi_i(0,0,t)\psi_j(0,0,0) \sim -\frac{9}{64\pi}\text{sign}(t) \pd_z \pd_i\pd_jW(\phi(0,0,0))  \,. \ee
is not associative. 

The chiral OPE (in the $z$ direction) at $t=0$ simply vanishes, suggesting that the chiral algebra of local operators at $t=0$ may still be rather simple. 
The disagreement between the topological and chiral OPE, though, likely indicates the existence of non-trivial higher operations which 
encode the quasi-isomorphism of the two OPE's. 

Finally, the above calculation shows how a bulk deformation $\int O^{(2)} dz$ modifies the OPE in the topological direction, with 
$O$ being a polynomial in the $\phi$ fields. This modification should be encoded by an higher operation of cohomological degree $-2$, in the same way as the 
deformation of the BRST differential is encoded by the secondary bracket of cohomological degree $-1$. 
 
\subsection{Classification of boundary conditions}
\label{sec:W-class}

We next consider more closely the types of $Q$-preserving boundary conditions that can exist in the presence of a superpotential. 
As above, suppose we have bulk chiral multiplets $(\mbf \Phi,\mbf \Psi)$ with scalars $\phi$ valued in a vector space $V$, and a superpotential $ W: V\to \C$.

Consider the basic boundary condition labelled by a $U(1)_R$-invariant subspace $L\subset V$,
\be \label{L-bcW} \mbf \Phi\big| \in  \underset{r}{\oplus}\, \alg^{\bullet,(r/2)} \otimes L^{(r)}\,,\qquad \mbf \Psi\big| \in   \underset{r}{\oplus}\,\alg^{\bullet,(1-r/2)} \otimes L^{\perp (-r)} [1]\,. \ee
In the absence of a superpotential, this boundary condition ensures that the half-space action $S = \int_{\C\times \R_-} \mbf \Psi \d' \mbf \Phi$ satisfies the master equation $\{S,S\}_{\rm BV}=0$. In particular, boundary terms in the $\{S,S\}_{\rm BV}$ have the form $\int_{\C\times \{0\}} \mbf \Psi\mbf \Phi\big|$, and vanish given \eqref{L-bcW}. If we add a superpotential, the half-space action becomes
\be S = \int_{\C\times \R_-} \big[ \mbf \Psi \d' \mbf \Phi  - W(\mbf \Phi)\big]\,, \ee
and now the BV bracket of $S$ with itself acquires an additional term
\be \{S,S\}_{\rm BV} = - \int_{\C\times \R_-} \d' W(\mbf \Phi) = - \int_{\C\times \{0\}} W(\mbf \Phi)\big|\,. \ee
This will vanish as long as the restriction of the superpotential to $L$ is a constant
\be \boxed{W\big|_L = \text{const}}\,. \label{W-0} \ee
Since $L$ is $U(1)_R$-invariant (and $W$ has R-charge 2), this further implies that $W\big|_L = 0$.

It is possible to violate the condition $W\big|_L=0$ if we add additional 2d $\CN=(0,2)$ boundary multiplets.  This is familiar from the study of the ``Warner problem'' in 2d $\CN=(2,2)$ theories with B-type boundary conditions \cite{Warner, KapustinLi}, whose 3d $\CN=2$ analogue was discussed by \cite{GGP-fivebranes, BrunnerSchulzTabler}. In the twisted formalism, the analysis goes as follows.

We know from \eqref{EJ2d} that a 2d theory with chiral multiplets $\mbf C,\mbf{\wt C}$, fermi multiplets $\mbf\Gamma,\mbf{\wt \Gamma}$, and E and J terms has an action $S_{2d}$ with $\{S_{2d},S_{2d}\}_{\rm BV} = 2 \int_\C E(\mbf C)\cdot J(\mbf C)$. If we couple bulk chiral multiplets on a half-space to such a 2d theory, by allowing the E and J terms to depend holomorphically on boundary values $\mbf \Phi|$ as well as $\mbf C$, then the master equation for the full bulk-boundary system is satisfied when
\be   \boxed{W\big|_L= 2E\cdot J}\,. \label{W-EJ} \ee
(Again, the two sides may differ by a constant, but preserving $U(1)_R$ symmetry requires the constant to vanish.)

We proceed to describe the boundary chiral algebras in the two cases \eqref{W-0} and \eqref{W-EJ}.

\subsection{Boundary algebra for $W\big|_L=0$}
\label{sec:W-bdy}

Consider a theory with boundary condition of type \eqref{W-0}. Being more explicit, suppose we have $n$ individual bulk chiral multiplets ($n=\text{dim}\,V$), split into two sets
\be (\mbf \Phi^i,\mbf\Psi_i)_{i=1}^m\,,\qquad (\mbf{\hat\Phi}^{\hat i},\mbf{\hat\Psi}_{\hat i})_{\hat i=m+1}^{n}\,, \ee
so that the $(\mbf \Phi,\mbf \Psi)$ are given Neumann b.c. and $(\mbf{\hat \Phi},\mbf{\hat \Psi})$ are given Dirichlet b.c.:
\be \label{NDW0} \mbf \Psi_i\big| = 0  \quad(i=1,...,m)\,,\qquad \mbf{\hat\Phi}^{\hat i}\big| = 0\quad ({\hat i}=m+1,...,n)\,.\ee%
This is the boundary condition associated to a subspace $L\subset V$ defined by $\hat\phi=0$. We also have a polynomial bulk superpotential $W(\phi,\hat \phi)=W(\phi^1,...,\phi^m,\hat\phi^{m+1},...,\hat\phi^{n})$ that satisfies $W\big|_L=0$, meaning $W(\phi,\hat\phi)|_{\hat \phi=0}=0$.

Recall that the bulk SUSY/BRST transformations are deformed by a superpotential to
\be \begin{array}{ll} Q\,\mbf \Phi^i = \d'\mbf \Phi^i &\qquad Q\,\mbf \Psi_i = \d'\mbf \Psi_i + \pd_i W(\mbf \Phi,\mbf {\hat \Phi}) \\[.1cm] 
Q\,\mbf {\hat\Phi}^{\hat i} = \d'\mbf {\hat\Phi}^{\hat i} &\qquad Q\,\mbf {\hat \Psi}_{\hat i} = \d'\mbf {\hat \Psi}_{\hat i} + \pd_{\hat i} W(\mbf \Phi,\mbf {\hat \Phi}) \end{array}
\ee
where $\pd_iW$ and $\pd_{\hat i}W$ are shorthand for $\pd W/\pd\mbf \Phi^i$ and $\pd W/\pd\mbf{\hat \Phi}^{\hat i}$, respectively. The constraint $W\big|_L=0$ ensures that the boundary condition  \eqref{NDW0} is preserved by the deformed action of $Q$.

We expect the $Q$-cohomology of boundary local operators to be generated by the bottom components $\phi^i,\hat\psi_{\hat i}$ of the superfields  $\mbf \Phi^i\big|, \mbf{\hat \Psi}_{\hat i}\big|$ that survive at the boundary, together with their $\pd_z$ derivatives. 
(Just like in the bulk, the higher components `cancel' in cohomology against the $\pd_{\bar z}$ and $\pd_t$ derivatives of lower components.)
We still have $Q\,\phi^i=0$, while
\be \label{QpsiW} Q\,\hat \psi_{\hat i} = \pd_{\hat i}W\big|_L\, := \Big(\frac{\pd}{\pd \hat \phi^{\hat i}}W(\phi,\hat \phi) \Big) \Big|_{\hat\phi = 0}\,. \ee
In addition, any boundary operators $\hat\psi_{\hat i}$ that are not $Q$-closed \emph{in the bulk} (\emph{i.e.} for which $\frac{\pd}{\pd \hat \phi^{\hat i}}W(\phi,\hat \phi)\neq 0$ \emph{before} setting $\hat \phi=0$) may acquire a singular OPE. In Section  \ref{sec:W-bulk-OPE}, we showed explicitly that singular terms in the bulk OPE of $\hat\psi_{\hat i}$ operators vanish. Repeating the calculation in the presence of a boundary condition leads to a different result; we will show momentarily that, at the boundary,
\be \label{OPE-W} \hat \psi_{\hat i}(z) \hat\psi_{\hat j}(0) \sim \frac{1}{z} \pd_{\hat i}\pd_{\hat j} W\big|_L\,, \ee
where $ \pd_{\hat i}\pd_{\hat j} W\big|_L$ is shorthand for $\Big( \frac{\pd^2}{\pd \hat \phi^{\hat i}\pd\hat\phi^{\hat j}} W(\phi,\hat \phi)\Big)\Big|_{\hat\phi=0}$.

The structure of this OPE is highly constrained by conservation of R-charge and twisted spin, together with properties of the bulk-boundary map. The bulk-boundary map requires $\hat \psi_{\hat i}(z) \hat\psi_{\hat j}(0)$ to be nonsingular whenever either $\hat \psi_{\hat i}$ or $\hat \psi_{\hat j}$ are Q-closed in the bulk. This implies that singular terms in the OPE must be proportional to $\pd_{\hat i}W\pd_{\hat j}W|_L$ or $\pd_{\hat i}\pd_{\hat j}W|_L$, or higher derivatives. In addition, if $\hat \phi^{\hat i}$ has R-charge $r_{\hat i}$, then the R-charges and spins of a few relevant components of the OPE are
\be \begin{array}{c|@{\quad}c@{\quad}c@{\quad}c@{\quad}c}
& W & \pd_{\hat i} & \hat\psi_{\hat i} & 1/z \\\hline
R & 2 & - r_{\hat i} &1- r_{\hat i} & 0 \\
J & 1 & - r_{\hat i}/2 & 1- r_{\hat i}/2 & 1 \end{array}
\ee
Thus,  $\hat \psi_{\hat i}(z) \hat\psi_{\hat j}(0)$ has $(R,J) = \big(2-r_{\hat i}-r_{\hat j},\, 2-(r_{\hat i}+r_{\hat j})/2\big)$, which matches perfectly with $\frac{1}{z}\pd_{\hat i}\pd_{\hat j}W$; whereas a term of the form $\frac{1}{z^k} \pd_{\hat i}W\pd_{\hat j}W$ is ruled out. No higher powers of $W$ can enter singular terms, as long as they depend only on $z$ and not $\bar z$.

Altogether, we propose that the $Q$-cohomology of boundary local operators can be modeled as the cohomology of a chiral algebra strongly generated by $\phi(z),\hat \psi(z)$, with OPE \eqref{OPE-W}, and differential \eqref{QpsiW},
\be \label{alg-W0}  \CV_\pd \simeq H^\bullet  \big\langle\!\!\big\langle \phi^i(z),\hat\psi_{\hat i}(z) \;\big|\; \hat\psi_{\hat i}(z) \hat\psi_{\hat j}(0) \sim \tfrac{1}{z}\pd_{\hat i}\pd_{\hat j}W|_L,\, Q(\hat\psi_{\hat i}) =\pd_{\hat i}W|_L  \big\rangle\!\!\big\rangle\,.\ee

In this description, we are still \emph{neglecting} potential higher operations. 
 We actually expect such operations to arise when the superpotential is of cubic or higher order. In Section \ref{sec:A-comment} we will explain one possible route to producing a dg model for $\CV_\pd$ that captures them. 

\subsubsection{Example 1: massive chiral}
\label{sec:W0-eg-m}

We describe two simple examples. First, consider a single 3d chiral multiplet $(\mbf \Phi,\mbf \Psi)$, with a quadratic superpotential
\be W(\mbf \Phi) = \tfrac12 m \mbf \Phi^2\,, \ee
for some $m\in \C$.
This is a complex mass term. Note that we must have $R(\phi) = 1$ for unbroken R-symmetry.

Physically, we expect the bulk theory to be trivial at scales below $m$. It is easy to see that the bulk algebra in the holomorphic twist is indeed trivial whenever $m\neq 0$\,: the bulk operators $\phi(z)$ and $\psi(z)$ (and their $\pd_z$ derivatives) could potentially contribute; but the superpotential sets
\be Q\,\psi = m\,\phi\,; \ee
thus both $\phi$ and $\psi$ disappear from the cohomology, and $\CV\simeq \C$. 

Boundary conditions for this theory may still carry ``edge modes,'' generating a nontrivial boundary algebra $\CV_{\pd}$, even though the bulk is trivial. The simplest boundary condition is Dirichlet, setting
\be \mbf \Phi\big|=0 \quad\Rightarrow \quad W\big|=0\,. \ee
The general prescription \eqref{alg-W0} above says that the boundary algebra is generated by $\psi(z)$ (and its $\pd_z$) derivatives. This operator is $Q$-closed on the boundary, since
\be Q\,\psi(z) = \pd_\phi W(\phi)\big|_{\phi=0} = m\,\phi\big|_{\phi=0}  = 0\,.\ee
However, there is a nontrivial boundary OPE
\be \psi(z)\psi(0) \sim \frac{1}{z}\pd_\phi^2 W(\phi)\big|_{\phi=0} = \frac{m}{z}\,.\ee

Thus, in $Q$-cohomology (or physically in the infrared) the Dirichlet b.c. gives rise to a trapped fermionic 2d edge mode, with the standard 2d free-fermion OPE.

\subsubsection{Example 2: XYZ model}
\label{sec:XYZ-NDD}

For a more interesting example, consider the ``XYZ model,'' a 3d theory of three chiral multiplets $(\mbf X,\mbf \Psi_X)$, $(\mbf Y,\mbf \Psi_Y)$,  and $(\mbf X,\mbf \Psi_Y)$, with cubic superpotential 
\be W = \mbf{XYZ}\,. \ee
We have set the coefficient to `$1$'; it can be considered part of the normalization of (say) $\mbf X$.
The bulk algebra is nontrivial. Following Section \ref{sec:W-bulk}, we see it is generated by $X(z),Y(z),$ $Z(z),\psi_X(z),\psi_Y(z),\psi_Z(z)$ with
\be Q\, \psi_X =  Y Z\,,\qquad Q\,\psi_Y = X Z\,,\qquad Q\,\psi_Z = X Y\,.\ee

An interesting boundary condition is Neumann for $\mbf X$ and Dirichlet for $\mbf Y,\mbf Z$,
\be \text{NDD}: \quad \mbf \Psi_X\big|=0\,,\quad \mbf Y\big|=\mbf Z \big| = 0\,.\ee
This clearly sets $W\big|_L=0$ (where the subspace $L$ is cut out by $Y=Z=0$). Our prescription \eqref{alg-W0} identifies the boundary algebra as being generated by $X(z),\psi_Y(z),\psi_Z(z)$, which are all $Q$-closed on the boundary, since
\be Q\,\psi_Y  = XZ\big| = 0\,,\qquad Q\,\psi_Z = XY\big| = 0\,. \ee
However, there is a nontrivial boundary OPE
\be \psi_Y(z)\psi_Z(0) \sim \frac{1}{z} \phi(0)\,. \ee

\subsubsection{The boundary OPE}
\label{sec:W-OPE}

We can adapt the calculation of Section \ref{sec:W-bulk-OPE} to compute the boundary OPE $\hat \psi_{\hat i}(z,\bar z,0)\hat\psi_{\hat j}(0,0,0)$.  We again consider a two-point function of the operators $\hat {\mbf \Psi}_{\hat i},\hat {\mbf \Psi}_{\hat j}$, in the presence of arbitrary background fields. There are still no non-perturbative corrections, since `instantons' would be constant along the half-infinite $t$ direction, and thus have infinite action.
A single tree-level Feynman diagram contributes, involving an insertion of the bulk vertex $W(\mbf \Phi,\hat{\mbf \Phi})$, connected by propagators to $\hat {\mbf \Psi}_{\hat i}(0,0,0)$ and $\hat {\mbf \Psi}_{\hat j}(z,\bar z,0)$. Two modifications are necessary, however.

The obvious modification is to restrict $\hat {\mbf \psi}_{\hat i}$ to be inserted at $t=0$ and to only integrate the vertex $W(\mbf \Phi(w,\bar w,s),\hat{\mbf \Phi}(w,\bar w,s))$ over a half-space with $s\in \R_-$ (say).

In addition, we must use bulk-boundary propagators to connect $\hat{\mbf \Psi}$'s to the $\hat{\mbf \Phi}$'s in the $W$ vertex. To determine the bulk-boundary propagator, we use the classic method of images.
Observe that the theory for chiral multiplets $(\hat{\mbf \Phi}, \hat{\mbf \Psi})$ on a half-space $\C\times \R_{\leq 0}$ with the Dirichlet b.c. $\hat{\mbf \Phi}\big|=0$ is equivalent to a theory on a full space $\C\times \R$, but where fields are equivariant under under the reflection operator $t\mapsto -t$, which we denote by $\rho$. The equivariance conditions are
\be  \begin{array}{l} \rho^* \hat{\mbf \Psi} = \hat{\mbf \Psi} \\[.1cm]
\rho^*\hat{\mbf \Phi} = -\hat{\mbf \Phi} \end{array} \ee
(The fields are differential forms, and $\rho^*$ denotes pullbacks under the reflection.) Correspondingly, the bulk-boundary propagator is obtained from the bulk propagator \eqref{prop-bulk} by (anti-)symmetrizing:
\begin{align} P_{\pd \hat i}{}^{\hat j}(z,\bar z,t;z',\bar z',t') &= \frac{1}{4}\big[ P_{\hat i}{}^{\hat j}(z,\bar z,t;z',\bar z',t')+ P_{\hat i}{}^{\hat j}(z,\bar z,-t;z',\bar z',t') \\
 &\qquad - P_{\hat i}{}^{\hat j}(z,\bar z,t;z',\bar z',-t')- P_{\hat i}{}^{\hat j}(z,\bar z,-t;z',\bar z',-t')\big]\,. \notag \end{align}

When the dust clears, we find that the OPE $\hat \psi_{\hat i}(z,\bar z,0)\hat\psi_{\hat j}(0,0,0)$ is computed by
\begin{align} & \int_{\C_{w,\bar w}\times \R_{s\leq 0}} P_{\pd \hat i}{}^{\hat k}(0;w,\bar w,s)P_{\pd \hat j}{}^{\hat \ell}(z,\bar z,0;w,\bar w,s) \pd_k\pd_\ell W(\mbf \Phi(w,\bar w,s),\hat{\mbf \Phi}(w,\bar w,s)) \d w^{-1} \notag \\
 & \hspace{.5in} = \frac{9}{2^8\pi^2}  \int_{\C_{w,\bar w}\times \R_{s\leq 0}}   \frac{\pd_{\hat i} \pd_{\hat j} W(\phi(w,\bar w,s),\hat \phi(w,\bar w,s))}{(|w|^2+s^2)^{\frac32}(|z-w|^2+s^2)^{\frac32}}
   s \bar z\, \d s \d \bar w \d w\,. \label{bdy-OPE-int}
\end{align}
As in the bulk calculation, we explore the singularity structure as $(z,\bar z)\to (0,0)$ by expanding $\pd_{\hat i} \pd_{\hat j} W$ as a series in $w$, noting that $\pd_{\bar w}$ and $\pd_s$ derivatives will give $Q$-exact contributions to the OPE.

The leading singularity comes from the constant term
\be \frac{9}{2^8\pi^2}  \pd_{\hat i} \pd_{\hat j} W\big|_L(0)  \int_{\C_{w,\bar w}\times \R_{s\leq 0}}   \frac{\bar z s}{(|w|^2+s^2)^{\frac32}(|z-w|^2+s^2)^{\frac32}}
    \d s \d \bar w \d w\,.   \ee
This no longer vanishes, since we are just integrating over a half-space.
The integrand (and measure) scales with a factor of $\lambda^{-1}$ under $(z,\bar z,w,\bar w,s)\mapsto (\lambda z,\bar\lambda \bar z,\lambda w,\bar \lambda \bar w, |\lambda|s)$. This now implies that the integral will evaluate to $f(z,\bar z) z^{-1}$, where the function $f$ is invariant under $(z,\bar z)\mapsto (\lambda z,\bar\lambda \bar z)$, and is therefore a constant.  Using a standard Feynman trick to evaluate the integral, we find $f=-2\pi$, whence
\be \hat \psi_{\hat i}(z,\bar z,0)\hat \psi_{\hat j} \sim -\frac{9}{2^7\pi}\, \frac{1}{z}  \pd_{\hat i} \pd_{\hat j} W\big|_L(0) \,.\ee

Subsequent terms in the expansion of $\pd_{\hat i} \pd_{\hat j} W$ do not lead to singularities. The linear term led to some complications in the bulk, which are absent on the boundary dues to setting $t=0$. On the boundary, the linear term gives the integral
\be \frac{9}{2^8\pi^2}  \pd_z\pd_{\hat i} \pd_{\hat j} W\big|_L(0)  \int_{\C_{w,\bar w}\times \R_{s\leq 0}}   \frac{w\bar z s}{(|w|^2+s^2)^{\frac32}(|z-w|^2+s^2)^{\frac32}}
    \d s \d \bar w \d w\,.   \ee
This is invariant under $(z,\bar z,w,\bar w,s)\mapsto (\lambda z,\bar\lambda \bar z,\lambda w,\bar \lambda \bar w, |\lambda|s)$, hence must evaluate to a function $g(z,\bar z)$ that is invariant under $(z,\bar z)\mapsto (\lambda z,\bar\lambda \bar z)$, which is necessarily a constant. Numerics indicate that $g = -\pi$.
 Higher-order terms give integrals that have well-defined, zero limits as $(z,\bar z)\to (0,0)$, for the same reason as in the bulk calculation. Altogether, we find that the 2-point function for small $z$ is
 \be \hat \psi_{\hat i}(z,\bar z,0)\hat \psi_{\hat j} =  -\frac{9}{2^7\pi}\, \frac{1}{z}  \pd_{\hat i} \pd_{\hat j} W\big|_L(0) -\frac{9}{2^8\pi} \pd_z \pd_{\hat i} \pd_{\hat j} W\big|_L(0) + O(|z|^{1-\epsilon}) \quad \text{(any $\epsilon > 0$)}\,. \ee
 The singular part leads to the OPE \eqref{OPE-W} as claimed, after absorbing the factor $ \frac{-9}{2^7\pi}$ into $W$.

\subsection{Boundary algebra for $W\big|_L=2E\cdot J$}
\label{sec:W-bdy-EJ}

We'll now describe the chiral algebra associated to the second type of boundary condition \eqref{W-EJ}, where additional boundary matter is required to preserve SUSY/BRST symmetry. We focus our attention on the case of boundary fermions ($\CN=(0,2)$ fermi multiplets).

To be explicit, let us label the bulk chiral multiplets as before,
\be (\mbf \Phi^i,\mbf\Psi_i)_{i=1}^m\,,\qquad (\mbf{\hat\Phi}^{\hat i},\mbf{\hat\Psi}_{\hat i})_{\hat i=m+1}^{n}\,, \ee
with $(\mbf \Phi,\mbf \Psi)$ given Neumann b.c. and $(\mbf{\hat \Phi},\mbf{\hat \Psi})$ given Dirichlet b.c.,
\be \label{NDW1} \mbf \Psi_i\big| = 0  \quad(i=1,...,m)\,,\qquad \mbf{\hat\Phi}^{\hat i}\big| = 0\quad ({\hat i}=m+1,...,n)\,.\ee
The subspace $L\subset V$ corresponds to $\hat \phi=0$. Now the bulk superpotential $W(\mbf \Phi,\hat{\mbf\Phi})$ does not vanish when $\hat{\mbf \Phi}=0$. Rather we introduce a collection of boundary fermi multiplets $\{\mbf \Gamma^\alpha,\wt{\mbf \Gamma}_\alpha\}_{\alpha=1}^N$, with polynomial E and J terms $E^\alpha(\mbf \Phi|)$, $J_\alpha(\mbf \Phi|)$ depending on the restrictions of the bulk superfields $\mbf \Phi^i$ to the boundary, such that
\be W(\mbf\Phi|,0) = 2 \sum_\alpha E^\alpha(\mbf \Phi|) J_\alpha(\mbf \Phi|)\,. \ee

If $W$, $E$, and $J$ all vanish, this is simply a collection of free 2d fermions tensored with a basic boundary condition for free bulk chirals. The boundary chiral algebra is generated by the bottom components of the bulk superfields $\phi^i(z),\hat \psi_{\hat j}(z)$ that survive at the boundary (with trivial OPE), and by $\Gamma^\alpha(z),\wt\Gamma_\alpha(z)$ with the standard 2d OPE \eqref{OPE2d-gen}, $\Gamma^\alpha(z)\wt\Gamma_\beta(0) \sim \delta^\alpha{}_\beta\frac{1}{z}$.

When $W,E,J$ are nonzero, it is still only the bottom components (and their $\pd_z$ derivatives) that can contribute to the boundary chiral algebra; higher components cancel in cohomology with $\pd_{\bar z}$ and $\pd_t$ derivatives as usual. The SUSY/BRST transformations are deformed to
\begin{subequations} \label{V-EJ}
\be  \begin{array}{l@{\qquad}l}Q\,\phi^i = 0\,, & Q\,\Gamma^\alpha = E^\alpha(\phi)\,, \\[.1cm]
 Q\,\hat\psi_{\hat i} = \pd_{\hat i} W\big|_L\,, & Q\,\wt\Gamma_\alpha = J_\alpha(\phi)\,, \end{array} \label{Q-EJ} \ee
where $\pd_{\hat i}W\big|_L = \frac{\pd}{\pd\hat \phi^{\hat i}} W(\phi,\hat \phi)\big|_{\hat \phi=0}$ as usual. We do not expect instanton corrections, as we did not add bosonic boundary degrees of freedom which could have classical field configurations. The transformations of the boundary fermions are just those of \eqref{Q2d-gen}, with $\phi$ playing the role of 2d chirals $C$, and no $\wt C$ (since the restrictions of 3d chirals to the boundary have no $\wt C$ counterpart).

The OPE may also be deformed. We claim that the only correction is the one we found already in Section \ref{sec:W-bdy}, so that the only singular OPE's are
\be \hat\psi_{\hat i}(z)\hat\psi_{\hat j}(0) \sim \frac{1}{z} \pd_{\hat i}\pd_{\hat j}W\big|_L\,,\qquad \Gamma^\alpha(z)\wt\Gamma_\beta(0) \sim\frac{1}{z}  \delta^\alpha{}_\beta\,. \label{OPE-EJ} \ee
\end{subequations}
To see that there cannot be perturbative corrections to this structure due to couplings between the superfields $\mbf \Gamma^\alpha, \wt{\mbf\Gamma}_\alpha$ and $\mbf \Phi$, consider (say) an arbitrary Feynman diagram involving the term $\int_\C \mbf \Gamma^\alpha J_\alpha(\mbf \Phi)$. If such a diagram contributes to the OPE between two boundary operators, it must have at least one propagator connecting some fields in $J_\alpha(\mbf \Phi)$ to a boundary vertex. This is necessarily a bulk-boundary propagator between a $\mbf \Phi^i$ and $\mbf \Psi_i$. However, both ends of this bulk-propagator are restricted to the boundary, forcing it to vanish. Therefore the amplitude of any such diagram is zero.
There also cannot be non-perturbative corrections to the OPE's, as they would involve `instanton' configurations of the bosonic $\phi$ field, which extend infinitely into the bulk, and have infinite action.

Altogether, we find that the boundary algebra $\CV_\pd$ may be constructed as the cohomology of a chiral algebra strongly generated by  $\phi^i(z),\hat \psi_{\hat i}(z), \Gamma^\alpha(z),\wt\Gamma_\alpha(z)$, with OPE \eqref{OPE-EJ} and differential \eqref{Q-EJ}.

We have restricted (and will restrict) our attention to boundary conditions that only involve 2d fermi multiplets $(\mbf \Gamma,\wt{\mbf\Gamma})$. Adding boundary chirals $(\mbf C,\wt{\mbf C})$ is possible as well, and was important for some of the dualities of \cite{GGP-fivebranes, DGP-duality, DP-4simplex}. A perturbative analysis of chiral algebras in the presence of boundary chirals may be performed along the same lines as Section \ref{sec:2d}; however, non-perturbative effects may also play an interesting role in this case.

\subsubsection{Example: XYZ with Neumann b.c.}
\label{sec:XYZ-NNN}

As a simple example, consider the XYZ model of Section \ref{sec:XYZ-NDD}. If we give Neumann b.c. to all three bulk chiral multiplets, we must add boundary fermion(s) with E and J terms to factorize the superpotential $W=XYZ$. There is no symmetric way to do this while preserving $U(1)_R$ symmetry.

The simplest option is to introduce a single boundary fermi multiplet $(\mbf \Gamma,\wt{\mbf \Gamma})$, and to choose a splitting of the three bulk chirals into $1+2$, say $\{X\}$ and $\{Y,Z\}$, setting
\be E = \mbf X\,,\qquad J = \tfrac12 \mbf Y\mbf Z\,. \ee
For Neumann b.c. on all bulk chirals and this choice of boundary interaction, the boundary chiral algebra becomes
\be \CV_\pd \simeq  H^\bullet\big\langle\!\!\big\langle X(z),Y(z),Z(z),\Gamma(z),\wt\Gamma(z)\,\big|\, 
 \Gamma(z)\wt\Gamma(0) \sim \tfrac1z\,;\, Q\,\Gamma=X\,,\; Q\,\wt\Gamma = \tfrac12 YZ \big\rangle\!\!\big\rangle\,.\ee

\subsubsection{Flips with a superpotential}
\label{sec:flipW}

In \cite{DGP-duality} it was explained how the introduction of a boundary Fermi multiplet with linear $E$ or $J$ terms has the effect of changing the boundary condition of a bulk chiral from Neumann to Dirichlet. This `flip' operation was described at the level of chiral algebras in Section \ref{sec:flip1}, in the case of a single bulk chiral. We can now generalize flips to multiple bulk chiral multiplets, coupled by a superpotential.

Suppose we have bulk chiral multiplets $\{\mbf \Phi^i,\mbf\Psi_i\}_{i=1}^m$ with Neumann b.c. and $\{\hat{\mbf \Phi}^{\hat i},\hat{\mbf\Psi}_{\hat i}\}_{\hat i=m+1}^n$ with Dirichlet b.c., and a bulk superpotential $W(\mbf \Phi,\hat{\mbf\Phi})$ such that $W(\mbf \Phi,0)=0$. The boundary algebra $\CV_\pd$ is given by \eqref{alg-W0} from Section \ref{sec:W-bdy}.

Now let us ``flip'' this boundary condition by introducing a boundary fermi multiplet $(\mbf\Gamma,\wt{\mbf \Gamma})$ with coupling $E=\mbf \Phi^m|$. In the description \eqref{V-EJ}, the new boundary algebra $\CV_\pd[\Gamma]$ is now generated by $\phi(z),\hat \psi(z),\Gamma(z),\wt\Gamma(z)$ with
\be  \begin{array}{l@{\qquad}l}Q\,\phi^i = 0\,, & Q\,\Gamma= \phi^m\,, \\[.1cm]
 Q\,\hat\psi_{\hat i} = \pd_{\hat i} W(\phi,\hat \phi)\big|_{\hat \phi=0}\,, & Q\,\wt\Gamma=0\,, \end{array} \qquad \hat\psi_{\hat i}(z)\hat\psi_{\hat j}(0) \sim \pd_{\hat i}\pd_{\hat j}W(\phi,\hat \phi)\big|_{\hat \phi=0}\,,\quad \Gamma(z)\wt\Gamma(0) \sim \frac{1}{z}\,. \label{flipW1}  \ee

We \emph{expect} that, at the level of cohomology, this algebra is equivalent to that of another elementary boundary condition where $\{\mbf \Phi^i,\mbf\Psi_i\}_{i=1}^{m-1}$ are given Neumann b.c. and $(\mbf \Phi^m,\mbf \Psi_m),\{\hat{\mbf \Phi}^{\hat i},\hat{\mbf\Psi}_{\hat i}\}_{\hat i=m+1}^n$ are given Dirichlet b.c. (In other words, the $m$-th bulk chiral multiplet has changed from $N$ to $D$.)
Denoting the elements of this new algebra $\CV_\pd'$ with primes (to distinguish them from elements of $\CV_\pd[\Gamma]$), we find from \eqref{alg-W0} that it is generated by $\{\phi'{}^i\}_{i=1}^{m-1}, \psi_m,\{\hat \psi_{\hat j}'\}_{\hat j=m+1}^n$, with
\be \begin{array}{l} Q \,\psi_m' = \pd_m W(\phi',\hat\phi')\big|_{\phi'{}^m=\hat\phi'=0} \\[.1cm]
 Q\, \hat\psi_{\hat i}' = \pd_{\hat i} W(\phi',\hat\phi')\big|_{\phi'{}^m=\hat\phi'=0} \end{array}
 \quad \psi_m'(z)\hat\psi_{\hat i}'(0) \sim \frac{1}{z}\pd_m\pd_{\hat i}W\big|_{\phi'{}^m=\hat\phi'=0}\,,\quad
 \hat \psi_{\hat i}'(z)\hat\psi_{\hat j}'(0) \sim \frac{1}{z}\pd_{\hat i}\pd_{\hat j}  W\big|_{\phi'{}^m=\hat\phi'=0}\,.
 \label{flipW2} \ee

We specifically expect a map between \eqref{flipW1} and \eqref{flipW2} that induces an isomorphism on $Q$-cohomology. Physically, this map would be induced by the RG flow that integrates out the boundary fermi multiplet $(\Gamma,\wt\Gamma)$. Some parts of the map are easy to see. For instance, in the cohomology of \eqref{flipW1}, both operators $\Gamma$ and $\phi^m$ are removed; while the boundary fermion $\wt\Gamma$ has exactly the same R-charge and twisted spin as $\psi_m'$. Thus, we expect to identify $\psi_m'$ with $\wt\Gamma$, just like we did in Section \ref{sec:flip1}. However, the complete map is nontrivial. It can be determined by carefully matching SUSY/BRST transformations and the boundary OPE.

We can start with the bosonic generators $\phi^i$ and $\phi^i{}'$ for $i\leq m-1$. These operators are central in the algebras $\CV_\pd[\Gamma]$ and $\CV_\pd'$, respectively; and the operator $Q$ acts trivially on all of them.
Moreover, each $\phi^i$ for $i\leq m-1$ could have a completely independent R-charge $r_i$. If we are to give a universal map between the two algebras, this forces us to identify $\phi^i=\phi^i{}'$ for $i\leq m-1$. From now on we will assume this identification and omit the primes. 

We also expect that $\psi_m'$ in \eqref{flipW2} is identified with $\wt \Gamma$ in \eqref{flipW1}, as indicated above. We should check compatibility with the action of $Q$ and the OPE. We have $Q\,\wt\Gamma=0$, while $Q \,\psi_m' = \pd_m W(\phi,\hat\phi)\big|_{\phi{}^m=\hat\phi=0}$. The latter vanishes, since $W(\phi,\hat \phi)\big|_{\hat \phi=0}$ is constant (by assumption) as a function of all the $\phi^1,...,\phi^m$. Similarly, we have $\wt\Gamma(z)\wt\Gamma(0)\sim 0$, whereas $\psi_m'(z)\psi_m'(0)\sim \frac{1}{z}\pd_m^2W\big|_{\phi^m=\hat\phi=0}$. The latter again vanishes because $W(\phi,\hat \phi)\big|_{\hat \phi=0}$ is constant.

It is with the $\hat\psi_{\hat i}$ and $\hat\psi_{\hat i}'$ that the map becomes nontrivial. Compare:
\be Q\, \hat\psi_{\hat i}' = \pd_{\hat i} W\big|_{\phi{}^m=\hat\phi=0} \quad\text{vs.}\quad Q\, \hat\psi_{\hat i} =  \pd_{\hat i} W\big|_{\hat\phi=0}\,. \label{psipsi'} \ee
Let us expanding $W$ as a function of $\phi^m$, 
\be W(\phi,\hat\phi) = \sum_{\ell\geq 0} (\phi^m)^\ell W_\ell(\phi^1,...,\phi^{m-1},\hat \phi)\,, \ee
so that the RHS of \eqref{psipsi'} takes the form
\be  Q\, \hat\psi_{\hat i} =  \pd_{\hat i} W\big|_{\phi{}^m=\hat\phi=0} + \sum_{\ell\geq 1}  (\phi^m)^\ell W_\ell\big|_{\hat \phi=0} \ee
We can get the two sides of \eqref{psipsi'} to match by identifying $\hat \psi_{\hat i}'$ with $\hat \psi_{\hat i}- \Gamma \sum_{\ell \ge 1} (\phi^m)^{\ell-1} \partial_{\hat i} W_\ell \big|_{\hat \phi=0}$. It then follows from $Q\,\Gamma=\phi^m$ that
\be Q\Big( \hat \psi_{\hat i}- \Gamma \sum_{\ell \geq 1} (\phi^m)^{\ell-1} \partial_{\hat i} W_\ell \big|_{\hat \phi=0}\Big)
 = Q\,\hat\psi_{\hat i} - \sum_{\ell \geq 1} (\phi^m)^\ell  \partial_{\hat i} W_\ell \big|_{\hat \phi=0}
  = \pd_{\hat i} W\big|_{\phi{}^m=\hat\phi=0}\,,\ee
in agreement with $Q\,\hat\psi_{\hat i}'$ on the LHS.

Beautifully, the identification of $\hat \psi_{\hat i}'$ with $\hat \psi_{\hat i}- \Gamma \sum_{\ell \ge 1} (\phi^m)^{\ell-1} \partial_{\hat i} W_\ell \big|_{\hat \phi=0}$ is also compatible with OPE's, up to $Q$-exact terms. For example, consider
\begin{equation} 
	\psi_m'(z)\hat\psi_{\hat i}'(0) \sim  \tfrac{1}{z} \partial_m \partial_{\hat i} W\big|_{\phi^m=\hat\phi=0}\,.  \label{W-checkOPE}
\end{equation}
Given our proposed identifications, this should agree with  $\wt\Gamma(z) \cdot\big(\hat \psi_{\hat i}-\Gamma \sum_{\ell \ge 1} (\phi^m)^{\ell-1} \partial_{\hat i} W_\ell \big|_{\hat \phi=0} \big)(0)$, and indeed:
\begin{align} 
	 & \wt\Gamma(z) \cdot	\left(\hat \psi_{\hat i}- \sum_{\ell \ge 1} (\phi^m)^{\ell-1} \Gamma \partial_{\hat i} W_\ell \big|_{\phi^m=\hat \phi=0} \right)(0)  \sim  \frac{1}{z}\sum_{\ell \ge 1} (\phi^m)^{\ell-1} \partial_{\hat i} W_\ell \big|_{\phi^m=\hat \phi=0}   \\
	& \hspace{.5in} 
	= \frac{1}{z} \pd_{\hat i}W_1\big|_{\hat \phi=0} + \frac{1}{z}\phi^m \sum_{\ell\geq 2} (\phi^m)^{\ell-2}\pd_{\hat i}W_\ell \big|_{\phi^m=\hat \phi=0} \notag \\
	&\hspace{.5in} = \frac{1}{z} \partial_m \partial_{\hat i} W\big|_{\phi^m=\hat \phi=0} + Q\Big(\Gamma \sum_{\ell\geq 2} (\phi^m)^{\ell-2}\pd_{\hat i}W_\ell \big|_{\phi^m=\hat \phi=0}\Big)\,, \notag 
\end{align}
which is equal to the RHS of \eqref{W-checkOPE} in $Q$-cohomology. Similarly, it is easy to see that the OPE's $\hat\psi_{\hat i}'(z)\hat\psi_{\hat j}'(0)$ will agree with $\big(\hat \psi_{\hat i}-\Gamma \sum_{\ell \ge 1} (\phi^m)^{\ell-1} \partial_{\hat i} W_\ell \big|_{\hat \phi=0} \big)(z)\big(\hat \psi_{\hat j}-\Gamma \sum_{\ell \ge 1} (\phi^m)^{\ell-1} \partial_{\hat j} W_\ell \big|_{\hat \phi=0} \big)(0) = \hat \psi_{\hat i}(z)\hat\psi_{\hat j}(0)$ up to terms that are proportional to $\phi^m$, and are therefore $Q$-exact.

Altogether, we have found that the map from \eqref{flipW2} to \eqref{flipW1} defined by
\begin{align} \label{flip-map}
	\psi_m' & \mapsto \wt \Gamma\,, \notag \\
	\hat \psi_{\hat i}' & \mapsto \hat \psi_{\hat i} - \Gamma \sum_{\ell \ge 1} (\phi^m)^{\ell-1} \partial_{\hat i} W_\ell \big|_{\hat \phi=0}  \\
	\phi'{}^i & \mapsto \phi^i \text{ if } i \leq m-1  \notag
\end{align}
induces an isomorphism on cohomology, identifying the algebras $\CV_\pd[\Gamma]\simeq \CV_\pd'$.

\subsection{A comment on higher boundary operations}
\label{sec:A-comment}

We  have stated that our  computation of the boundary algebra of operators when we have a superpotential misses some higher structures that one would expect  to be present. This problem occurs when the superpotential contains terms of degree three and higher in those bulk chiral fields which are given Dirichlet boundary conditions.  

The analysis above suggests a way to correct  this problem.  We can always turn some Dirichlet boundary conditions into  Neumann boundary conditions  at  the  price  of  introducing extra boundary complex fermions.  We can  do this until we are in a situation where the  superpotential is at most  quadratic as a function of those chirals  which have Dirichlet boundary conditions.  Then, we do not  expect  there to be an $A_\infty$ algebra of boundary operators.  Instead, the algebra of  boundary operators is a differential-graded vertex algebra; it is the algebra denoted $\widehat \CV_\pd$ and promised in Section \ref{sec:dg}.   

What is happening here is an example of  ``strictification'': we have replaced an $A_\infty$ algebra by a strict algebra at the price of making the algebra larger, by the introduction of boundary  fermions.   Conversely, we expect that the cohomology of a differential graded vertex algebra will have an $A_\infty$ vertex algebra  structure (whatever  that is), and that  this $A_\infty$ structure should coincide with one coming from a direct field theory analysis.

\section{Gauge fields with Neumann boundary conditions}
\label{sec:gauge-N}

Our next goal is to describe chiral algebras for gauge theories. We recall from Section \ref{sec:bdy-sum}, that the basic weakly-coupled boundary conditions for a gauge multiplet 
are labelled by the subgroup $H$ of the gauge group $G$ which survives at the boundary. In terms of the bulk superfields $(\mbf A,\mbf B)$, the boundary conditions require 
$\mbf A\big|$ to be an $H$ connection and $\mbf B\big|$ to lie in the annihilator of the corresponding Lie subalgebra $\mathfrak{h} \in \mathfrak{g}$. 

These basic boundary conditions can be combined with any boundary condition $\mc{B}$ for the matter fields, possibly including 2d degrees of freedom that will only couple to the gauge fields, subject to an anomaly cancellation constraint that we review momentarily. In this section and in Section \ref{sec:gauge-D} we propose how to compute the resulting boundary chiral algebras $\mc{V}_\partial[H;\mc{B}]$ from the boundary chiral algebra $\mc{V}_\partial[\mc{B}]$ of the matter fields, employing in various ways the action on $\mc{V}_\partial[\mc{B}]$ of the group of holomorphic gauge transformations. 

An important source of complexity in the analysis of gauge theory operator algebras is the existence of ``disorder'' operators, which generally cannot be expressed as polynomials in the elementary fields of the theory. Bulk 3d $\CN=2$ gauge theories have long been known to contain BPS monopole operators (disorder operators associated to non-trivial configurations of the gauge multiplet), \emph{cf.} \cite{AHISS, BKW1, BKW2}. They contribute to the 3d index \cite{Kim-index, IY-index, KW-index}, and thus must contribute to the bulk chiral algebra. On boundary conditions, two types of disorder operators contributing to $\CV_\pd$ may appear:
\begin{itemize}
\item Boundary monopole operators \cite{BDGH, DGP-duality}, associated to nontrivial configurations of the 3d gauge multiplet, when the boundary condition is Dirichlet-like ($G\neq H$); and
\item ``Monopole-vortex'' operators, involving nontrivial configuration of both the 3d gauge multiplet and 2d chiral multiplets supported on the boundary (or other 2d degrees of freedom with a continuous spectrum).
\end{itemize}
We will avoid the latter entirely in this paper, by \emph{only} allowing Fermi multiplets (\emph{a.k.a.} chiral fermions) as auxiliary 2d degrees of freedom. We also will not attempt to construct bulk chiral algebras for gauge theories.
\footnote{In principle, the the bulk algebra of a gauge theory could be obtained by taking the derived center of the boundary algebra on a Neumann boundary condition; or by generalizing some of the computations involving cohomology of moduli spaces of $G$-bundles from Section \ref{sec:gauge-D}. It would be interesting to develop this in future work.} %
We \emph{will} discuss boundary monopole operators on Dirichlet-like boundary conditions later, in Section \ref{sec:gauge-D}.


In this Section we will discuss the case of Neumann boundary conditions $\mc{N}$ for the gauge multiplet, \emph{i.e.} $H=G$ (and with only Fermi multiplets or other compact CFT's as auxiliary 2d degrees of freedom). These do not support boundary disorder operators and can be treated with more conventional methods.%
\footnote{It is conceivable, though, that the derived, algebraic geometric formulation of our final answer will apply to more general auxiliary chiral algebras.} %
We anticipate our final answer in a concise form: the boundary chiral algebra $\mc{V}_\partial[\CN;\CB]$ for Neumann boundary conditions consists of $G(\CO)$-invariant operators (in a derived sense) in $\mc{V}_\partial[\mc{B}]$,
\be \mc{V}_\partial[\CN;\CB] = \mc{V}_\partial[\mc{B}]^{G(\CO)}\,, \ee
where $G(\CO)$ is a group of holomorphic gauge transformations acting in the infinitesimal neighborhood of a boundary operator insertion.
The analysis will let us describe several interesting examples of dual boundary algebras, induced by 3d $\CN=2$ XYZ $\leftrightarrow$ SQED duality and Aharony dualities. We give a complete proof of isomorphism of boundary algebras for XYZ $\leftrightarrow$ SQED duality in Section \ref{sec:SQEDXYZduality}.

\subsection{Boundary anomalies}
\label{sec:anomaly}

Since Neumann boundary conditions preserve boundary gauge symmetry, all boundary gauge anomalies must cancel for consistency of the quantum theory. A complete analysis of boundary anomalies was carried out in \cite[Sec. 2.4-2.5]{DGP-duality}, generalizing \cite{GGP-fivebranes}. We recall some relevant results:
\begin{itemize}
\item A bulk $G$ gauge multiplet with Neumann b.c. has boundary anomaly $+\sh$.
\item A bulk $G$ gauge multiplet with Dirichlet b.c. has boundary anomaly $-\sh$.
\item A bulk Chern-Simons term at level $k$ contributes boundary anomaly $k$.
\item Bulk chiral multiplets in representation $V$ of $G$ with Neumann b.c. have anomaly $-\frac12T_V$.
\item Bulk chiral multiplets in representation $V$ of $G$ with Dirichlet b.c. have anomaly $+\frac12T_V$.
\end{itemize}

The conventions here are as follows. When saying ``the boundary anomaly is $n$,'' we mean that the 2d boundary anomaly polynomial for $G$ is $n\text{Tr}(F^2)$, where the Killing form `$\text{Tr}$' is normalized to be the trace in the fundamental representation for type-A groups. In addition, $h$ denotes the dual Coxeter number, with $\sh=0$ if $G$ is abelian. $T_V$ denotes the quadratic index of a representation $V$, normalized so that $T_{\mathfrak g}=2\sh$, and $T_\C=n^2$ for $G=\C^*$ ($G_c=U(1)$) and $\C$ the 1d representation of weight $n$.

Note that the above anomalies are \emph{half} of what one would expect for purely 2d $\CN=(0,2)$ multiplets. In the same conventions, a 2d $G$ gauge multiplet has anomaly $2h=T_{\mathfrak g}$; a 2d chiral in representation $V$ has anomaly $-T_V$; and a 2d fermi in representation $V$ has anomaly $+T_V$. This is consistent with interval compactifications. For example, a 3d chiral multiplet compactified on an interval with Neumann b.c. on both ends becomes a purely 2d chiral, and the two boundary anomalies $-\frac12 T_V-\frac12 T_V$ add to give the anomaly $-T_V$ of a 2d chiral.

When bulk gauge fields have Dirichlet b.c., breaking $G$ to a boundary flavor symmetry, the above anomalies are all 't Hooft anomalies, and need not be cancelled. Rather, they put interesting constraints on RG flows and dualities. 

In contrast, when bulk gauge fields have Neumann b.c. and $G$ gauge symmetry is preserved on the boundary, the total anomaly for $G$ must cancel. This can always be done by adjusting the bulk Chern-Simons level. Alternatively, the anomaly may be cancelled by adding extra 2d $\CN=(0,2)$ boundary matter that is charged under $G$.

\subsection{Boundary algebra for pure gauge theory}
\label{sec:pureG-N}

Given the anomalies above, we see that it is generally inconsistent at the quantum level to define a Neumann b.c. for pure $G$ gauge theory, unless $G$ is abelian. However, the problem is easily fixed: we can cancel the anomaly by introducing a bulk Chern-Simons term at level $k=-\sh$. (In the twisted action \eqref{actionWk} the term looks like $-\frac{\sh}{4\pi}\int \text{Tr}_{\text{fund}}(\mbf A\pd\mbf A)$, or more universally $-\frac{1}{8\pi}\int \text{Tr}_{\text{adj}}(\mbf A\pd\mbf A)$.)

We will assume in this section that we have such a gauge theory, with Neumann b.c. $\mbf B\big|=0$ and cancelled anomaly. We would like to determine the boundary chiral algebra. 

Naively, we would expect the boundary algebra to be generated by the fermionic bottom component $\c(z)$ of the twisted superfield $\mbf A$ that survives on the boundary, and its $\pd_z$ derivatives. (As usual, the higher components of $\mbf A$ are not $Q$-closed, and cancel in cohomology against $\pd_t$ and $\pd_{\bar z}$ derivatives of lower components.) The action of $Q$ on $\c$, from \eqref{Qsuperfields}, is
\be Q\, \c = - i\c^2\,. \ee
If we expand $\c=\sum_a\c^aT_a$ into a basis for the complex Lie algebra $\mathfrak g$, and use the structure constants $[T_a,T_b]=if^c{}_{ab}T_c$, this becomes the more familiar BRST formula
\be Q\, \c^a = \tfrac12 f^a{}_{b c}\c^b\c^c\,. \label{Qc} \ee
We also see that it is impossible to generate a singular $c(z)c(0)$ OPE. The bulk-boundary propagator connects the boundary operator $\mbf A$ with $\mbf B$ in the bulk, but there are no vertices involving $\mbf B$, so there are no Feynman diagrams that can contribute to a perturbative OPE. There are also no non-perturbative corrections, as they would come from boundary monopole operators, which only exist in the presence of Dirichlet b.c.

Thus, our first guess for the boundary algebra is that it is computed by the cohomology
\be \label{GN-1} \CV_\pd \overset{?}= H^\bullet \big\langle \!\!\big\langle \c(z)\,\big|\, Q\,\c = -i\c^2 \big\rangle\!\!\big\rangle\,.\ee
This is almost correct. We can understand what is wrong with this guess (and how to fix it) in two different ways, one more physical and one more mathematical.

Physically, it was explained \eqref{clambda} that the $\c$-ghost is cohomologous to a gaugino in the full 3d $\CN=2$ theory. More precisely, the \emph{derivative} of $\c$ is cohomologous to the gaugino $\lambda_-$,
\be D_z \c \sim \lambda_-\,,\quad D_z^2\c  \sim D_z\lambda_-\,,\; \text{etc.} \ee
In well established computations of the half-index for Neumann boundary conditions \cite{GGP-walls, GGP-fivebranes, YoshidaSugiyama}, it is only gauge-invariant polynomials formed from the gaugino and its higher modes that contribute. (We revisit this computation in Section \ref{sec:GN-index}.)

We are thus motivated to modify \eqref{GN-1} by including only higher modes of $\c$, and restricting to invariants for the group $G$ of global gauge transformations,%
\footnote{Physically, the derivatives in \eqref{GN-2} should actually be covariant $D_z$ derivatives. However, in the twisted formalism, there is no distinction between covariant and non-covariant derivatives. (The connection in the twisted formalism only has $A_{\bar z}$ and $A_t$ components.) A practical way to understand this is that, when describing the abstract structure of the algebra, it does not matter whether covariant or non-covariant derivatives are used. The connection never appears; nor does the $F_{zz}$ curvature component, since it automatically vanishes.}
\be \label{GN-2} \CV_\pd = H^\bullet\big(\C\big[ \{\pd_z^n \c\}_{n\geq 1}\big]^G,\, Q\,\c = -i\c^2 \big)\,.\ee
In other words, we take the free graded-commutative algebra generated by $\pd_z\c,\pd_z^2\c,...$, restrict to the subalgebra of $G$-invariants, and take cohomology with respect to the differential induced from $Q\,\c=-i\c^2$ (even though $\c(z)$ itself no longer exists).

There is a more sophisticated way to see that \eqref{GN-2} is the correct description of the boundary algebra, working directly in the twisted formalism. Roughly speaking, the $\c$-ghost is introduced in order to impose gauge invariance in a derived way, compatible with cohomological operations. 
However, taking invariants with respect to constant (global) gauge transformations is already compatible with cohomological operations, so the corresponding zero-mode of the $\c$-ghost should be removed. The algebra \eqref{GN-2} appears mathematically as the algebra of derived invariants of the trivial representation $\C$, with respect to the group $G(\CO)$ of holomorphic gauge transformations in the infinitesimal neighborhood of a point on the boundary:
\begin{align} \CV_\pd &\simeq  \C^{G(\CO)}\quad \text{(derived)} \label{GN-3} \\
  &:= \text{Ext}^\bullet_{G(\CO)\text{-mod}}(\C,\C) \notag \end{align}
We  explain this in more detail in Sections \ref{sec:gG} -- \ref{sec:GO}, as the mathematical ideas involved may not be familiar to physicists. 

\subsubsection{Derived invariants of groups vs. algebras}
\label{sec:gG}


We digress to review a few mathematical facts about taking invariants.

Suppose that $V$ is a vector space --- such as the space of local operators in QFT --- equipped with the action of a Lie algebra $\mathfrak g$. The naive operation of taking $\g$-invariants by hand, \emph{i.e} simply restricting to the subspace $V^\g \subseteq V$ of $\g$-invariant elements, is typically not compatible with cohomological operations on $V$. For example, if $Q_V$ is some $\mathfrak g$-invariant differential acting on $V$, it is \emph{not} generally true that taking cohomology commutes with taking invariants:
\be [H^\bullet(V,Q_V)]^\g \neq H^\bullet(V^\g,Q_V)\,. \ee
Mathematically, one says that the operation of taking $\mathfrak g$-invariants is not an exact functor in the category of $\mathfrak g$-modules; it need not preserve kernels and cokernels of maps among modules.

In order to define an operation that \emph{is} naturally compatible with cohomology, one must take derived invariants. This means the following. In the category of $\g$-modules, ordinary invariants may be expressed as $V^\g = \text{Hom}_{\g\text{-mod}}(\C,V)$ (\emph{i.e.} maps from the trivial representation to $V$ that commute with the $\g$ action). Derived invariants are defined as
\be V^\g_{\rm der} := \text{Ext}^\bullet_{\g\text{-mod}}(\C,V)\,. \label{der-Vg} \ee
This is computed explicitly by using the Chevalley-Eilenberg cochain complex, better known in physics as introducing a $\c$ ghost.  Namely,
\be V^\g_{\rm der}  = H^\bullet( \C[\c]\otimes V, Q_{\rm CE})\,, \ee
where $\C[\c]$ denotes the exterior algebra generated by the components $\c^a\in \g^*[1]$ of the fermionic ghost%
\footnote{Quick comment on $\g$ vs. $\g^*$: In physics, the $\c$ ghost is a $\g$-valued field. The operators formed from the $\c$ ghost are functionals of this field, and are therefore elements of $\g^*$. We can see this explicitly when ``expanding $\c$ into components'': in a formula $\c = \c^aT_a$, the generators $T_a$ are elements of $\g$ and the coefficients $\c^a$, which are the actual local operators, are elements of $\g^*$. This agrees with the standard mathematical formulation of the Chevalley-Eilenberg cochain complex, in terms of $\g^*$.\label{foot:CE}}, %
and the Chevalley-Eilenberg differential has a familiar form
\be Q_{\rm CE} \, \c^a = \tfrac12 f^a{}_{bc}\c^b\c^c\,,\qquad Q_{\rm CE}\,v = \c\cdot v \quad\text{(for $v\in V$)}\,, \ee
where $\c\cdot v$ denotes the action of $\g$ on $v$.

Now suppose that $V$ itself comes equipped with a differential $Q_V$, which commutes wit hthe action of $\g$.  Then, we can compare the cohomology computed in two different orders: the cohomology of $V_{\rm der}^\g$ with respect to $Q_V$, or the derived invariants of $H^\bullet(V,Q_V)$ with respect to $\g$.  In general, these are not exactly the same, if we give $H^\bullet(V,Q_V)$ the naive $\g$-action.  However, it is possible to give $H^\bullet(V,Q_V)$ an $L_\infty$-action of $\g$ so that the derived invariants coincide with the cohomology of the derived invariants of $V$:
\be  [H^\bullet(V,Q_V)]^\g_{\rm der}  H^\bullet(V_{\rm der}^\g,Q_V)\,, \ee
These are both equal to the more symmetric $H^\bullet(\C[\c]\otimes V,Q_V+Q_{\rm CE})$, where the differentials are simply added.   

In QFT, when we are working in the BV formalism and/or in a twist of a supersymmetric theory, the space of local operators is naturally equipped with a differential. Gauge-invariants should be taken in a derived manner in order to be compatible with cohomology. This is what the standard BRST procedure does.

There is a catch, however, when dealing with a group rather than an algebra. If $V$ is a representation of a reductive group $G$ (or equivalently, a compact real group $G_c$), the operation of taking ordinary $G$-invariants turns out to already be exact:
\be V^G  = V^G_{\rm der}\,.\ee
Alternatively, $\text{Hom}_{G\text{-mod}}(\C,V)  = \text{Ext}^\bullet_{G\text{-mod}}(\C,V)$ (there is no higher cohomology in the Ext functor). In this case, there is no need to introduce a $\c$-ghost. Indeed, it would be wrong to do so; derived invariants for a reductive group are \emph{not} computed by the Chevalley-Eilenberg cochain complex.

The general upshot for QFT is that, in order to compute gauge-invariant local operators, one should not introduce a ghost for constant gauge transformations. This is actually a familiar, if underappreciated, feature. Perturbatively, local operators are functions on the (analytic) infinite jet space of the fields at a point $p$ (in other words, functions of fields and arbitrary derivatives). In a theory with gauge group $G$, the group $\mathcal G_p$ of gauge transformations acting on local operators at $p$ is infinite-dimensional, but decomposes as a semi-direct product
\be \mathcal G_p \simeq G \ltimes \mathcal G_p'\,, \label{G-semidirect} \ee
where $G$ contains constant gauge transformations near $p$ and $\mathcal G_p'$ contains all derivatives of gauge transformations at $p$.  If the group $G$ is compact (or complex reductive, as in our twisted formalism), its derived invariants are the same as ordinary invariants.
On the other hand, $\mathcal G_p'$ behaves like a Lie algebra, and its derived invariants must be taken using ghosts. 
Altogether, the derived $\mathcal G_p$-invariants of local operators at $p$ are computed by
 introducing all nontrivial derivatives of a ghost field $\c$ (but not $\c$ itself), restricting to $G$-invariants by hand, and taking Chevalley-Eilenberg/BRST cohomology.

In the case relevant for us, a Neumann boundary condition for 3d $\CN=2$ pure $G$ gauge theory in the twisted formalism, there are no (non-exact) local operators on the boundary aside from those formed from the ghost field. We are thus computing derived gauge-invariants of the trivial representation $\C$. We do this by considering all derivatives of the $\c$ ghost at a point on the boundary. However, the $\bar z$ and $t$ derivatives are $Q$-exact, so we are left with the $z$-derivatives, precisely encapsulated in the algebra \eqref{GN-2}.

\subsubsection{Geometric description}
\label{sec:GO}

Another derivation of the algebra \eqref{GN-2} uses a version of the state-operator correspondence adapted to the holomorphic twist. The following is a somewhat schematic translation of the problem into algebro-geometric language, which seems to give a physically sensible result.

The $Q$-cohomology of the space of boundary local operators should be equivalent to the $Q$-cohomology of the space of states on an infinitesimal hemisphere surrounding a point on the boundary. In order to be compatible with a transverse holomorphic foliation of spacetime $\C\times \R_-$, we deform the hemisphere into an infinitesimal disc $D$. Algebraically, we take $D$ to be a `formal disc,' meaning its algebraic functions are formal Taylor series,
\be D = \text{Spec}(\CO)\,,\qquad \CO:=\C[\![z]\!]\,. \label{defD}\ee
We think of this disc as parallel to the boundary, and identified with the boundary except at its origin.

In pure $G$ gauge theory in the twisted formalism, the solutions to the equations of motion on a disc (identified almost everywhere with the boundary) are holomorphic $G$-bundles.  On $D$, we will think of them as algebraic $G$ bundles. Every $G$ bundle on $D$ can be trivialized, and has an isomorphism group $G(\CO)$ of `residual gauge transformations,' where
\be G(\CO) = \text{Maps}(D,G)= \text{the algebraic group $G$ defined over formal Taylor series $\CO$}\,. \ee
(If $G$ is a matrix group, then $G(\CO)$ is the corresponding group of matrices with Taylor-series entries.)
Therefore, the moduli space of solutions to the equations of motion on $D$ may be described as a stack $\text{\{pt\}}/G(\CO)$.

In the holomorphic twist, the space of quantum states on $D$ is the $H_{\bar \pd}^{(0,\bullet)}$ Dolbeault cohomology of the space of solutions to the equations of motion.%
\footnote{Closely related computations of the $Q$-cohomology of the Hilbert space of 3d $\CN=2$ gauge theories on Riemann surfaces were performed in \cite{BullimoreFerrari}. It explained carefully therein that the Hilbert space is Dolbeault cohomology valued in a particular sheaf, depending on the fermionic matter of the theory, and twisted by a power of the canonical bundle due to Chern-Simons terms. In pure gauge theory on a disc with Neumann b.c. and a Chern-Simons level $-\sh$, the overall twist cancels out, so we are left with the cohomology of the structure sheaf as in \eqref{HpG}.} %
 Algebraically, this becomes cohomology of the structure sheaf of the stack $\text{\{pt\}}/G(\CO)$, which is precisely the derived $G(\CO)$-invariants of the trivial representation:
\be H^\bullet(\text{\{pt\}}/G(\CO), \CO_{\text{\{pt\}}/G(\CO)}) = H^\bullet(\{\text{pt}\},\CO_{\{\text{pt}\}})^{G(\CO)}_{\rm der} = \C^{G(\CO)}_{\rm der}  \label{HpG} \ee
This recovers the description \eqref{GN-3}.

Explicitly, the decomposition \eqref{G-semidirect} for $G(\CO)$ takes the form
\be G(\CO) \simeq G\ltimes(\text{Id}+z\, G(\CO))\,. \ee
The subgroup $\text{Id}+z\, G(\CO)$ of non-constant gauge transformations is pro-unipotent, and completely equivalent to its Lie algebra $z\g(\CO)$. This is because the exponential map relating $\text{Id}+z\, G(\CO)$ and $z\g(\CO)$ truncates to define an algebraic isomorphism at any finite order in $z$. Thus, the derived $G(\CO)$ invariants of the trivial representation are computed using the Chevalley-Eilenberg complex built from derivatives of the $\c$-ghost
\be \C[\{\pd_z^n \c\}_{n\geq 1}] \simeq \text{Sym}^\bullet \big(z\g(\CO)[1]\big)\,, \ee
and restricting to $G$-invariants by hand, as in \eqref{GN-2}.

\subsubsection{Character}
\label{sec:GN-index}

We can reproduce the well-known half-index of Neumann b.c., as the graded character of the boundary algebra \eqref{GN-2}. 

We first consider the space of all polynomials in the odd variables $\partial_z^n \c$ $(n\geq 1)$, and take its character as a $U(1)_J\times G$ representation. The character should be unaffected by further taking $Q$-cohomology.

To set up conventions, let $q$ denote the usual $U(1)_J$ fugacity. 
As for $G$, recall that the character of any $G$-representation is a $G$-invariant function of $G$.  The ring of such  functions is the same as the ring of $W$-invariant polynomials on the torus $T \subset G$. For any weight $w$ of $G$, we let $s_w : T \to \C^*$ be the corresponding rank one representation of $T$.  In particular, for any root $\alpha$, we have functions $s_\alpha : T \to \C^*$. The character of the space of polynomials in $\partial_z^n \c$ is readily seen to be
\begin{equation} 
	(q;q)_\infty^{\op{rank}(G)}  \prod_{\alpha \in \op{roots}(G) }(qs_\alpha;q)_\infty\,,
\end{equation}
in terms of the $q$-Pochhammer symbol \eqref{def-qPoch}.

We then project to $G$-invariants ``by hand.'' 
To accomplish this for the character, we should integrate over the compact group with the Haar measure. This the same as performing a contour integral over the compact torus $T_c$ against the Vandermonde determinant 
\begin{equation} 
	\frac{1}{\abs{\op{Weyl}(G)}} \prod_{\alpha \in \op{roots}(G)} (1-s^\alpha). 
\end{equation}
Therefore we find that the character of the algebra of boundary operators with Neumann boundary conditions is
\begin{equation} 
	\frac{\prod_{n \ge 0} ( 1-q^{n+1})^{\op{rank}(G)}}{\abs{\op{Weyl}(G)}}   \oint_{s} \frac{\d s}{2 \pi i s}  \prod_{\alpha \in \op{roots}(G) }(1 - s^\alpha) \prod_{n \ge 0} (1-s^\alpha q^{n+1})\,,
\end{equation}
in agreement with \cite{GGP-fivebranes, DGP-duality, YoshidaSugiyama}.

\subsection{Adding matter}
\label{sec:GN-matter}

For simplicity, we will describe the calculation of the boundary algebra for gauge theories with 3d chiral matter, a bulk superpotential, standard Neumann/Dirichlet matter boundary conditions and perhaps extra boundary chiral fermions, as detailed in Section \ref{sec:W}. We will then explain how the prescription is adapted to general matter theories and boundary conditions. 

It is very easy to modify the prescription of Section \ref{sec:W} in the presence of bulk gauge $G$ fields with Neumann b.c. We just take derived invariants of the boundary algebra associated to the matter fields, under the action of $G(\CO)$. In other words, we introduce derivatives of a $\c$ ghost.

Let's describe how this works concretely. Suppose we have a theory with both gauge fields and matter, in some representation $V$ of the gauge group $G$. Suppose we have a $G$-invariant superpotential $W:V\to \C$. We choose a sub-$G$-representation $L\subseteq V$ and give Neumann b.c. to the $L$-valued chiral multiplets $(\mbf \Phi^i,\mbf \Psi_i)$, and Dirichlet b.c. to the remaining chiral multiplets $(\hat{\mbf\Phi}^{\hat i},\hat{\mbf\Psi}_{\hat i})$. We add boundary fermi multiplets $(\mbf\Gamma^\alpha,\wt{\mbf\Gamma}_\alpha)$ with E and J terms satisfying $W\big|_L = 2E\cdot J$. 
We then give Neumann b.c. to the bulk gauge multiplet $(\mbf A,\mbf B)$, and adjust the bulk Chern-Simons level to cancel any boundary gauge anomaly.

The boundary algebra is generated by the bottom components $\phi(z),\hat\psi(z), \Gamma(z),\wt\Gamma(z)$ and their $\pd_z$ derivatives, as well as the derivatives $\{\pd_z^n\c\}_{n\geq 1}$ of the $\c$-ghost. (In the full physical theory, these would all be covariant derivatives.) We restrict by hand to combinations of these operators invariant under constant $G$ transformations.

The singular OPE's are exactly as before:
\be \Gamma^\alpha(z)\wt\Gamma_\beta(0) \sim \frac{1}{z}\delta^\alpha{}_\beta\,,\qquad
\hat\psi_{\hat i}(z)\hat\psi_{\hat j}(0) \sim \frac{1}{z}\pd_{\hat i}\pd_{\hat j} W\big|_L\,. \ee
Note that the derivatives of $\c(z)$ have non-singular OPE's with all other fields. There are no Feynman diagrams involving the superfield $\mbf A$ and any bulk or boundary matter fields, since the propagator connects $\mbf A$ with $\mbf B$, whereas only $\mbf A$ couples to matter at interaction vertices.

The action of $Q$ is induced by the transformations of bottom components of superfields from \eqref{Qsuperfields}
\be \label{Qc-full} \begin{array}{ll} Q\,\phi = -i\c\cdot \phi\,,\quad &Q\,\hat\psi = dW\big|_L -i\c\cdot \psi\,,  \\[.2cm]
 Q\,\Gamma = -i\c\cdot\Gamma+E\,,\quad &Q\,\wt\Gamma = -i\c\cdot\wt\Gamma+J\,, \end{array}\qquad
   Q\,\c = -i\c^2\,,\ee
where `$\c\cdot$' denotes the action of $\c\in \g[1]$ on various charged matter fields. We say that the action is ``induced'' from \eqref{Qc-full} in the following sense: Using \eqref{Qc-full} one can compute transformations of all derivatives and normal-ordered products of the basic operators $\phi,\hat\psi,\Gamma,\wt\Gamma,\c$. 
Because of the singular terms in the OPE, the BRST transformations of normal-ordered products can contain unexpected terms. We will see some examples below. 

A priori, the transformations will involve the zero mode of $\c$. However, once we restrict to products of operators that are invariant under the global $G$ action, only derivatives of $\c$ will remain. We will be left with a well-defined action of $Q$ on the $G$-invariant products of $\{\pd_z^n\phi,\pd_z^n\hat\psi,\pd_z^n\Gamma,\pd_z^n\wt\Gamma,\pd_z^{n+1}\c\}_{n\geq 0}$.

It is important to observe that the action of $Q$ is built directly from the action of the group $G(\mc{O})$ of holomorphic gauge transformations on 
the boundary chiral algebra of the matter theory: 
\be \label{gaugec-full} \begin{array}{ll} \delta_{\alpha}\,\phi = -i\alpha\cdot \phi\,,\quad &\delta_{\alpha}\,\hat\psi = dW\big|_L -i\alpha\cdot \psi\,,  \\[.2cm]
 \delta_{\alpha}\,\Gamma = -i\alpha\cdot\Gamma+E\,,\quad &\delta_{\alpha}\,\wt\Gamma = -i\alpha\cdot\wt\Gamma+J\,, \end{array}\ee
The extension to general matter theories and boundary conditions with boundary chiral algebra $\mc{V}_\partial[\mc{B}]$ is obvious: we build
the  Chevalley-Eilenberg complex from derivatives of the $\c$-ghost and $\mc{V}_\partial[\mc{B}]$
\be \text{Sym}^\bullet \big(z\g(\CO)[1]\big) \otimes \mc{V}_\partial[\mc{B}]\, \ee
and restrict to $G$-invariants by hand.
\footnote{A possible concern, which we will not explore further here, is that the presence of gauge fields weakens the argument we gave to exclude instanton corrections involving 
configurations of bulk fields which are constant in the direction normal to the boundary. In the presence of gauge fields, the bulk fields would only be covariantly constant, and 
the divergence of the instanton action should be revisited.}

\subsection{Example: SQED}

Let's apply the above prescription to an interesting boundary condition for SQED that appeared in \cite[Sec 6]{DGP-duality}. 

The bulk theory has $G=\C^*$ (or $G_c=U(1)$) and two fundamental chiral multiplets $(\mbf \Phi,\mbf\Psi)$, $(\wt{\mbf \Phi},\wt{\mbf \Psi})$, such that $\mbf\Phi,\wt{\mbf\Phi}$ have gauge charges $+1,-1$, respectively. More abstractly, the vector space $V$ is $\C^2$. There is no Chern-Simons term for $G$. However, there is a mixed Chern-Simons coupling between $G$ and a topological flavor symmetry $\C^*_T$. There is also a flavor symmetry $\C^*_A$ for which $\mbf\Phi,\wt{\mbf\Phi}$ have charges $+1,0$.

We consider a Neumann b.c. on the gauge multiplet, together with Neumann b.c. for both chirals. Using the anomaly calculation summarized from Section \ref{sec:anomaly}, we see that this boundary condition has gauge anomaly $-\tfrac12\times 2=-1$. Thus the boundary condition is inconsistent on its own. To cancel the anomaly, we add a boundary fermi multiplet $(\mbf \Gamma,\wt{\mbf \Gamma})$ of gauge charge $(+1,-1)$.

One can also calculate mixed anomalies between gauge and flavor symmetries (see \cite{DGP-duality}). The mixed anomaly involving $\C^*_A$ vanishes automatically. The mixed anomaly involving the topological symmetry $\C^*_T$ vanishes if we additionally give $(\mbf \Gamma,\wt{\mbf \Gamma})$ charges $(-1,+1)$ under $\C^*_T$ .

Altogether, we find a boundary chiral algebra generated by gauge-invariant combinations of $\phi(z),\wt \phi(z),\Gamma(z),\wt \Gamma(z)$ and derivatives of $\c(z)$, with standard OPE
\be \Gamma(z)\wt\Gamma(z) \sim \frac 1 z\,, \label{SQED-OPE} \ee
and action of $Q$ induced from (absorbing a factor of $i$ into $\c$),
\be \begin{array}{ll}
 Q\,\phi = \c\phi\,, & \qquad Q\,\Gamma = \c\Gamma\,, \\[.2cm]
 Q\,\wt\phi = - \c\wt\phi\,, & \qquad Q\,\wt\Gamma = -\c\wt\Gamma\,, \end{array} \qquad Q\,\c = 0\,.\ee
The charges of various generators under $\C^*$ gauge symmetry, our two $\C^*$ flavor symmetries, R-symmetry, and twisted spin are:
\be \begin{array}{c|cccccc|c}
& \phi & \wt \phi & \Gamma & \wt \Gamma & \c & \pd_z & Q \\\hline
\text{gauge} & 1 & -1 & 1 & -1 & 0 & 0 & 0 \\\hline
A & 1 & 1 & 0 & 0 & 0 & 0 & 0 \\
T & 0 & 0 & -1 & 1 & 0 & 0 & 0 \\\hline
R & 0 & 0 & 0 & 0 & 1 & 0 & 1 \\
J & 0 & 0 & \frac12 & \frac12 & 0 & 1 & 0
\end{array} \ee

We can give a simple example of BRST transformation of a composite field, $:\!\Gamma\wt\Gamma\!:$. Classically, this product is gauge-invariant (under both constant and non-constant gauge transformations). Quantum-mechanically, we use the definition $:\!\Gamma\wt\Gamma\!:(0) = \lim_{z\to 0} \Big( \Gamma(z)\wt\Gamma(0) - \frac1 z\Big)$ of normal-ordering to find an anomalous gauge transformation
\begin{align} \quad Q\big[:\!\Gamma\wt\Gamma\!:(0)\big] &= 
\lim_{z\to 0} \Big( Q\Gamma(z)\wt\Gamma(0) - \Gamma(z) Q\wt\Gamma(0) - Q\frac1 z\Big) \notag \\
 &= \lim_{z\to 0} \Big( \c(z)\Gamma(z)\wt\Gamma(0) + \Gamma(z)\c(0)\wt\Gamma(0)\Big) \label{GGcalc} \\
  &= \lim_{z\to 0} [\c(z)-\c(0)] \Gamma(z)\wt \Gamma(0)
  = \lim_{z\to 0} \frac{\c(z)-\c(0)}{z} = \pd_z\c(0)\,. \notag \end{align}
Therefore, $Q(:\!\Gamma\wt\Gamma\!:) = \pd_z\c$. In particular, $:\!\Gamma\wt\Gamma\!:$ and $\pd_z\c$ are not in the BRST cohomology. 

It is natural to expect that the currents $:\!\Gamma\wt\Gamma\!:(z)$ and $\pd_z\c(z)$ and their derivatives will effectively cancel each other, 
leaving cohomology only in degree $0$. We will now sharpen this expectation into a precise mirror symmetry statement. 

\subsubsection{SQED $\leftrightarrow$ XYZ duality}
\label{sec:SQEDXYZduality}


The chiral algebra above is expected to be dual to the boundary chiral algebra of the XYZ model, with the boundary condition studied in Section \ref{sec:XYZ-NDD}. Recall that the XYZ algebra has just three generators $X(z),\psi_Y(z),\psi_Z(z)$, with trivial action of $Q$, and OPE
\be \psi_Y(z)\psi_Z(0) \sim \frac{X(0)}{z-w}\,. \label{XYZ-OPE} \ee
The charges under R-symmetry, twisted spin, and $\C^*_A\times \C^*_T$ flavor symmetry are
\be \begin{array}{c|cccc}
& X & \psi_Y & \psi_Z & \pd_z  \\\hline
A & 2 & 1 & 1 & 0  \\
T & 0 & -1 & 1 & 0  \\ \hline
R & 0 & 0 & 0 & 0  \\
J & 0 & \frac12 & \frac12 & 1 
\end{array} \ee
Notably, $R$-charges are all zero in this convention%
\footnote{Corresponding to these charge assignments, the bulk bosonic chirals $(X,Y,Z)$ have R-charges $(0,1,1)$. This ensures that the superpotential $W=XYZ$ has R-charge 2, as required.}%
: the entire chiral algebra sits in cohomological degree zero, which is compatible with a trivial action of $Q$. We also observe that $A$-charges are positive-definite.

It was explained in \cite[Sec 6]{DGP-duality} how several simple boundary operators should match across the duality. 
This leads us to conjecture that the equivalence of the XYZ and SQED boundary algebras is induced from the map
\be \label{XYZ-SQED-map} \rho:\quad \begin{array}{ccl} X &\mapsto& \phi\wt\phi \\[.1cm]
 \psi_Y &\mapsto& \Gamma\wt\phi \\[.1cm]
 \psi_Z &\mapsto& \wt\Gamma\phi \end{array}
\ee

There are a several things to check about this proposal. First, we need to make sure that \eqref{XYZ-SQED-map} is a map of quadruply graded dg chiral algebras --- \emph{i.e.} it preserves the four $\C^*$ symmetries, the action of $Q$, and the OPE. Preserving the symmetries is obvious. It is also clear that the three operators $\phi\wt\phi,\Gamma\wt\phi,\wt\Gamma\phi$ are globally $G$-invariant, hence belong to the SQED algebra. Preserving the action of $Q$ means that $\phi\wt\phi,\Gamma\wt\phi,\wt\Gamma\phi$ are all $Q$-closed. This is the same as saying that $\phi\wt\phi,\Gamma\wt\phi,\wt\Gamma\phi$ are invariant under non-constant gauge transformation, since the only role of $Q$ in the SQED algebra is to implement non-constant gauge invariance cohomologically. Since the pairs $(\phi,\wt\phi)$, $(\Gamma,\wt \phi)$, and $(\wt\Gamma,\phi)$ all having nonsingular OPE's, $Q$-closedness/gauge-invariance follows easily; explicitly, we compute:
\be Q(\phi\wt \phi) = \c\phi\wt\phi + \phi(-\c)\wt\phi = 0\,,\quad \begin{array}{l} Q(\Gamma\wt\phi) = \c\Gamma\wt\phi - \Gamma(-\c)\wt\phi = 0\,, \\  Q(\wt\Gamma\phi) = -\c\wt\Gamma \phi-\wt\Gamma(\c\phi) = 0\,. \end{array}  \ee
Finally, we check that the OPE \eqref{XYZ-OPE} is reproduced in the SQED algebra:
\be \label{YZ-OPE} (\Gamma\wt\phi)(z)\, (\wt\Gamma\phi)(0) = \phi(0)\wt\phi(z) \Gamma(z)\wt\Gamma(0) = \phi(0)\wt\phi(0)\frac{1}{z} + \text{regular} \sim \frac{(\phi\wt\phi)(0) }{z}\,, \ee
exactly as required.

The above properties guarantee that $\rho$ descends to a map between the cohomologies of the XYZ and SQED algebras, preserving all gradings. Much less trivial is the claim that $\rho$ is an \emph{isomorphism} on cohomology. 

First, we note on the XYZ side, all operators are in cohomological degree $0$.  For there to be any chance of having an isomorphism between the algebras of operators, the same statement would need to hold on the gauge theory side.  Let us prove this. We first note that we can view the normally-ordered product $:\!\Gamma\wt\Gamma\!:(z)$ as a $U(1)$ current, which we will call $\alpha$.  Let $M$ denote the vacuum module of the SQED algebra, before taking cohomology.  Let $M_0$ be the sub-module of $M$ consisting of those states which are in the kernel of the negative modes $\alpha_{-n}$ of $\alpha$. As with any module for the $U(1)$ current algebra, the entire module $M$ is obtained by applying raising operators $\alpha_n$ to $M_0$:
\begin{equation} 
	M = M_0 [\alpha_1,\alpha_2,\dots].  
\end{equation}
The ghost $\partial_z \c$ has trivial OPE with every operator.  This means that the entire vacuum module is a tensor product of those operators which do not contain ghosts, with the exterior algebra generated by the ghosts states $\partial_z^k \c$. If we denote by $N_0 \subset M_0$ the subspace of states which do not contain any ghosts and are in the kernel of $\alpha_{-n}$, we have
\begin{equation} 
	M = N_0 [ \partial_z \c, \partial_z^2 \c, \dots, \alpha_1,\alpha_2,\dots ].  
\end{equation}
That is, the entire module is obtained from $N_0$ by tensoring with an exterior algebra on the variables $\partial^k_z \c$, and a symmetric algebra on the variables $\alpha_k$, $k > 0$. 

The BRST operator in this basis can be quite complicated, but one of the terms simply sends $\alpha_k \to \frac{1}{k!} \partial^k_z \c$, as we saw above.  We can compute the cohomology of the vacuum module $M$ by using a spectral sequence whose first term  only retains this piece of the BRST operator. If we do so, on the first page of the spectral sequence all the ghosts $\partial_z^k \c$ will cancel with the modes $\alpha_k$, so that we only have $N_0$.  Because this sits in cohomological degree $0$, there is no room for any further differentials, and the entire BRST cohomology of the vacuum module $M$ is isomorphic to $N_0$.   

Concretely, $N_0$ is the subspace of all operators build from $\phi,\til{\phi},\Gamma,\til{\Gamma}$ of charge $0$ under the global gauge $U(1)$ action,  which are in the kernel of the negative modes of $:\!\Gamma\wt\Gamma\!:(z)$.   To prove that the chiral algebras match, it remains to show that the space of such operators is isomorphic to the space of boundary operators for the $XYZ$ model.     
	 
Let us explicitly check that our answer for the cohomology on the SQED side matches with the XYZ side, at low charges for the $\C^*_A$ flavour symmetry.  Firstly, at charge $0$, on the XYZ side the only operator is the identity. We can match this feature in SQED.
At $A=0$, we can not have any operators built from $\phi$ or $\til{\phi}$. The analysis above tells us that the operators that survive BRST cohomology are those in the vacuum module of the boson  $:\!\Gamma\wt\Gamma\!:(z)$, which are also in the kernel of all the negative modes of the same current.  The only such operator is the identity.  

Similarly, in SQED, there are two operators of charge $1$ under $\C^\ast_A$, which correspond to $\Gamma \til{\phi}$ and $\til{\Gamma} \phi$.  It is not hard to see that these are the only operators present on the SQED side.  Indeed, these operators have only a $1/z$ singularity in the OPE with the current $\alpha = :\!\Gamma\wt\Gamma\!:(z)$, so they are in the kernel of $\alpha_{-n}$ for $n > 0$.  Any other operator of charge $A = 1$ on the SQED side must be obtained by applying $\alpha_n$, for $n > 0$, to the operators $\Gamma \til{\phi}$ and $\til{\Gamma} \phi$.  

We can proceed in this manner to check that our map of chiral algebras, from the XYZ algebra to the SQED algebra, is an isomorphism at low charges for $A$.

To complete the proof that our map of chiral algebras is an isomorphism, we can procede as follows.  Vanishing of higher BRST cohomology implies that, in both the XYZ and SQED models, fermionic operators have odd $A$-charge, and bosonic operators have even $A$-charge.  This tells us that when we take the index, including the fugacity for $A$, there can be no cancellation among the operators.   It was shown in \cite{DGP-duality} that the indices match.  We conclude that the vector space of operators of a given quantum numbers is the same  in the XYZ and SQED models.  To show that the map of chiral algebras that we have constructed is an isomorphism, it remains to show that it is either surjective or injective.  

We will show that it is injective. The kernel of the map is some subspace of the space of boundary operators of the XYZ model, which is a chiral algebra ideal. Being a chiral algebra ideal means that it is closed under the application of any mode of an operator.  We need to show that our ideal is zero.  

An element of the ideal is a sum of words in $X, \psi_Y, \psi_Z$ and their derivatives.  By applying repeatedly modes of $\psi_Y$, we can remove all $\psi_Z$ dependence from such a word. Then, by applying repeatedly the modes of $\psi_Z$, we remove all $\psi_Y$ dependence. We conclude that, if our ideal is non-zero, it contains an element which is a sum of words of just $X$ and its derivatives.  

However, an element of this form can not possibly be in the kernel of the map from the XYZ boundary algebra to the SQED boundary algebra. This is because $X$ gets sent to $\phi \til{\phi}$, and in the SQED algebra, there are no linear relations among the words built from the operator $\phi \til{\phi}$ and its derivatives.

\subsection{Example: SQCD and the detYZ model.}

The SQED-XYZ duality has a generalization to a $U(N)$ gauge theory with $N$ fundamental chiral flavours \cite{AHISS,Aharony-duality}. The dual description involves an $N \times N$ matrix $M$ of chiral fields together with two singlet chiral fields $Y$ and $Z$ and superpotential 
\begin{equation}
W= Y Z \det M
\end{equation} 
There are some conjectural pairs of boundary conditions as well \cite{DGP-duality}. The simplest Neumann-type boundary conditions on the SQCD side involves auxiliary chiral fermions 
$\Gamma,\wt\Gamma$ which transform in the determinant and inverse determinant representations of the $U(N)$ gauge group. 

This boundary condition is expected to be dual to Neumann boundary conditions for $M$, Dirichlet for $Y$ and $Z$. 
It is natural to propose an identification 
\be \label{detYZ-SQCD-map} \rho:\quad \begin{array}{ccl} M &\mapsto& \phi\wt\phi \\[.1cm]
 \psi_Y &\mapsto& \Gamma \det \wt\phi \\[.1cm]
 \psi_Z &\mapsto& \wt\Gamma\det \phi \end{array}
\ee
which is compatible with the expected OPE 
\be \psi_Y(z)\psi_Z(0) \sim \frac{\det M(0)}{z-w}\,. \label{detYZ-OPE} \ee
It is far less obvious, of course, that this map extends to a quasi-isomorphism. 

\subsection{Example: Level-rank dualities.}
There is an expected level-rank duality between pure Chern-Simons gauge theories $U(N)_{-k-N,-k}$ and $SU(k)_{k+N}$.
The duality is expected to extend to boundary conditions \cite{DGP-duality}, with Neumann b.c. for $U(N)_{-k-N,-k}$ coupled to $Nk$ 2d chiral fermions 
being dual to Dirichlet boundary conditions for $SU(k)_{k+N}$.

At the level of chiral algebras, the former involves the $\c$ ghosts combined to the $Nk$ 2d chiral fermions. It is well-known that 2d chiral fermions can be expressed as 
an extension of a product of WZW models $\text{WZW}[U(N)_k] \times \text{WZW}[SU(k)_N]$. We expect the $G(\mc{O})$ derived invariants of the $\text{WZW}[U(N)_k]$ 
chiral algebra to consist of the identity field only, so that the boundary chiral algebra reduces to the $\text{WZW}[SU(k)_N]$ chiral algebra. This is the expected answer, see 
Section \ref{sec:gauge-D}.

\section{Gauge theory with Dirichlet boundary conditions}
\label{sec:gauge-D}

A general strategy for dealing with disorder operators is to employ a state-operator map and study the space of states of the theory on a surface surrounding the operator insertion. In a semiclassical quantization, one has to 
deal carefully with moduli spaces of classical solutions of the equations of motion, which may be singular or infinite-dimensional. In the current setup, the analysis can be made tractable using somewhat refined notions from algebraic geometry. We will describe here the boundary chiral algebras $\mc{V}_\partial[H;\mc{B}]$ as vector spaces, 
with particular attention to the other extreme case of Dirichlet boundary conditions $\mc{D}$, where the gauge group becomes trivial at the boundary. 
We will not attempt to compute their vertex algebra structure. 

Dirichlet boundary conditions in gauge theory are the first example in which non-perturbative corrections enter the boundary chiral algebra. These corrections are due to monopole operators, which can exist on a Dirichlet boundary; for an extended physical discussion of such boundary disorder operators, see \cite[Sec. 4.1]{BDGH}. They are also analogous to the boundary 't~Hooft lines of 4d gauge theory considered in \cite[Sec 3.6]{Witten-fivebranes}.

In this section, we will give a mathematical definition of the \emph{vacuum module} of the chiral algebra of a gauge theory with matter and with Dirichlet b.c.\ for the gauge fields, including boundary monopole operators.  Our definition of the module is derived by geometric quantization of the phase-space of the twisted theory on a disc with Dirichlet boundary conditions, and is expressed in terms of an infinite-dimensional algebraic variety called the affine Grassmannian. We do not currently know how to define the chiral algebra structure on the vacuum module, although we formulate a mathematical conjecture to this effect.

We begin with pure gauge theory. Even the perturbative analysis of the boundary chiral algebra is nontrivial. It was predicted in \cite{DGP-duality} that pure 3d $\CN=2$ gauge theory with bulk Chern-Simons level $k$ gives rise perturbatively to a Kac-Moody algebra at level $k-\sh$ on a Dirichlet b.c.,
\be \CV_\pd[\mathcal D]^{\rm pert} \simeq \widehat G_{k-\sh}\,, \ee
where $\sh$ is the dual Coxeter number. This is compatible with the summary of anomalies from Section \ref{sec:anomaly}: a Dirichlet boundary condition carries a boundary 't Hooft anomaly $k-\sh$, for the boundary $G$ flavor symmetry. Of course, the Kac-Moody algebra has singular OPE's. We will explain how they arise from bulk-boundary Feynman diagrams, in a manner similar to the analysis of pure matter theories with a bulk superpotential (Section \ref{sec:W}).

The full, nonperturbative boundary algebra in pure gauge theory at level $k\geq \sh$ was conjectured in \cite{DGP-duality} to be the Wess-Zumino-Witten algebra,
\be \CV_\pd[\mathcal D] \simeq \text{WZW}[G_{k-\sh}]\,.\ee
This is a module for the Kac-Moody algebra;  for positive effective level $k -  \sh > 0$ and semisimple $G$, it is a quotient of the Kac-Moody vacuum module. 

In Section \ref{sec:mon-geom} we will give a geometric construction of the full boundary algebra for gauge theory with Dirichlet boundary conditions, coupled to arbitrary matter. Our analysis is based on a state-operator correspondence and geometric quantization. The method is closely analogous to the BFN construction of monopole operators in 3d $\CN=4$ theories \cite{Nak-I, BFN-II}, as well as to work of \cite{BullimoreFerrari} on 3d $\CN=2$ Hilbert spaces. We explain how to extract the character of the full boundary algebra from this geometric construction in Section \ref{sec:mon-char}.  In this case of pure gauge theory, this reproduces a formula for WZW characters from \cite{DGP-duality}, and in the case of gauge theory with matter, it reproduces a half-index computation of \cite{DGP-duality}.

\subsection{Perturbative algebra in pure gauge theory}

With Dirichlet boundary conditions for the gauge multiplet, the superfield $\mbf A$ is set to zero on the boundary.  The superfield $\mbf{B}$ is non-zero on the boundary. Perturbatively, the boundary chiral algebra is generated by the bottom component $B(z)$ of $\mbf{B}$ and its $z$-derivatives. We recall from Section \ref{sec:twisted} that this is a coadjoint ($\mathfrak g^*$) valued field, which corresponds in the physical theory to the complexified curvature $\frac{1}{g^2}(F_{zt}+iD_z\sigma)$, as in \eqref{def-Fzt}, dualized by the Killing form.

We note that the local operators formed from $B$ are valued in the complex Lie algebra $\g$ (the dual of $\g^*$). Explicitly, if we choose a dual basis $T^a$ for $\g^*$ and expand $B = \sum_a B_a T^a$, the local operators $B_a$ will be elements of $\g$. The operators $B_a$ all have R-charge zero and twisted spin $J=1$.

In the bulk, the action of $Q$ on the bottom component $B$ may be read off from \eqref{Qsuperfields}:
\be Q\, B = -i[\c,B] -\tfrac{k}{2\pi}\pd_z\c\,. \ee
We expect a quantum correction to shift $k\to k-\sh$ in this formula.
In the physical theory, this is a well-known correction due to gauginos with real mass proportional to $k$, at least when $k > \sh$ \cite{AHISS}.

On the boundary, the Dirichlet b.c. sets $\c\big|=0$, compatible with breaking of gauge symmetry. Thus, in the boundary algebra we simply have $Q\,B=0$. Moreover, since $B$ is $Q$-closed on the boundary but not in the bulk, it can have a singular boundary OPE.

The boundary OPE of $B(z)$ with itself indeed turns out to be nontrivial. It gets perturbative quantum corrections that may be computed via Feynman diagrams, in close analogy with our analysis of superpotential corrections in Section \ref{sec:W-OPE}. 

For gauge theory in the twisted formalism, the propagator connects superfields $\mbf A$ and $\mbf B$
\be \mbf A^a\, \raisebox{.2cm}{\underline{\hspace{1in}}}\; \mbf B_a \ee
There are two interaction vertices,
\be \includegraphics[width=1.5in]{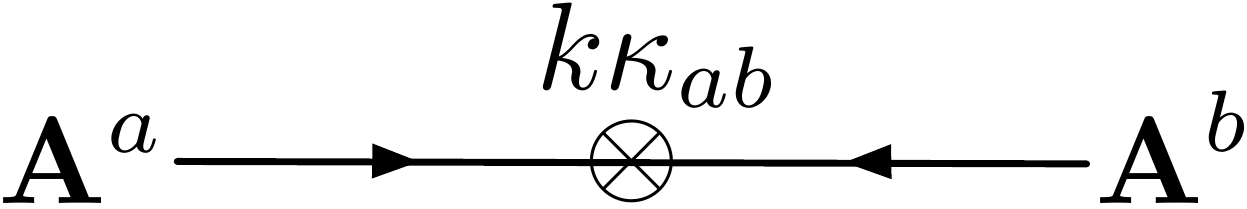}\,,\hspace{.5in} \raisebox{-.42in}{\includegraphics[width=1.5in]{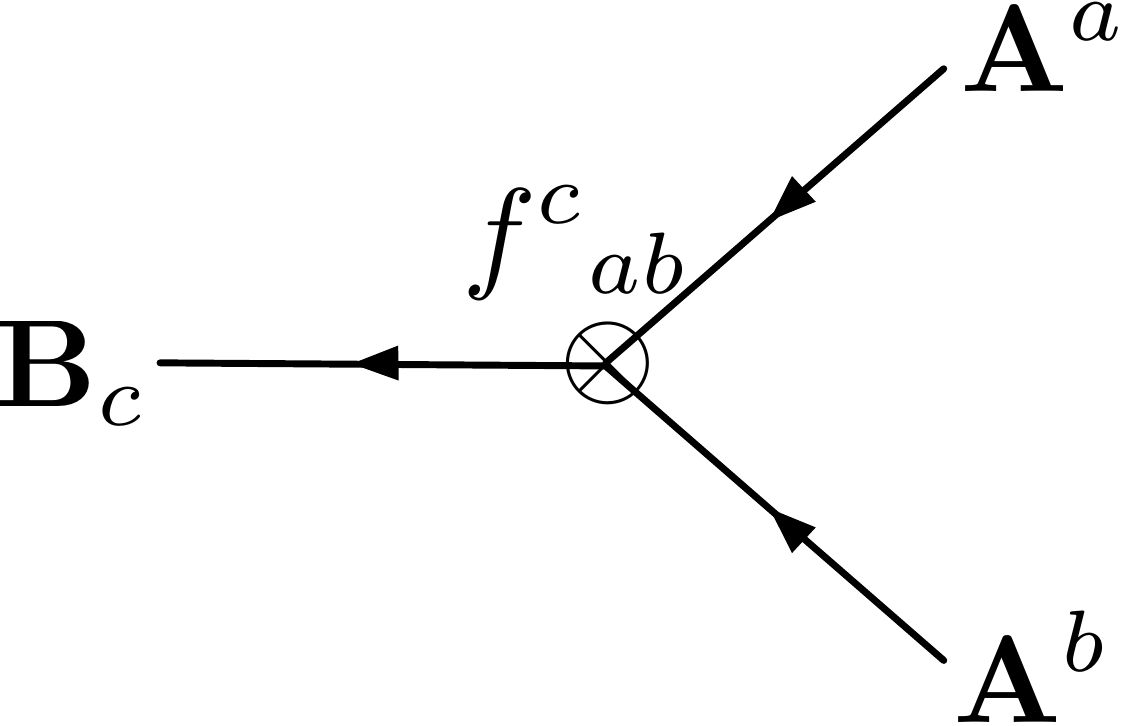}} \ee
coming from the Chern-Simons term $k \mbf A\pd \mbf A$ (which uses the Killing form $\kappa_{ab}$) and the nonabelian part of the kinetic term $\mbf B[\mbf A,\mbf A] \in \mbf B F(\mbf A)$.

We refer to \cite{GwilliamWilliams2019} for a detailed and careful analysis of the Feynman diagrams of this theory with Dirichlet boundary conditions, in the gauge where $\dbar^\ast \mbf{B}= 0$, $\dbar^\ast \mbf{A}= 0$.   The results of \cite{GwilliamWilliams2019} show that, in this gauge, the theory is one-loop exact (i.e.\ the only Feynman diagrams that contribute have at most one loop in the bulk). This is in contrast to the standard choice of gauge in Chern-Simons theory, where diagrams at arbitrary loops can contribute.  Further, Gwilliam and Williams in \cite{GwilliamWilliams2019} show that the theory is finite (this is not obvious in this gauge). 

Because of these results, we find that there are just three Feynman diagrams that can be drawn that contribute to the 2-point function of boundary operators $\mbf B_a(0,0,0)$ and $\mbf B_b(z,\bar z,0)$, in an arbitrary background of the bulk $\mbf A$ and $\mbf B$ fields (subject to the boundary conditions):
\be \raisebox{-.6in}{\includegraphics[width=5in]{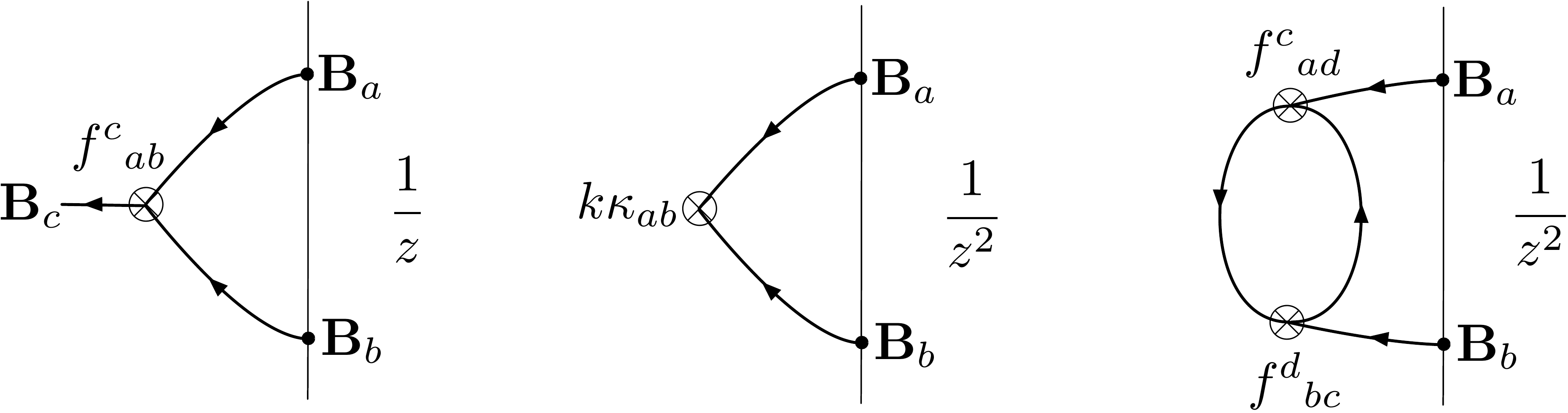}} \ee
Dimensional analysis together with conservation of twisted spin shows that the diagrams contribute at orders $1/z$, $1/z^2$, and $1/z^2$, respectively, to the OPE. 

We expect that a careful calculation (which we do not do) leads to the boundary OPE
\be B_a(z)B_b(0) \sim  \frac{(k-\sh)\kappa_{ab}}{z^2} + \frac{if^c{}_{ab}}{z}B_c(0)\,. \label{KM-OPE} \ee
The coefficient of $\frac{1}{z}$ is computed by a Feynman diagram analysis identical to that in section \ref{sec:W-OPE}.  The coefficient of $k / z^2$ comes from a rather similar analysis.  The shift by the dual Coxeter number comes from the third diagram, with $f^c{}_{ad}f^d{}_{bc} = 2\sh \kappa_{ab}$.  See theorem 6.1 of \cite{GwilliamWilliams2019}, who show that this third diagram has the effect of modifying the coefficient of the $\mbf{A} \partial \mbf{A}$ term in the Lagrangian by a term proportional to $\sh$. Unfortunately, Gwilliam and Williams did not compute the constant of proportionality; the computation of the relevant amplitude is a little non-trivial, because we are on a manifold with boundary.   

Assuming the Feynman diagrams work as expected, this algebra is the Kac-Moody algebra at level $k - \sh$.   

This perturbative analysis applies for any value of $k$, even those for which $\abs{k} < \sh$. In this range, supersymmetry is expected to be spontaneously broken.  This, however, is a non-perturbative effect: we expect that if $\abs{k} < \sh$, boundary monopoles will contribute operators of non-zero R-charge (cohomological degree) which cancel cohomologically with the perturbative operators.  We will find some hint of this below.

\subsection{Geometric construction of monopole operators for pure gauge theory} 
\label{sec:mon-geom}

There are further, non-perturbative, corrections to the boundary algebra when we have Dirichlet boundary conditions for gauge fields. To describe the complete space of local operators on the boundary, we can use a holomorphic version of a state-operator correspondence, much as we did in Section \ref{sec:GO} for Neumann boundary conditions.

We surround (putative) local operators on the boundary with an infinitesimal disc $D$, identified with the boundary everywhere except the origin $z=\bar z=0$. Algebraically, we model $D$ as a formal disc \eqref{defD}, whose algebraic functions are formal Taylor series, denoted $\CO=\C[\![z]\!]$. We also introduce the formal punctured disc $D^* = D\backslash \{0\}$, whose algebraic functions are formal Laurent series, denoted $\CK = \C(\!(z)\!)$. Altogether, the notation is:
\be D = \text{Spec}(\CO)\,,\quad D^*=\text{Spec}(\CK)\,;\qquad \CO=\C[\![z]\!]\,,\quad \CK = \C(\!(z)\!)\,. \ee
We expect the complete space of local operators to be equivalent to the $Q$-cohomology of the Hilbert space of the theory on $D$, with fields constrained to obey the Dirichlet boundary condition on $D^*$.

There are two (ultimately equivalent) ways to construct the desired Hilbert space. One is to borrow results of \cite{BullimoreFerrari} on Hilbert spaces of twisted 3d $\CN=2$ gauge theories, on smooth and compact Riemann surfaces --- and to guess a generalization of the results to the setting of formal discs.
Alternatively, we may rederive the Hilbert space directly in the twisted formalism. We follow the first approach for now, and explain the derivation in the twisted formalism in Section \ref{sec:twisted-HD}. 

Given a 3d $\CN=2$ gauge theory and a smooth compact Riemann surface $\Sigma$, the authors of \cite{BullimoreFerrari} carefully reduced the theory on $\Sigma\times \R_t$ to 1d $\CN=2$ B-type quantum mechanics. The quantum mechanics localizes to $Q$-fixed points of the equations of motion on $\Sigma$. In the case of pure 3d $\CN=2$ gauge theory with group $G_c$, the $Q$-fixed points may be identified with holomorphic $G$-bundles. Thus one obtains an effective 1d quantum mechanics with target $\text{Bun}_G(\Sigma)$, the moduli space of holomorphic $G$-bundles. The $Q$-cohomology of the Hilbert space is then identified with Dolbeault cohomology
\be \CH(G,\Sigma) \simeq H^\bullet_{\bar\pd}(\text{Bun}_G(\Sigma), K^{\frac12})\,, \ee
taking coefficients in the square root of the canonical bundle on $\text{Bun}_G(\Sigma)$.%
\footnote{We are using algebraic/sheaf notation for cohomology on the RHS. Analytically, the RHS would be written as $H^{(0,\bullet)}_{\bar\pd}(\text{Bun}_G(\Sigma), K^{\frac12})$, \emph{i.e.} it is the cohomology of the $\bar\pd$ operator acting on $(0,p)$ forms valued in $K^{\frac12}$.}

The presence of a 3d Chern-Simons coupling at level $k$ modifies the construction, further twisting by the $k$-th power of a line bundle $\CL$ whose first Chern class generates the singular (or de Rham) cohomology $H^2(\text{Bun}_G(\Sigma),\Z)$ on each connected component.%
\footnote{If $G$ has trivial center, then $\text{Bun}_G$ is connected and $H^2(\text{Bun}_G(\Sigma),\Z)=\Z$, so one can simply say that $c_1(\CL)$ generates. In general, $c_1(\CL)$ is a multiple of the K\"ahler form, normalized to restrict to a generator of $H^2$ on each connected component.} %

The canonical bundle is $K \simeq \CL^{-2\sh}$, so that altogether $K^{\frac12}\otimes \CL \simeq \CL^{k-\sh}$, and
\be \CH(G_k,\Sigma) \simeq  H^{\bullet}_{\bar\pd}(\text{Bun}_G(\Sigma), \CL^{k-\sh})\,. \label{HSigma} \ee
For $k\geq \sh$, one might recognize this as the space of conformal blocks of the $G_{k-\sh}$ WZW model \cite{BeauvilleLaszlo, FriedanShenker}, which famously reappeared in the geometric quantization of compact (non-superymmetric) Chern-Simons theory \cite{Witten-Jones, ADW, Hitchin-GQ}.   In fact, a consequence of the twisted formalism is that the twist of 3d $\CN=2$ pure gauge theory, at Chern-Simons level $k > \sh$, is equivalent to ordinary Chern-Simons theory at level $k - \sh$.

Interestingly, for $|k|< \sh$ the physical 3d $\CN=2$ theory is expected to spontaneously breaks supersymmetry \cite{Ohta, BHKK}, and correspondingly the space \eqref{HSigma} should be empty.  This is not at all obvious and would be worth exploring further. We leave a proper discussion of the $k<-\sh$ case to future work.  

To describe local operators on the boundary, we extend this construction from a compact Riemann surface $\Sigma$ to the formal disc $D$. We are led to look at the moduli space  $\text{Bun}_G(D|D^*)$ of algebraic $G$-bundles on $D$, trivialized on $D^*$ so as to obey the Dirichlet boundary condition. Such a bundle is characterized by its transition function from a neighborhood of the origin on $D$ to the complement of the origin on $D^*$; the transition function $g(z)$ is an algebraic gauge transformation on $D^*$, \emph{i.e.} an element $g(z)\in G(\CK)$ where
\be   \begin{array}{ccl} G(\CK) &= &\text{\{Maps\,:\;$D^*\to G$\}} \\
 &=& \text{$G$ defined over formal Laurent series $\C(\!(z)\!)$}\,. \end{array}
\ee
The bundles must be counted modulo isomorphisms, which consist of algebraic gauge transformations that extend over the origin, \emph{i.e.} elements of \{Maps\,:\;$D\to G$\} $=G(\CO)$. Altogether, our moduli space  is the one-sided quotient
\be \text{Bun}_G(D|D^*) \simeq G(\CK)/G(\CO)  = \text{Gr}_G\,, \ee
called the affine Grassmannian for $G$ and commonly denoted $\text{Gr}_G$.%
\footnote{The double quotient $G(\CO)\backslash G(\CK)/G(\CO) = G(\CO)\backslash \text{Gr}_G$, also known as the space of Hecke modifications, has appeared in the physics of bulk 't Hooft lines in 4d Yang-Mills \cite{KapustinWitten, Kapustin-hol}, and in the BFN construction of 3d $\CN=4$ Coulomb branches \cite{Nak-I,  BFN-II}. The one-sided quotient is appearing in the construction of boundary monopole operators here for a similar reason.}

Naively, the vector space of local operators that we are looking for is the Dolbeault cohomology of $\text{Gr}_G$ with values in $K^{\frac12}\otimes \CL^k$ (or, algebraically, sheaf cohomology). Some care is needed in making sense of this, because $\text{Gr}_G$ is infinite-dimensional! We claim that one should consider not Dolbeault cohomology in the usual sense, but rather Dolbeault \emph{homology}.  We define the Dolbeault homology with coefficients in a holomorphic vector bundle $\mathcal E$ 
to be the linear dual of the Dolbeault cohomology with coefficients in the dual bundle $\mathcal E^\vee$,
\begin{equation} 
	H_{n,\bar\pd} (\op{Gr}_G, \mathcal E) := 	H^n_{\bar \pd}(\op{Gr}_G, \mathcal E^\vee)^\vee. 
\end{equation}
Intuitively, Dolbeault cohomology behaves like functions, or functions twisted by a bundle; Dolbeault homology behaves like distributions, twisted by a bundle.

Following the analysis of Bullimore-Ferrari \cite{BullimoreFerrari}, but using a disc with Dirichlet boundary conditions instead of a closed surface, we propose that the space of local operators on the boundary of $3d$ $\CN=2$ pure gauge theory with Chern-Simons level $k$ is the Dolbeault homology of $\op{Gr}_G$ with coefficients in the line bundle $\mc{L}^{\sh-k}$.
In section \ref{sec:mon-char} we will calculate the character of the Dolbeault homology, and find that it reproduces the known formulae for $1/2$ indices of 3d $\CN=2$ theories \cite{DGP-duality}. 

When $k > \sh$, the twist of 3d $\CN=2$ pure gauge theory at level $k$ is ordinary Chern-Simons at level $k - \sh$. This leads us to expect that our definition of boundary local operators in terms of Dolbeault homology  matches with the vacuum module of the WZW model. This is indeed true, and known to specialists in geometric representation theory \cite{Kumar1987, Mathieu1988, CliffKremnitzer2020}, 
\be H^\bullet_{\bar\pd}(\text{Gr}_G, \mc{L}^{\otimes k-\sh} )^\vee \simeq \text{WZW}(G_{k-\sh})\,. \label{HGrG}  \ee

The OPE is not obvious in this description. Schematically, it arises from considering $G$-bundles on pairs of small discs embedded in a larger disc. The algebraic version of this moduli space is known in mathematics as the Beilinson-Drinfeld Grassmannian.

We leave a proper discussion of the $k<-\sh$ case to future work.

\subsubsection{Hilbert space in the twisted formalism}
\label{sec:twisted-HD}

We may also arrive at the description \eqref{HGrG} of the Hilbert space on a disc (with Dirichlet boundary conditions) via a first-principles computation in the twisted formalism. We explain how this goes.

We first construct a phase space on the disc, which is the symplectic manifold solutions to the equations of motion for the twisted action, satisfying Dirichlet boundary conditions. Then we apply geometric quantization to find the Hilbert space. Since everything is complexified in the twisted formalism, phase space will turn out to be a \emph{holomorphic} symplectic manifold, and we will apply a holomorphic version of geometric quantization.

We will analyze the solutions to the equations of motion in the gauge in which $A_t = 0$.  The equations of motion then say that
\be \begin{array}{rl}
	\partial_t B &= 0\\[.1cm]
	(\partial_{\zbar} + A_{\zbar}) B &= 0\\[.1cm]
	\partial_t A_{\zbar} &= 0\,.
\end{array} \ee
That is, the fields $A,B$ are independent of $t$. The gauge transformations that are independent of $t$ still act. Under these gauge transformations, $A_{\zbar}$ transforms as a $(0,1)$ connection in the $z,\zbar$ plane, and $B$ transforms as a Higgs field. 

Altogether, we find that the moduli space of solutions to the equations of motion on a closed surface $\Sigma$ (which lives at a constant value of $t$) is the Hitchin moduli space $\mc{M}^{Higgs}(\Sigma)$ of Higgs bundles on $\Sigma$.  

We are interested in the solutions to the equations of motion on a disc $D$ whose boundary $\pd D$ lies on the Dirichlet boundary condition at $t=0$. The value of $t$ over the interior of any such disc varies, but since all field configurations are independent of $t$, this does not matter. The field $A$ then describes a holomorphic $G$-bundle on $D$, which is trivialized on $\partial D$.  The field $B$ describes a Higgs field for this bundle. Dirichlet boundary conditions do not constrain~$B$.  

The usual argument that shows that the Hitchin moduli space is holomorphic-symplectic on a  closed surface continues to apply in this situation. Gauge-equivalent classes of variations of the gauge field $A$ are given by the relative Dolbeault cohomology $H^{(0,1)} (D, \partial D; \g_P)$ with coefficients in the adjoint bundle.  Dolbeault cohomology relative to the boundary is the same as compactly supported Dolbeault cohomology.  Serre duality \cite{Serre1955} provides a non-degenerate pairing between $H^{(0,1)}_c(D, \g_P)$ (where the subscript $c$ indicates compact support)  with $H^{(1,0)}(D,\g_P)$, which is the space of first-order variations of the field $B$.  In fact, what we have shown is that is that our moduli space is the cotangent bundle of the moduli of holomorphic bundles on $D$ trivialized at the boundary.

The trivialization of the $G$-bundle on the boundary of $D$ is provided by a section of the bundle on $\partial D$.  It is convenient to assume that this trivialization is analytic, and in fact analytically extends to a  trivializaton on the punctured disc $D^\ast$ with a finite-order pole at $0$.  The moduli space of bundles with a trivialization on $D^\ast$ is a dense subspace of the moduli space of bundles trivialized on the boundary, so this assumption will not affect our analysis in any important way.

The moduli of $G$-bundles on $D$, trivialized on the punctured disc $D^\ast$, is then the complex-analytic version of the affine Grassmannian discussed above.  Our assumption that the trivialization on $D^\ast$ has finite-order poles at the origin implies that this complex-analytic version of the affine Grassmannian in fact coincides with the algebro-geometric one.

The space of local operators --- which quantizes this phase space --- should then be the Dolbeault cohomology of the affine Grassmannian, with coefficients in an appropriate line bundle. The line bundle, as above, is $\mc{L}^{\sh - k}$.

\subsubsection{Correlation functions of boundary operators}
Here we will explain how to define correlation functions of boundary operators using the affine Grassmannian, following Beilinson-Drinfeld \cite{BeilinsonDrinfeld2004} (although working much less carefully!).

The correlation functions of a collection of boundary local operators on a surface $\Sigma$ should be an element of the Hilbert space on $\Sigma$.  Equivalently, if we put our theory on an interval $[-1,0] \times \Sigma$, with Dirichlet boundary conditions at $0$, and choose a state at $-1 \times \Sigma$, we can define correlation functions that are numbers.

In our conventions, the Hilbert space at $0 \times \Sigma$ is Dolbeault (co)homology
\be  H^\bullet_{\bar\pd}(\op{Bun}_G(\Sigma), \CL^{ k - \sh } )^\vee = H_{\bullet,\bar\pd}(\text{Bun}_G(\Sigma), \CL^{ \sh - k} )\,. \ee
The Hilbert space at $-1\times \Sigma$ is the dual vector space, simply $H^\bullet_{\bar\pd}(\op{Bun}_G(\Sigma), \CL^{ k - \sh } )$. 

To produce correlation functions of operators inserted at any collection of points $p_1,\dots,p_n \in \Sigma$, we need to define maps 
\begin{equation} 
	\bigotimes_{i=1}^n H_{\bullet,\bar\pd} \left( \op{Gr}_G(p_i) , \CL^{\sh - k}     \right)  \to H_{\bullet,\bar\pd} (\op{Bun}_G(\Sigma), \CL^{\sh - k} ).  
\end{equation}
Here $\op{Gr}_G(p_i)$ is the affine Grassmannian based at $p_i$: it is the moduli of $G$-bundles on a small disc around $p_i$, trivialized in the complement of $p_i$, just as in \eqref{HGrG}.

Equivalently, $\op{Gr}_G(p_i)$ is the moduli of bundles on the entire surface $\Sigma$ trivialized away from $p_i$.  Further, 
\begin{equation} 
	\op{Gr}_G(p_1) \times \dots \times \op{Gr}_G(p_n) = \op{Bun}_G(\Sigma\,|\, \Sigma\! \setminus\! \{p_1,\dots,p_n\} )  
\end{equation}
is the moduli of bundles on $\Sigma$, trivialized away from the points $p_i$. Thus there is a natural map
\begin{equation} 
	m : \op{Gr}_G(p_1) \times \dots \times \op{Gr}_G(p_n) \to \op{Bun}_G(\Sigma)  
\end{equation}
defined by forgetting the trivialization on the complement of the points $p_i$. Dolbeault homology is covariantly functorial, since it is the complex analytic version of distributions. We will get an induced map on Dolbeault homology twisted by the appropriate line bundles as long as the line bundles on each side match up, \emph{i.e.} as long as
\begin{equation} 
	m^* \mc{L}_{\Sigma} = \boxtimes \mc{L}_{p_i} \,. 
\end{equation}
Here $\mc{L}_{\Sigma}$, $\mc{L}_{p_i}$ indicate the line bundle on $\op{Bun}_G(\Sigma)$ or on $\op{Gr}_G(p_i)$.  This is a standard fact in the theory of the affine Grassmannian: see \cite{Zhu2016}, Theorem 4.2.1.

\subsection{Including bulk matter}

Suppose we have a 3d $\mathcal N=2$ theory with $G$-symmetry, so that it can be coupled to $\mathcal N=2$ gauge theory with Chern-Simons level.  Let us choose some boundary condition for the bulk fields, resulting in some boundary chiral algebra $\mc{V}$.  Our goal in this section is to construct a new chiral algebra obtained by coupling our matter theory to $G$-gauge theory with Dirichlet boundary conditions.  

We are able to describe the vacuum module of this new chiral algebra, and also describe correlation functions.  A rigorous construction of the structure maps of this new chiral algebra (i.e.\ the OPE) is partly conjectural, but we hope is accessible to connoiseurs of Beilinson-Drinfeld style factorization algebras.  

Our description will be in terms of the affine Grassmannian, as above.  A point $P \in \op{Gr}_G$ is a $G$-bundle on the disc $D$, trivialized on the punctured disc $D^*$.  The first thing we need to explain is how every such $P$ gives rise to a module $\mc{V}_P$ for the boundary chiral algebra of our matter theory.

We will explain in concrete terms how this works when our bulk theory is given by a collection of chiral fields $(\mbf\Phi,\mbf\Psi)$ transforming in some representation $V$ of $G$ of $R$-charge $r$,  and we choose Neumann boundary conditions.  In this case, the presence of the bundle $P \in \op{Gr}_G$  on the disc $D$ means that the fields $\mbf{\Phi}$ on the boundary are sections of the associated bundle $V_P$ on the disc:
\begin{equation} 
	\mbf{\Phi} \in \Omega^{0, \bullet}(D, K_D^{\tfrac{r}{2} }\otimes V_P^{(r)}).  
\end{equation}
Operators are built in the usual way, as functions of these fields. 

To make this more concrete, let us introduce some special points in the affine Grassmannian, which are the fixed points for the action of $T \times \C^*$, where $T$ is the maximal torus of $G$ and $\C^*$ rotates $z$.  Given any cocharacter $m : \C^* \to T$, we can view $m$ as defining an algebraic loop in $T$, and so an algebraic loop in $G$. Thus $m$ defines an element of $G(\mathcal K)=G(\!(z)\!)$.  Since there is a quotient map $G(\mathcal K) \to \op{Gr}_G$, we get a corresponding point in the affine Grassmannian, which we call $P_m$. 

Let us decompose the representation $V$ into weight spaces $w$.  Then, the associated bundle $V_{P_m}$ to this bundle on the disc simply gives sections of $V$ poles or zeroes at the origin according to their weights:
\begin{equation} 
	\mbf{\Phi} \in \Omega^{0, \bullet}(D, \oplus_w K_D^{\tfrac{r}{2} }\otimes V_w \otimes \Oo (-\ip{m,w}) ) .  
\end{equation}
Changing the order of pole or zero of the field $\mbf{\Phi}$ results in a module $\mc{V}_m$ for the chiral algebra $\mc{V}$ which is a spectral flow of the vacuum module. 

Clearly, this construction applies for any Lagrangian 3d $\CN=2$ gauge theory with $G$-symmetry and a $G$-invariant boundary condition.  It produces, for each $P \in \op{Gr}_G$, a module $\mc{V}_P$ for the boundary chiral algebra $\mc{V}$.  

Let us treat $\mc{V}_P$ and $\mc{V}$ as dg chiral algebras, before taking BRST cohomology. Then, as we vary $P$, $\mc{V}_P$ forms a cochain complex of infinite-dimensional holomorphic vector bundles on $\op{Gr}_G$. We have sketched the construction of a module $\mc{V}_{\op{Gr}_G}$ for the dg chiral algebra $\mc{V}$ which lives in the category of complexes of complexes of vector bundles on the affine Grassmannian $\op{Gr}_G$. 

The cohomology groups of $\mc{V}_P$ may, in principle, jump as we vary $P$. Thus, the cohomology groups $H^\bullet(\mc{V}_P)$ must  be treated as quasi-coherent sheaves, and the total cohomology is a module for the graded chiral algebra $H^\bullet(\mc{V})$ in the category of quasi-coherent sheaves on $\op{Gr}_G$.  

Now let us state our proposal for the vacuum module of the chiral algebra obtained by coupling to gauge theory with Dirichlet boundary conditions. We let $k_{\text{eff}}$ be the \emph{effective} Chern-Simons level, as in \cite[Eqn (3.22)]{DGP-duality} --- it is the shift of the bare Chern-Simons level by $-\sh$ as well as contributions from the matter fields.    Our proposal is that the space of local operators is 
\begin{equation} 
	H_{\bullet,\dbar}( \op{Gr}_G,  \mc{V}_{\op{Gr}_G} \otimes \mc{L}^{-k_{\text{eff}}} ).  
\end{equation}
That is, we take the same definition we used for the vacuum module of pure gauge theory, except now we take homology with coefficients in the sheaf of modules $\mc{V}_{\op{Gr}_G}$, twisted by $\mc{L}^{-k_{\text{eff}}}$. 

We do not show rigorously how to give our proposed vacuum module the structure of a chiral algebra, although we expect that the Beilinson-Drinfeld factorization machinery will be able to show this.  Let us formulate this as a conjecture, using a slightly different description of the bundle of modules $\mc{V}_{\op{Gr}_G}$.   
\begin{conjecture}
	Let $\mc{V}$ be any vertex algebra (without a Virasoro current) with an action of the group $G(\mc{O})$.  Form the associated bundle $G(\mc{K}) \times_{G(\mc{O})} \mc{V}$ on the affine Grassmannian $\op{Gr}_G$ ( this is an alternative description of the bundle $\mc{V}_{\op{Gr}_G}$ given above).  

	Then, the Dolbeault homology with coefficients in $\mc{V}_{\op{Gr}_G}$ has the structure of a vertex algebra. 
\end{conjecture}

\subsection{The character of the algebra of boundary operators with Dirichlet boundary conditions}
\label{sec:mon-char}

In this section, we will compute the character of our proposed vacuum module for a theory with Dirichlet boundary conditions for the gauge fields. We do this by using localization on the affine Grassmannian.  The answer is in perfect agreement with the computations of \cite{DGP-duality}, providing a highly non-trivial check of our proposal. 

 We will compute the character equivariantly under the action of $T \times \C^*_q$, where $T \subset G$ is the Cartan.   The fixed points in the affine Grassmannian for the action of $T \times \C^*_q$ are labelled by cocharacters $m : \C^* \to T$.  We view any such $m$ as an element of the loop group
\begin{equation} 
	m(z)\in T(\mathcal K) \subset G(\mathcal K)\,,  
\end{equation}
 and so as a point in the affine Grassmannian $\text{Gr}_G = G(\mathcal K)/G(\mathcal O)$. As a familiar example, in the case that $G= SU(2)$, the cocharacters are given by
 \begin{equation} 
	 m(z) = \begin{pmatrix}
		 z^k & 0 \\
		 0 & z^{-k} 
	 \end{pmatrix}
 \end{equation}
These fixed points on the affine Grassmannian correspond to boundary monopole field configurations for the maximal torus $T \subset G$.

The localization formula for the Dolbeault homology will give us a sum over fixed points.  To understand how this works, let us recall the standard localization formula computing the Dolbeault cohomology of some variety $X$ with an action of some torus $T$ with isolated fixed points $p$, and some equivariant vector bundle $V$.   The character of $T$ acting in $H^\bullet_{\dbar}(X, V)$ is given by a sum over the fixed points, where each fixed point contributes the character of $T$ acting on holomorphic sections of $V$  on a small neighbourhood of $p$.  

Holomorphic sections of $V$ on a small neighbourhood of $p$ have a Taylor expansion as elements of $V_p \otimes \what{\Sym}^\ast T^\ast_p X$, so that the character formula is
\begin{equation} 
	\chi_T  H^\bullet_{\dbar}(X, V) = \sum_{p} \chi_T (V_p)\chi_T(   \Sym^\ast T^\ast_p X ).  
\end{equation}
In our context, we are interested in Dolbeault homology, which we defined to be the linear dual of Dolbeault cohomology with coefficients in the dual $V^\vee$.  We find the character of Dolbeault homology is
\begin{equation} 
	\chi_T  H_{\bullet,\dbar}(X, V) = \sum_{p} \chi_T (V_p) \chi_T( \Sym^\ast T_p X), 
\end{equation}
i.e.\ we have replaced the cotangent bundle by the tangent bundle.

As before, let us fix some bulk matter theory with a $G$-action and a $G$-invariant boundary condition.  We let $\mc{V}$ be the corresponding boundary vertex algebra, and for each point $P \in \op{Gr}_G$, we let $\mc{V}_P$ be the module discussed above, which is the fibre of a sheaf of modules $\mc{V}_{\op{Gr}_G}$ on the affine Grassmannian. 

We are interested in computing, by localization, the character of 
\begin{equation} 
	H_{\bullet,\dbar}(\op{Gr}_G, \mc{V}_{\op{Gr}_G} \otimes \mc{L}^{-k_{\text{eff}}}).  
\end{equation}
By the localization formula above, this becomes the sum
\begin{equation} 
	\sum_{m : \C^* \to T} \chi_{T \times \C_q^\times}( \mc{L}_m)^{-k_{\text{eff}}} 
	   \chi_{T \times \C^*_q} ( \mc{V}_m)  \chi_{T \times \C^*_q} \left(\Sym^\ast T_m \op{Gr}_G \right). 
\end{equation}
In this formula, the sum is over cocharacters $m : \C^* \to T$, $\mc{L}_m$ indicates the fibre of the determinant line bundle at $m$, $\mc{V}_m$ the spectral-flow module associated to $m$,  and $T_m \op{Gr}_G$ is the tangent space at $m$ to the affine Grassmannian.

Note that unlike in the calculation of the bulk index, all coweights of $G$, and not just the dominant coweights, contribute to the calculation. This is because the Dirichlet boundary conditions for the gauge field break all gauge symmetry at the boundary, so that two monopole configurations related by an element of the Weyl group are distinct.   

We can understand the terms in this sum as being monopole operators for the Abelian group $T$ dressed by perturbative operators.  The bare monopole operator is the $\delta$-distribution at $m \in \op{Gr}_G$, and the dressed operators are obtained by differentiating some number of times.  Each derivative is an element of $T_m \op{Gr}_G$.   The space $\mc{V}_m$ are the operators built from the matter fields in the presence of the monopole $m$. 

The description of boundary monopole operators in terms of monopoles for the Abelian subgroup $T$ only works because introducing fugacities for $T$ into the index has the effect of giving a VEV to a diagonal gauge field.  This breaks the gauge group to $T$, and the off-diagonal components of the gauge field have mass.

To write our formula completely explicitly, we need to compute the three factors $\chi_{T \times \C^*} ( \mc{L}_m )$, $\chi_{T \times \C^*} ( \mc{V}_m)$, $\chi_{T \times \C^*} ( \Sym^\ast T_m \op{Gr}_G)$ that appear at each term in the sum.

\subsubsection{Contribution from the tangent space of the affine Grassmannian}
Let us first analyize $T_m \op{Gr}_G$ as a representation for $T \times \C^*$.

For every root $\alpha$ we let $\g_\alpha \subset \g$ be the root space.  The tangent space at $m$ to $\op{Gr}_G$ is 
\begin{equation} 
	T_m \op{Gr}_G =  z^{-1} \mf{t} [z^{-1}] \oplus \bigoplus_{\alpha} z^{(\alpha,m)-1} \g_\alpha[z^{-1}].
\end{equation}
To see this, consider a first-order version $m +\eps \phi$, for $\phi \in \g((z))$. Applying an infinitesimal gauge transformation by $X(z) \in \g[[z]]$ sends $\phi$ to
\begin{equation} 
	\phi(z) + [m(z), X(z)]. 
\end{equation}
If $\phi_\alpha(z)$, $X_\alpha(z)$ are the terms in the $\alpha$ root space, then this expression adds $z^{(\alpha,m)} X_\alpha(z) $ to $\phi_\alpha(z)$.  Since $X_\alpha(z)$ is an arbitrary series with no poles at $z = 0$, this means we can remove the terms in the series $\phi_\alpha(z)$ starting with $z^{(\alpha,m)}$ by a gauge transformation.   The action of $\C^*_q$ sends $z \mapsto q^{-1}z$ (this convention is necessary so that operators of positive spin are counted with positive powers of $q$).  

From this, we find that 
\begin{equation} 
	\Sym^\ast ( T_m \op{Gr}_G ) =  \Sym^\ast ( z^{-1} \mf{t} [z^{-1}] ) \bigotimes_{\alpha}\Sym^\ast \left(   z^{(\alpha,m)-1} \C[z^{-1}]   \right)\,,
\end{equation}
where $\mathfrak t$ denotes the Cartan of $\mathfrak g$,
so that 
\begin{equation} 
	\chi_{T \times \C^*} (\Sym^\ast T_m \op{Gr}_G) = \prod_{n > 0} \frac{1}{(1-q^n)^{\op{rank}(G)}}   \prod_{\alpha \in \op{roots}(G)}\frac{1}{1 - s^\alpha q^{n + \ip{m,\alpha} } }  
\end{equation}
where $s$ is a fugacity for the $T$-symmetry. Pleasingly, this matches one of the factors in the formula suggested for the boundary index in this case, namely Equation 3.31 of \cite{DGP-duality}. 

\subsubsection{The index of the matter}
Next, let us compute the contribution of $\mc{V}_m$ to the index.  We will assume our matter theory is Lagrangian.   Then, boundary operators are described by words in the boundary fields and their derivatives, where ``boundary fields'' means those not set to zero on the boundary.  For concreteness, one can imagine, as above, bulk chirals living in some representation $V$  of $G$, with Neumann boundary conditions.  The analysis applies to any Lagrangian theory, however.  

The presence of the boundary monopole means that a field fields charged under $T$ with weight $w$ are acquire poles or zeroes at the origin of order $(m,w)$. For example, if we have bulk chirals with Neumann boundary conditions in some representation $V$ of $G$, then the boundary fields are $\mbf{\Phi}\in \Omega^{(r/2), \bullet} (D,V)$ (where $D$ is a disc in the boundary).  If we decompose $V$ into weight spaces $V_w$, then the fields which live in $V_w$ become sections of $\Oo(-(m,w))$ in the presence of the boundary monopole.

We can trivialize the $\Oo(-(m,w))$ by the section $z^{- (m,w)}$.   This identifies $\mc{V}_m$ with $\mc{V}$, but in a way that does not preserve spin.   Instead, operators in $\mc{V}$ that have weight $w$ under $T$ have their spin shifted by $-(m,w)$.  

 Let us assume that the bulk matter representation $V$ has one-dimensional weight spaces, and let $\Phi_w$ be the field of the corresponding weight.  Let us assume, for simplicity, that the twisted spin $r$ of the field is zero.

Then, the boundary operators built from the bulk field $\Phi$ are generated as an algebra by the linear functions $\Phi \mapsto \partial_z^k \Phi(0)$. This vector space is 
\begin{equation} 
	\otimes_{w} \Sym^\ast (V^\vee_w \otimes \C[\partial_z]) 
\end{equation}
where the tensor product is over the weights of $V$.  Note that $V^\vee_w \otimes \C[\partial_z]$ is the linear dual of $V_w \otimes \C[[z]]$. 

In the presence of a boundary monopole, the field in $\Phi_w$ has Taylor expansion in $V_w \otimes \C[[z]] z^{(m,w)}$.  Dually, the operators which are linear functions of $\Phi_w$ live in $V^\vee_w \otimes \partial_z^{(m,w)}\C[\partial_z]$, and all operators built from $\Phi_w$ live in the symmetric algebra of this space. 

Operators which are linear functions of $\Phi_w$ have weight $-w$, and we have shifted their spin by $w \cdot m$.  Since the shift in the spin of any composite operator is determined by that on the operators linear on the fields, we find that for any operator $\mc{O}$ of weight $w$, the presence of the boundary monopole shifts its spin by $-(m,w)$.  

Writing this in terms of indices, we find that 
\begin{equation} 
	\chi_{T \times \C^*} (\mc{V}_m)( s^{w_1}, \dots, s^{w_r} , q) = \chi_{T \times \C^*} (\mc{V}) (s^{w_1} q^{-(m,w_1)}, \dots s^{w_r} q^{-(m,w_r)}  , q ).    
\end{equation}
In this expresion, we have chosen a basis $w_1,\dots,w_r$ for the weight lattice.  We can write the transformation of $s$ in a more short-hand way by $s \mapsto s q^{-m}$. This matches the matter contribution of equation 3.31 of \cite{DGP-duality}, up to the sign of $m$. 

Although we have explained this argument in the simple case of bulk chiral fields with Neuman boundary conditions, the argument is quite general. 
\subsubsection{The index of the line bundle}
The final, and trickiest, step in our computation is the contribution from the line bundle $\mc{L}$. To understand this, we need some details about how $\mc{L}$ is constructed.

We will use the definition of the line bundle presented in \cite{Zhu2016}. Every representation $R$ of $G$ can be thought of as a vector bundle on $BG$, and so gives rise to a class $p_1(R)$ in $H^4(BG)$.  Since $H^4(BG) = \Z$, we can choose some $R$ so that $p_1(R)$ generates $H^4(BG)$.  In the case of $G = SU(n)$, $R$ will be the fundamental representation.    We note that, by the definition of the dual Coxeter number $\sh$, we have
\begin{equation} 
	\op{Tr}_{adjoint} t^2 = 2 \sh \op{Tr}_R t^2. 
\end{equation}

A point on the affine Grassmannian is a $G$-bundle on a disc trivialized on the punctured disc.  We get an associated $R$-bundle $R_P$ on the disc, and we can take sections $\Gamma(D,R_P)$.  In the case that $P$ is associated to a cocharacter $m$ of $T$, this space of sections has a simple description. If $R_w$ are the weight spaces of $R$, the space of sections is
\begin{equation} 
	\oplus_w R_w [[z]] z^{(m,w)}.  
\end{equation}

There is a map
\begin{equation} 
	\Gamma(D,R_P) \to R[[z]] 
\end{equation}
obtained as the composition of the linear operators
\begin{equation} 
	\Gamma(D,R_P) \to \Gamma(D^*, R_P) = R((z)) \to R[[z]] 
\end{equation}
where the last map discards the polar part of a series.

The map $F : \Gamma(D,R_P) \to R[[z]]$ is Fredholm, and the fibre of the line bundle $\mc{L}$ at $P$ is defined, in section 2.4 of \cite{Zhu2016}, following \cite{Faltings2003} to be the Fredholm determinant:
\begin{equation} 
	\mc{L}_P = \op{det} ( \op{Ker} F)^{-1} \otimes \op{det} ( \op{Coker} F). 
\end{equation}
There are two possible sign choices here, giving the line bundle $\mc{L}$ and its inverse.  We want an ample line bundle, which has a section.  If we take $\mc{L}$ as defined, then $\op{det} F$ defines a section. Faltings \cite{Faltings2003} uses this section to demonstrate that the line bundle is ample. 

We are interested in the character of this under the action of $\C^*_q$ which gives $z$ weight $-1$, in the case that $P$ is associated to a cocharacter $m$.  In that case, the Fredholm determinant is clearly a product over all weight spaces of $R$. Each weight for which $(m,w)$ is negative contributes the inverse of the determiant of the vector space with basis
\begin{equation} 
	z^{(m,w)},\ z^{(m,w) + 1},\ \dots, z^{-1}. 
\end{equation}
Recall that $z$ has weight $q^{-1}$.  The character of the inverse of the determinant of this vector space is $q^{-\tfrac{1}{2} (-(m,w))(-(m,w) + 1)}= q^{-\tfrac{1}{2}((m,w))((m,w) - 1)}$.   

Similarly, if $(m,w) > 0$, the contribution is the determinant of the vector space with basis $1,\dots, z^{(m,w) - 1}$. This contributes the same factor $q^{-\tfrac{1}{2}((m,w)) ((m,w) - 1)}$.  We find that the contribution is independent of the sign of $(m,w)$.  

Note that
\begin{equation} 
	\sum_{w} (m,w) = \op{Tr}_{R} m = 0 
\end{equation}
because $G$ is simple.  It follows that the character of $\mc{L}_m$ under $\C^*_q$ is, summing over all weights,
\begin{equation} 
	q^{-\tfrac{1}{2} \sum_{w} (m,w)^2}. 
\end{equation} 
Next, let us compute the character under the action of $H$. We will proceed as above. For weights $w$ such that $(m,w) > 0$, we have the the determinant of a vector space of dimension $ (m,w)$, acted on by $T$ with weight $w$.  This contributes $s^{w ((m,w))}$.  Similarly, if $(m,w) < 0$, we find the inverse of the determinant of a vector space of dimension $-(m,w)$, acted on with $T$ with weight $w$.  This also contributes $s^{w ((m,w))}$.  So, in sum, we find $s^{\sum_w w ((m,w))}$, where the sum is over weights of $R$ with multiplicity.

Let us simplify this formula. The inner product on $\g$ we use to define the Chern-Simons level is given by $\ip{t_a,t_b} = \op{Tr}_R(t_a t_b)$ for $t_a,t_b \in \g$.    This inner product gives an inner product on the  Cartan $\mf{t}$, and so allows us to turn a co-root into a weight.  The co-root $m$ is sent to the weight $\sum \ip{m,w} w$ where the sum is over weights of $R$ with multiplicity.  If we map co-roots to weights in this way, then we can rewrite the character of the fibre of the line bundle $\mc{L}$ at $m$ is  $s^{ m} q^{-\tfrac{1}{2} m^2}$.   The factor that appears in the index is the character of $\mc{L}_m^{-k_{\text{eff}}}$, which is
\begin{equation} 
	s^{-k_{\text{eff}} m} q^{\tfrac{1}{2} k_{\text{eff}}  m^2}.   
\end{equation}

\subsubsection{The total index} 
Now let us put the three contributions we have determined together.  We find that the total index is
\begin{equation} 
	\left( \prod_{n > 0} \frac{1}{(1-q^n)^{\op{rank}(G)}} \right)	\sum_{m \in \op{coroots}(G)}  \prod_{n > 0}   \prod_{\alpha \in \op{roots}(G) }\frac{1}{1 - s^\alpha q^{n + \ip{m,\alpha} } } q^{ k_{\text{eff}} m^2 / 2} s^{-k_{\text{eff}} m} \mbb{I}_{matter} ( s q^{-m}, q) . 
\end{equation}
where $k$ is the Chern-Simons level and $m$ is the monopole charge. 

This is precisely the formula found in \cite{DGP-duality}, equation 3.31 (to match with that equation we must send $m \to -m$, which does not change the answer as we sum over monopole sectors).  If we have no Chern-Simons matter, this formula reproduces the character of the vacuum module of the chiral WZW model at level $k_{\text{eff}}$.

\subsection{Supersymmetry breaking}
It has been argued on physical grounds that $\CN=2$ pure gauge theory at level $\abs{k} < \sh$ has spontaneously broken supersymmetry.  This suggests that the Hilbert spaces and the spaces of bulk and boundary local operators of the twisted theory vanish.  

From a mathematical point of view, this is a non-trivial statement. Applied to the space of local operators, it tells us that, for $-2 \sh < k_{\text{eff}} < 0$, 
\begin{equation} 
	H_{\bullet,\dbar} (\op{Gr}_G, \mc{L}^{-k_{\text{eff}}}) = 0. 
\end{equation}
Since Dolbeault homology is the linear dual of Dolbeault cohomology, this implies
\begin{equation} 
	H^{\bullet,\dbar}(\op{Gr}_G, \mc{L}^{k_{\text{eff}}}) = 0 
\end{equation}
in the same range of $k_{\text{eff}}$. 

We were unable to find precisely this statement in the (extensive) math literature on this topic.  The closest we could find is the following. It follows from the results of S. Kumar \cite{Kumar1987} that the Dolbeault homology of the affine Grassmannian with coefficients in $\mc{L}^{k}$ is in a single degree.  From this we conclude that the vanishing of the index will imply vanishing of the space of boundary local operators.  The index formula does indeed vanish in the range $0>k>-h$, but is meaningless for $k\leq -h$ due to powers of $q$ which are unbounded from below. 

If $\Sigma$ is a closed surface, we would expect similarly that 
\begin{equation} 
	H^\bullet_{\dbar}(\op{Bun}_G(\Sigma), \mc{L}^{k_{\text{eff}}}) = 0 
\end{equation}
in the same range. If we take for $\op{Bun}_G(\Sigma)$ the moduli space of semi-stable $G$-bundles up to $S$-equivalence, this is not hard to show. The line bundle $\mc{L}$ is ample, and $\mc{L}^{- 2 \sh}$ is the canonical bundle.  By the Kodaira vanishing theorem, 
\begin{equation} 
	H^i( \op{Bun}_G(\Sigma), \mc{L}^{-2 \sh + k} ) = 0  
\end{equation}
for $k > 0$, $i > 0$.  By Serre duality, 
\begin{equation} 
	 H^i( \op{Bun}_G(\Sigma), \mc{L}^{ k} ) = 0  
\end{equation}
for $k < 0$, $i < \op{dim} \op{Bun}_G(\Sigma)$.  Therefore the cohomology vanishes in all degrees with coefficients in $\mc{L}^{\otimes k}$ as long as $-2 \sh < k < 0$, as desired.

If we treat $\op{Bun}_G(\Sigma)$ as a stack, the corresponding (much more difficult)  result was proved by  C. Teleman, \cite{Teleman1998}.  

\subsection{Mixed boundary conditions}
Next, we will briefly look at boundary conditions preserving a subgroup $H_c$ of the physical gauge symmetry $G_c$, to be complexified to an
$H$ subgroup in $G$. 

The crucial step, as usual, is to understand the disk phase space. We have a reduction of the structure group from $G$ to $H$ at the boundary. Correspondingly, we have a phase space
\be [H \backslash G](\CK)/G(\CO)\equiv \op{Gr}_{G,H}\ee
and we can build a sheaf $\mc{V}_{\op{Gr}_{G,H}}$ as 
\be \mc{V}_{\op{Gr}_{G,H}} = \left([H \backslash G](\CK)\otimes \mc{V} \right)/G(\CO)\ee
In conclusion, the answer should be 
\begin{equation} 
	H_{\bullet,\dbar}( \op{Gr}_{G,H},  \mc{V}_{\op{Gr}_{G,H}} \otimes \mc{L}^{-k^H_{\text{eff}}} ).  
\end{equation}
for an appropriate effective level $k^H_{\text{eff}}$.
\subsubsection{SQED $\leftrightarrow$ XYZ duality}
\label{sec:SQEDXYZduality2}

The simplest possibility we can consider is to combine Dirichlet boundary conditions for the gauge fields with Dirichlet boundary conditions for the 
chiral multiplets in SQED. The resulting chiral algebra is expected to be dual to the boundary chiral algebra of the XYZ model, with 
Dirichlet boundary conditions for $X$, Neumann for $Y$ and $Z$, combined with boundary free chiral fermions $\Gamma$, $\tilde \Gamma$. 
Recall that $Q \psi_X = \phi_Y \phi_Z$ but the OPE are otherwise trivial. 

For an Abelian gauge theory, the affine Grassmanian subtleties are absent and boundary monopole operators $M_n$ are labelled by their 
monopole charge $n$. The effective Chern-Simons coupling is such that the ``bare'' monopole operators $M_n$ have charge $n$ under the 
global $U(1)_\partial$ symmetry arising from the bulk gauge group as well as under the monopole charge $U(1)_{T}$. They also have twisted spin
$\frac{n^2}{2}$.

In particular, $M_{\pm 1}$ are natural candidates to match $\Gamma$, $\tilde \Gamma$. Naively, we can build a whole free fermion chiral algebra 
out of $M_n$ and $B$, just as it happens for the WZW$[U(1)_1]$ model appearing at a Dirichlet boundary of a pure $U(1)$ gauge theory 
with Chern-Simons level $1$. In particular, we can tentatively identify $B$ with $:\!\Gamma\wt\Gamma\!:(z)$.

Monopole operators behave as spectral flow modules for the $\psi$, $\tilde \psi$ boundary fields, meaning that the OPE of 
$\psi$ with $M_n$ starts at order $z^n$, while the OPE of $\tilde \psi$ with $M_{n}$ starts at order $z^{-n}$. The ``dressed'' monopole operators 
are built from $M_n$ and the appropriate modes of $\psi$, $\tilde \psi$. 

In particular, we have OPE 
\begin{equation}
\tilde \psi(z) M_1(0) \sim \frac{(\tilde\psi_{\frac12} M_1)(0)}{z} \qquad \qquad \psi(z) M_1(0) \sim \frac{(\psi_{\frac12} M_{-1})(0)}{z}
\end{equation}
The spin $J=0$ fields $(\tilde\psi_{\frac12} M_1)$ and $(\psi_{\frac12} M_{-1})$ can be identified with $\phi_X$ and $\phi_Y$.

Correspondingly, $\psi$ and $\tilde \psi$ match $\phi_X \Gamma$ and $\phi_Y \tilde \Gamma$. This gives them the correct OPE with $B$ = $M_n$. 
We leave a full analysis of this conjectural duality for future work.

\subsection*{Acknowledgements}

We thank Natalie Paquette and Sam Raskin for useful discussions. This research was supported in part by a grant from the Krembil Foundation. K.C. and D.G. are supported by the NSERC Discovery Grant program and by the Perimeter Institute for Theoretical Physics. Research at Perimeter Institute is supported in part by the Government of Canada through the Department of Innovation, Science and Economic Development Canada and by the Province of Ontario through the Ministry of Colleges and Universities. The work of T.D. was supported by NSF CAREER Grant DMS-1753077. T.D.'s work was performed in part at the Aspen Center for Physics, which is supported by NSF grant PHY-1607611, as well as during visits to the Perimeter Institute.

\bibliographystyle{JHEP}
\bibliography{3dTwists}

\end{document}